UNIVERSITY OF CALGARY

A Deep Learning Based Method for Fast Registration of Cardiac Magnetic Resonance Images

by

Benjamin Graham

A THESIS
SUBMITTED TO THE FACULTY OF GRADUATE STUDIES
IN PARTIAL FULFILLMENT OF THE REQUIREMENTS FOR THE
DEGREE OF MASTER OF SCIENCE

GRADUATE PROGRAM IN COMPUTER SCIENCE

CALGARY, ALBERTA
APRIL, 2024

© Benjamin Graham  2024

# Abstract

Image registration is used in many medical image analysis applications, such as tracking the motion of tissue in cardiac images, where cardiac kinematics can be an indicator of tissue health. Registration is a challenging problem for deep learning algorithms because ground truth transformations are not feasible to create, and because there are potentially multiple transformations that can produce images that appear correlated with the goal. Unsupervised methods have been proposed to learn to predict effective transformations, but these methods take significantly longer to predict than established baseline methods. For a deep learning method to see adoption in wider research and clinical settings, it should be designed to run in a reasonable time on common, mid-level hardware. Fast methods have been proposed for the task of image registration but often use patch-based methods which can affect registration accuracy for a highly dynamic organ such as the heart.

In this thesis, a fast, volumetric registration model is proposed for the use of quantifying cardiac strain. The proposed Deep Learning Neural Network (DLNN) is designed to utilize an architecture that can compute convolutions incredibly efficiently, allowing the model to achieve registration fidelity similar to other state-of-the-art models while taking a fraction of the time to perform inference. The proposed fast and lightweight registration (FLIR) model is used to predict tissue motion which is then used to quantify the non-uniform strain experienced by the tissue. For acquisitions taken from the same patient at approximately the same time, it would be expected that strain values measured between the acquisitions would have very small differences. Using this metric, strain values computed using the FLIR method are shown to be very consistent.

# Acknowledgements

I would like to acknowledge the support and guidance of my supervisor Dr. Christian Jacob, Dr. Usman Alim for allowing me to work with his lab and give me guidance in finishing my graduate studies while Dr. Jacob was away, and Dr. Alborz Amir-Khalili for his support in starting this thesis project and for providing some of the data that was used in the analysis in this thesis. I would like to thank my parents for their ongoing support in my academic career. I would like to thank the examination committee for their insightful and valuable feedback.

# Table of Contents

# List of Figures

# List of Tables

# List of Symbols, Abbreviations, and Nomenclature

| Symbol | Definition |
|--------|------------|
| 2D | 2 dimensional |
| 3D | 3 dimensional |
| ACDC | Automated Cardiac Disease Classifier |
| CNN | Convolutoinal Neural Network |
| CMR | Cardiac Magnetic Resonance |
| CPU | Central Processing Unit |
| DLNN | Deep Learning Neural Network |
| ED | End Diastolic |
| ES | End Systolic |
| GPU | Graphics Processing Unit |
| LGE | Late Gadolinium Enhanced |
| LV | Left Ventricle |
| ML | Machine Learning |
| MR | Magnetic Resonance |
| ReLU | Rectified Linear Unit |
| RF | Radio Frequency |
| RV | Right Ventricle |
| SSFP | Stead-State Free Precession |
| UKBB | UK Biobank |
| UMainz | Johannes Gutenburg University Mainz |
| $\phi$ | Displacement Tensor |
| $I_M$ | Moving Image |
| $I_F$ | Fixed Image |
| $I_W$ | Warped Image |
| $\Omega$ | Input image grid to the correlation coefficient |
| $L_{corrcoef}$ | Correlation coefficient loss |
| $L_{TV}$ | Total variation loss |
| $\omega_{corrcoef}$ | Correlation coefficient weight |

# Chapter 1

# Introduction

Cardiovascular disease is the leading cause of death in the United States, accounting for approximately 1 in every 2.7 deaths in 2006 [38]. Acute myocardial infarction, which is the blockage of the coronary arteries that supply the myocardium with oxygen-rich blood, affects approximately 3 million people worldwide annually [37]. More than 1 million deaths from myocardial infarction occur annually in the United States alone [37]. Currently, clinical assessment of acute myocardial infarction is performed using either echocardiography [48], or late gadolinium enhanced (LGE) magnetic resonance (MR) images [55]. The inherent limitations of these medical imaging techniques have created a demand for safe, novel methods for analysis of myocardial infarction with improved accuracy [36], [64], [40].

Cardiac magnetic resonance (CMR) allows for noninvasive assessment of the myocardium at high resolution with a large field of view [50]. The ability of CMR to acquire multi-modal images during a single examination has led to it being coined as "one-stop shopping" for medical imaging analysis [50]. MR imaging (MRI) produces images based on the interactions of the nuclear magnetic moments of hydrogen atoms with strong magnetic fields and therefore does not require the use of ionizing radiation. One of the most common CMR acquisition techniques for cardiac imaging is Steady State Free Precession (SSFP) due to its high signal-to-noise ratio (SNR) and relatively short acquisition time. One limitation of SSFP, however, is that it is unable to distinguish between healthy myocardial tissue and infarcted tissue. Currently, only LGE can visualize an infarction with CMR imaging [14]. An LGE acquisition involves injecting a patient with an intravenous gadolinium-based contrast agent which accumulates more in infarcted tissue compared to healthy tissue. While LGE is effective in analyzing scar tissue and edema, gadolinium is a heavy metal. As a result, the FDA issued guidelines limiting its use in 2017 [21], and work is being done to prevent and treat the deposits of gadolinium that have been found in the brain, bones, and organs of patients as a result of its use as a contrast agent [53]. Because of the inherent drawbacks of LGE and the popularity and safety of SSFP acquisitions, research efforts to aid in the visualization of cardiac infarctions with SSFP warrant further attention.

## 1.1 Hypothesis

This thesis proposes the Fast and Lightweight Registration (FLIR) model which is able to efficiently perform registration on cardiac MR images and is able to predict tissue deformation for the quantification of tissue strain. Designing an architecture that is able to quickly and efficiently perform registration could support the adoption of strain quantification as an alternative to LGE imaging.

## 1.2 Cardiac Motion Based Strain Quantification

To image areas of dead or scarred cardiac tissue, it is possible to infer the location of the unhealthy tissue by evaluating the motion of the tissue and using areas that experience abnormal motion as an indication of where unhealthy tissue might be. An established method to do this has already been developed for echocardiography. Myocardial deformation imaging originally became possible using tissue Doppler, which uses the Doppler effect in echocardiography to measure tissue motion [26], but recent developments have allowed for deformation tracking using speckle tracking. Ultrasound speckles are naturally occurring acoustic artifacts that occur as a result of interactions between ultrasound waves and cardiac tissue. Because the speckles are created due to an interaction with physical cardiac material, the speckles will move as cardiac tissue moves. Because the geometric shift of the speckles is correlated with the cardiac tissue; the strain of that tissue can be calculated [18]. Figure 1.1 illustrates how the speckles appear in an echocardiography image where the starting location of speckles is shown in green and the final position after motion is shown in red. Specialized speckle tracking software is used to track the motion of the speckles and quantify strain values.

Typically strain is measured in the radial, circumferential, and longitudinal directions which are shown as a projection of the left ventricle (LV) in Figure 1.2. Echocardiography is only able to create one 2D projection for each acquisition so generally, two acquisitions must be acquired, one viewing the heart along the side of the myocardium wall or long axis, and one viewing the heart from the base of the LV looking towards the apex, or short axis. An example long axis image is shown in Figure 1.3a and an example short axis image is shown in Figure 1.3b.

Cardiac tissue itself is composed of subendocardial and subepicardial fibers where subendocardial fibers contribute mostly to the longitudinal strain and subepicardial fibers contribute the most to the radial and circumferential strain. Longitudinal strain is typically the first indication of cardiomyopathy because the subendocardial fibers generally get compromised first [9]. By contrast, the subepicardial fibers generally stay preserved for longer meaning that abnormal radial and circumferential strain is a later indication of myocardium infarction. From this, we can assert that it is important to be able to effectively track the radial, circumferential, and longitudinal strain to determine the extent and progression of myocardium infarction [9]. There are some significant limitations to speckle tracking echocardiography; echocardiography only gives an image at one location which makes it very difficult to properly track the motion through the image plane even with multiple acquisitions, and because the speckles are artifacts that appear as a result of the acoustic wave created by the ultrasound transducer, it has been found that there is significant inter-manufacturer variability. [9] Fundamentally, ultrasound imaging measures tissue motion and doesn't directly measure tissue characteristics [19]. Additionally, ultrasound imaging requires multiple acquisitions to achieve volumetric imaging [31]. MR imaging, by comparison allows for direct assessment of tissue characteristics [32], and is also able to effectively create volumetric images [63]. This supports the motivation that strain quantification using MR images provides some benefits over speckle tracking echocardiography.

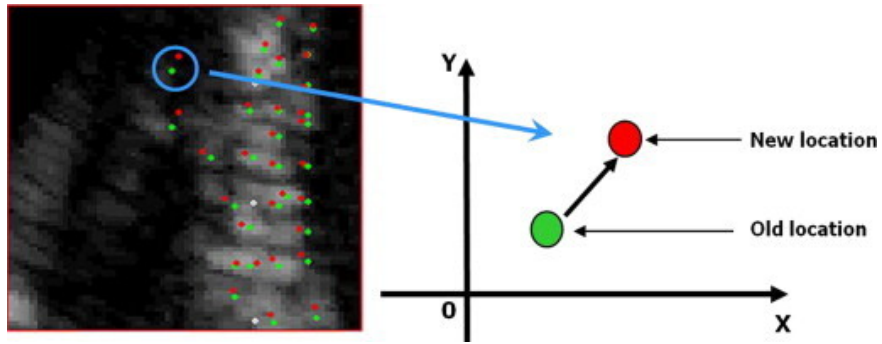

Figure 1.1: Illustration of speckles in an echocardiography image where the speckle's initial point is highlighted in green. The speckle's final position is highlighted in red [9].

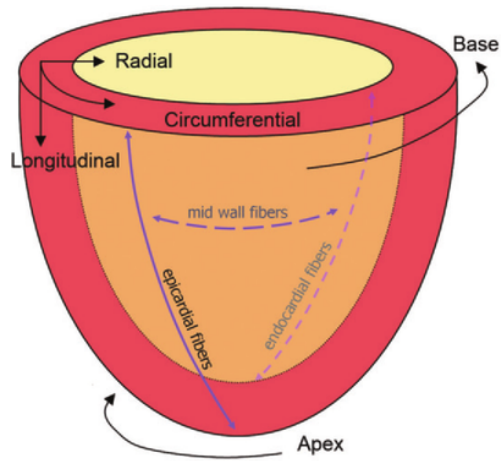

Figure 1.2: Illustration of an approximation of the left ventricle with the common directions along the myocardium wall that strain is calculated [29].

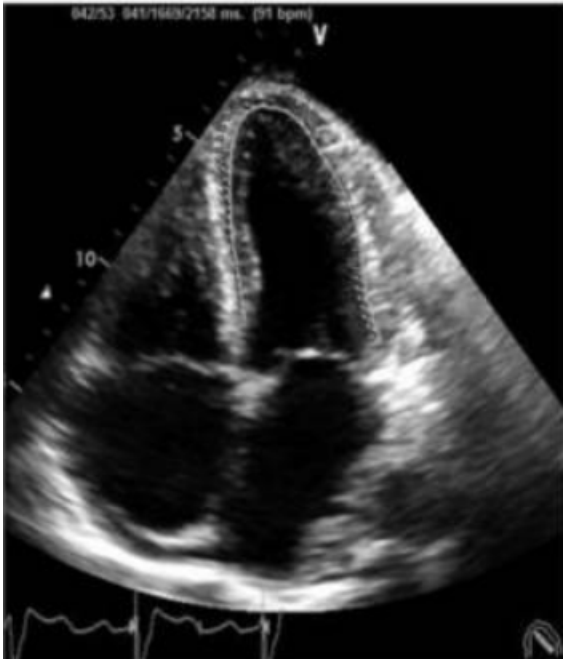 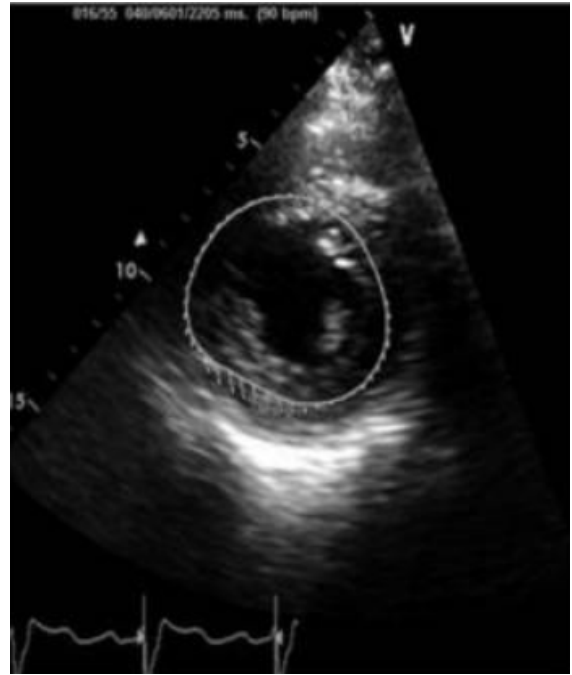

(a) Long axis echocardiography image [20]　　　　(b) Short axis echocardiography image

Figure 1.3: Sample long and short axis echocardiography acquisitions that would be required for strain quantification [20]

Using tissue deformation to calculate strain should be possible using any imaging modality that gives a time series of images. CMR is also able to provide an effective time series for a subject, but because it does not use acoustic waves like ultrasound, the Doppler effect cannot be used to quantify motion and speckles do not appear in CMR images. Instead, a method would need to be developed to track the motion of each pixel in the image. CMR can be used to produce a 3D volume by acquiring images of various slices of the subject and stacking them together. When referring to a 3D image created in such a way it is typical to refer to the image pixels as a voxel, which is the terminology that will be used in this thesis. The significant limitations to strain quantification with echocardiography mentioned previously are inherently solved by using CMR as well. Because CMR allows for the reconstruction of a 3D volume of the heart, this means that voxel displacement can be tracked far more accurately than with echocardiography. Displacements predicted using CMR should be tracked based on voxel signal intensity which also addresses the problem of inter-manufacturer variability. Because CMR is a non-quantitative method there are no standardized values for what signal intensity represents for MR imaging. Because of this, a model designed to work on CMR images should be able to predict deformations agnostic of the absolute values of the signal intensity which should mean that the model should be able to fit a variety of images and scanner parameters.

Deformation of the tissue can be quantified using a process called image registration. Registration is the process of taking two images and predicting a non-linear transformation that will convert one image to look like another image. Classical registration techniques are very demanding on computer hardware and take a very long time to compute which represents a significant bottleneck for its viability in research or clinical applications. For this reason, machine learning is being broadly researched for how it can be used as a registration method. Instead of needing to compute deformations at run time, a slow and difficult task even for high-end computers, a lot of the run time can instead be front-loaded into the training time of a machine learning model which makes the run time, or inference time, much faster.

Strain quantification based on deep learning-based registration of SSFP images, at the time of writing, is not FDA-approved for clinical use, however even as a research tool inference time is an important variable. For processing large datasets such as the UK BioBank (UKBB) dataset which features over 100,000 patient studies, slow-running models represent a barrier to entry for processing such data as it usually means a more powerful, and more expensive, machine is required to process the dataset in a reasonable time frame.

## 1.3 Thesis Contributions

Finding a fast and accurate way to estimate motion from a commonly obtained CMR acquisition such as SSFP would allow for strain quantification to be performed using an image acquisition that is already frequently used to assess cardiac function and would reduce the need for gadolinium, a harmful contrast agent [53]. Currently, the best-established way to quantify motion from a CMR SSFP acquisition is to use image registration as described previously. Image registration is a challenging problem for deep learning because ground truth transformations are not feasible to create, and because there are potentially multiple transformations that can produce images that appear correlated with the goal. In this thesis, a 3D registration model is proposed that can perform registration with a similar level of fidelity to other state-of-the-art methods but runs significantly faster at inference time. Many of the state-of-the-art deep learning-based registration methods can achieve improved fidelity by adding a significant amount of complexity to their network architectures which does improve fidelity but at the expense of inference time. As previously mentioned, inference time can represent a significant barrier to the acceptance of deep learning-based registration methods for strain quantification to be used for research and possible clinical acceptance of such models. The FLIR deep learning model is proposed that attempts to optimize for the mathematical operations that are used in running a deep learning model. A deep learning architecture is designed to optimize performance on cardiac MR SSFP images by utilizing domain knowledge to eliminate unnecessary operations and reduce the total number of operations that need to be performed on images at their full resolution. These improvements result in a deep learning model that achieves significantly faster inference times. Furthermore, our proposed method can achieve a similar improvement over VoxelMorph [7], a popular baseline method, as other state-of-the-art methods without dramatically increasing the inference time required.

## 1.4 Methodology

To quickly perform registration between two 3D volumes from the same CMR acquisitions, we can first assume that there should be minimal motion of the patient between the two volumes as significant patient motion results in image artifacts that would generally cause an acquisition to be reacquired or for specialized models to be used to adjust the images for the motion [70], [43]. Using the assumption that there should be minimal patient motion, the proposed model can forgo common pre-processing stages used by many state-of-the-art models to align images before performing registration. Next, the proposed model architecture itself is designed to be more time efficient in the convolution computation so that high-performance fidelity can be achieved without significantly increasing runtime. A preliminary experiment on the popular deep learning model, VoxelMorph, is performed to illustrate the strong effect on the runtime of using more layers that process the image at its largest resolution. Following this, a model is designed to use more layers that process the image at a lower resolution, and fewer layers at a high resolution. This should reduce the time required to run the model over a single iteration but will also decrease performance. By cascading multiple of these models together, however, high performance can be achieved while minimizing the additional runtime required. By using a cascading model architecture which has become popular in other state-of-the-art models, the proposed FLIR model can achieve a high level of registration fidelity comparable with the state-of-the-art while predicting tissue deformation in a much shorter time. Unlike other state-of-the-art models, however, FLIR can perform inference much faster by using the assumption that usable CMR SSFP acquisitions will have no patient motion to forgo common preprocessing steps, and by using an architecture that is designed to perform operations mostly on downsampled images which is shown to decrease inference time.

The evaluation of the model proposed in this thesis is based on the dice score, a commonly used similarity metric for comparing segmentations [12]. The datasets used for evaluation include segmentations of the LV myocardium, LV blood pool, and RV. By using the predicted deformation to warp a segmentation, the similarity between the segmentation of the goal image and the warped segmentation provides a consistent metric for measuring the fidelity of registration models as the warped segmentation will more closely match the segmentation of the goal image for a more accurate registration model.

An additional evaluation of the proposed registration network's effectiveness as part of a pipeline for strain quantification is performed. Strain values in this evaluation are computed on a data set of test-retest images. The test-retest data set includes two acquisitions for each patient. Any infarction present for a patient should be present in both image acquisitions so the ability of a method for evaluating strain can be determined by evaluating the difference between the strain values for the test and retest acquisitions. A 3D strain calculator is written to take as input the predicted deformations that are computed as a result of tracking tissue motion as a result of the FLIR model. The LV is assumed to be approximately a cylinder so the strain calculator will give computed strain values in both the Cartesian and cylindrical coordinate spaces, using both the Lagrangian and Eulerian frame of reference.

## 1.5 Overview

Chapter 1 presented a high-level discussion on the topic of volumetric strain quantification using CMR SSFP volumes, discussed the clinical value that could be added by reducing the need for gadolinium as a contrast agent, the challenges in quantifying motion using CMR SSFP data, the challenges in training a deep learning model for 3D dense image registration, and how the dense image registration can be used to quantify strain.

Chapter 2 provides a background on MRI acquisition methods, existing clinical methods for visualizing scar tissue, a background on registration as a process, a description of strain, and discusses how nonuniform strain can be calculated given a deformation. This chapter also gives a background on existing deep-learning registration methods that were used as the basis for the proposed network.

Chapter 3 describes the implementation of the proposed strain quantification method, including details of the implementation for the deep learning registration model and specific details on the strain calculator.

Chapter 4 details the experiments that were run, including data used, hyperparameters, and a description of the metrics that will be used to compare different models' registration performance as well as details of the experiment for evaluating strain performance. This chapter features analyses of the results of the evaluation of registration fidelity of the FLIR model compared to other state-of-the-art methods, and an analysis of the performance of the FLIR model when used as part of a strain quantification pipeline.

Chapter 5 summarizes the conclusions of the experiments and presents improvements of the model proposed in this thesis over state-of-the-art registration models. This chapter also discusses the limitations of the current work and the potential directions for future research.

## Chapter 2

# Background

This chapter describes the physical principles of CMR image acquisition, and myocardial infarction identification in CMR images. This is followed by a description of strain, including how strain is calculated for uniform deformation and the challenges with calculating strain for non-uniform deformation. This background information motivates the thesis objective: to develop a fast, optimized pipeline using deep learning-based image registration to predict 3D non-linear deformations for myocardial strain calculations.

## 2.1 Cardiac Magnetic Resonance Imaging

The cardiac magnetic resonance imaging modality provides clinicians with the ability to evaluate the structural and kinetic characteristics of the human heart. A CMR study generally takes 30-45 minutes to perform with the patient in the scanner. Because respiratory motion can significantly affect the quality of the produced images, the patient is required to perform multiple 7-15 second breath-holds and have a relatively regular heart rhythm [5].

MR images are acquired by taking advantage of the magnetic properties of the proton in a hydrogen atom, which exists in great abundance in the human body. The intrinsic nuclear magnetic moments of hydrogen nuclei are known to exist in two possible quantum states, designated as spin-up and spin-down. When hydrogen nuclei experience an externally applied magnetic field, $B_0$, these magnetic moments precess about the direction of the externally applied field at a characteristic frequency, known as the Larmor frequency, $\omega$. The Larmor frequency is given as $\omega = \gamma B$, where $\gamma$ is the gyromagnetic ratio proportional to the charge of the particle to its mass [49]. Because the spin-up or parallel state represents the lower energy state compared to the spin-down state, when describing the bulk behavior of the hydrogen atoms on the scale of the human body, slightly more protons will be aligned parallel to $B_0$ than anti-parallel. The common frame of reference looking at the interactions of protons in an externally applied magnetic field is to say $B_0$ points along the $z$-axis, with the plane formed by the $x$- and $y$-axes perpendicular to $B_0$ as shown in Figure 2.1.

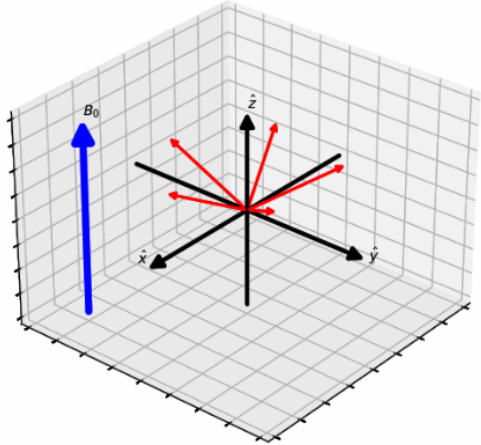

Figure 2.1: An illustration of a hydrogen proton precessing around a magnetic field $B_0$ aligned with the Z-axis.

The excess of proton magnetic moments precessing parallel to the direction of $B_0$ produces a net magnetization, $M_0$ in the same direction as $B_0$. The application of a second magnetic field, perpendicular to $B_0$, and rotating at the Larmor frequency causes $M_0$ to tip, or nutate away from the $z$-direction into the transverse $xy$-plane. At clinical field strengths of 1.5T and 3.0T, the Larmor frequency of hydrogen nuclei is 64 MHz and 128 MHz, respectively, which falls within the radio band of the electromagnetic spectrum. As a result, this secondary rotating magnetic field is often referred to as a radio-frequency (RF) pulse [61]. Following the application of this transient RF pulse, over time the excited protons return to their 'ground state', with $M_0$ once again aligned with $B_0$. The temporal evolution of the bulk magnetization from the excited state to the ground state is characterized by two simultaneously observable phenomena: T1 recovery and T2 relaxation. The T1 time constant represents the required time for the longitudinal regrowth of $M_0$ back to 63% of its original value along the $z$-axis. The T2 relaxation time describes the required time for the net magnetization in the transverse plane to dephase to 37% of its original value. The T1 and T2 measurements are illustrated in Figure 2.2 and Figure 2.3, respectively. Importantly, different tissues have different T1 and T2 relaxation times. Some sample typical T1 and T2 times are shown in Table 2.1 [49].

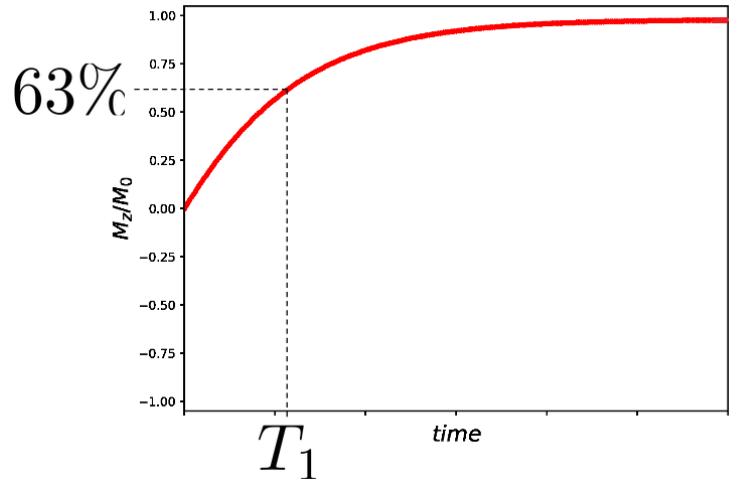

Figure 2.2: An illustration of how T1 time is measured in relation to net magnetization along Z axis.

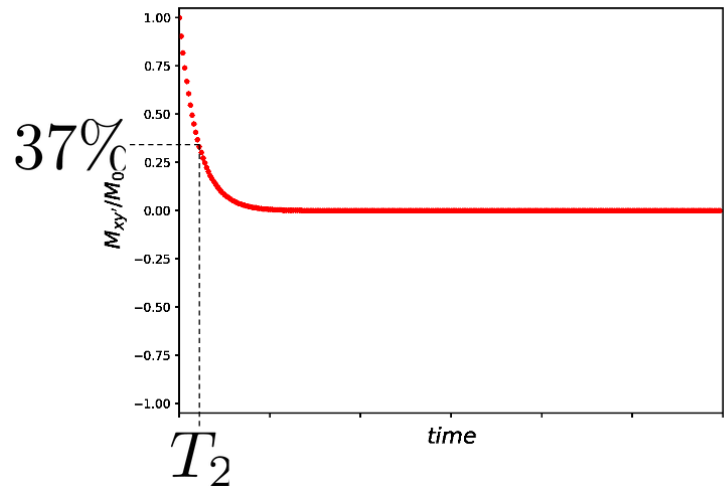

Figure 2.3: An illustration of how T2 time is measured in relation to net magnetization along X and Y axis.

| Tissue | T1 (ms) | T2 (ms) |
|---|---|---|
| Gray matter | 950 | 100 |
| White matter | 600 | 80 |
| Muscle | 900 | 50 |
| Cerebrospinal fluid | 4500 | 2200 |
| Fat | 250 | 70 |
| blood | 1400 | 200 |

Table 2.1: Sample T1 and T2 recovery times for various tissues at 1.5T [49].

Note that the T1 relaxation time is always longer than the T2 relaxation time. Because the excited bulk magnetization is still precessing, the decaying transverse magnetization varies with time. This changing magnetic field from the rotating magnetization induces a current in a pair of receiver coils in the scanner which is detected as a frequency signal. Each signal acquisition represents one line in 2D frequency space or $k$-space. Spatially varying gradient magnetic fields are applied to further localize the acquired signal across the object of interest, and acquire a 2D matrix of $k$-space lines. The inverse Fourier transform is then applied to the frequency domain signal to generate a usable medical image [49].

A 'pulse sequence' is a temporal description of the various RF pulses, magnetic field gradient pulses, and data sampling periods that are used to acquire data from a patient and reconstruct an image. Pulse sequences that use alternating gradient fields to acquire signal data are called gradient recalled echo (GRE) sequences. The time it takes to acquire one line of $k$-space is called the repetition time (TR) [2]. Motion from the chest due to breathing will introduce artifacts into an MRI image, thus patients are required to hold their breath during CMR imaging. For this reason, the repetition time needs to be short to reduce the amount of breath-hold time for the patient, which further improves the overall image quality. A shorter repetition time will also contribute to greater patient comfort.

### 2.1.1 Cine Steady-State Free Precession Images

First proposed for MR spectroscopy and later applied to MR imaging, Cine Steady-State Free Precession (SSFP) is a GRE sequence in which a steady, residual magnetization in the $xy$-plane is maintained between successive cycles [15]. The sequence is notable for its ability to create time-quality time-series images for the analysis of cardiac function. In general, SSFP is often the best sequence to assess cardiac function because it requires no contrast agent and requires short breath holds for the patient. As discussed in Section 2.1, the magnetization of hydrogen protons has two main components, $M_z$ and $M_{xy}$ that are independent of each other. Over several sequences, a steady-state equilibrium can be achieved with constant magnitudes of $M_z$ and $M_{xy}$ at the beginning of each cycle. This steady state can be achieved in tissues with a sufficiently long T2 interval, by keeping the TR shorter than T2. Because the TR must be shorter than the T2, closer to the order of the T1 time, to achieve a steady state, this contributes to SSFP being one of the most time-efficient sequences to obtain for the high-quality signal-to-noise ratio (SNR) provided [56, 45]. As we can see in Table 2.1, T1 and T2 times vary across different types of tissue. The health of the tissue, which is if it contains a scar, is necrotic, or healthy, does not significantly affect T1 or T2 times [14]. This is because T1 and T2 times are mostly affected by hydrogen proton density and scar or dead tissue will still have approximately the same hydrogen proton density as healthy tissue. This means that scar and dead tissue will not appear in images obtained with an SSFP GRE sequence.

### 2.1.2 Late Gadolinium Enhanced Images

Late Gadolinium Enhanced (LGE) permits optimal differentiation between normal and diseased myocardium with the use of extracellular gadolinium-based contrast agents and a special pulse sequence. Despite gadolinium being a toxic heavy ion, gadolinium-based contrast agents are rendered biologically inert through a chemical chelation process, and cannot normally cross cell membranes. In normal myocardium, tissue cells are densely packed and uptake of gadolinium into healthy tissue will be minimal. In tissue with acute myocardial damage, however, cellular membrane breakdown allows gadolinium to diffuse into what was previously intracellular space. This means that in damaged tissue there is increased gadolinium concentration which has a significantly shorter T1 than healthy cardiac tissue. Similarly, in cases of chronic myocardial damage, cardiac tissue cells have been replaced with a collagenous scar which also increases the intracellular space, allowing more gadolinium to enter the scar tissue [62].

To illustrate how the appearance of scars and infarctions is dependent on the pulse sequence, Figure 2.4 shows images from two sequences for the same subject where both sequences are enhanced with gadolinium. Figure 2.4a is a pulse sequence called a spin echo (SE), enhanced with gadolinium. Compared to the inversion recovery gradient echo (IR GRE) in Figure 2.4b, the black arrows on each point to regions of identified scar tissue. This shows how, even with a contrast agent, a scar and different tissues can appear dramatically differently in a CMR image based on the pulse sequence used.

LGE images, specifically using an IR GRE pulse sequence are highly effective for the identification of myocardium infarction, but the use of gadolinium as a contrast agent is controversial. It is now known to leave deposits in the brain, bones, and other organs of patients. Work is being done to prevent and treat specifically deposits of gadolinium in patients who have had it used as a contrast agent. In 2017, the FDA

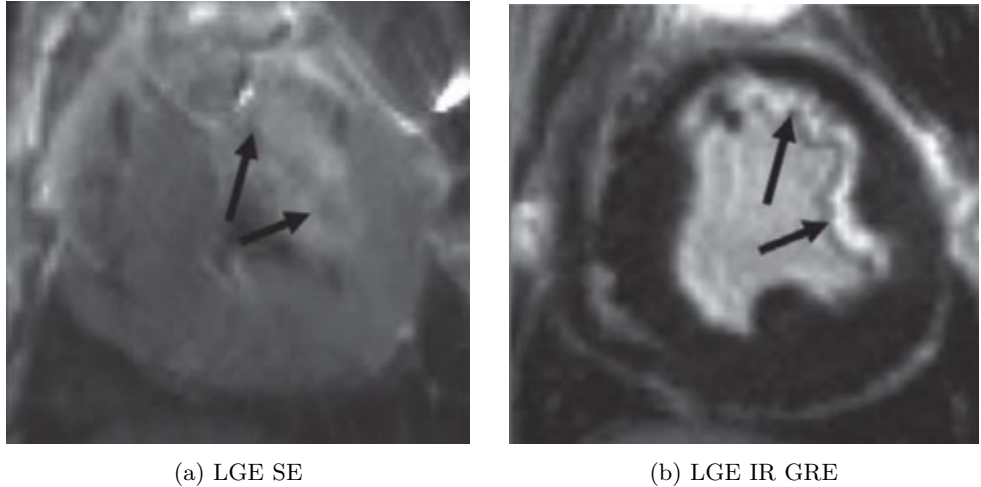

(a) LGE SE  (b) LGE IR GRE

Figure 2.4: LGE SE and LGE IR GRE sequences myocardium infarction identified with arrows [62]

issued guidelines recommending that the use of gadolinium as a contrast agent should be limited.

## 2.2 Speckle Tracking Echocardiography

There is an existing method for using strain to infer dead and scarred cardiac tissue. Strain quantification is an established technique used in echocardiography that has been used for over ten years and is seeing more and more use. To measure strain, the speckle tracking method is used, which is a non-invasive method, independent of the limits of the Doppler signal. The 'speckles' are acoustic artifacts generated by interactions with the acoustic wave from the ultrasound transducer used to generate the image and the myocardium. These artifacts can be followed in their displacement during the cardiac cycle. These speckles are statistically equally distributed throughout the myocardium and their size is about 20 to 40 pixels [48]. Recall Figure 1.1 where the speckles' original position is highlighted in green and their position after motion is highlighted in red. Using software, the speckles can be tracked from frame to frame and when the frame rate is known, the change in speckle position allows the tissue's velocity to be determined. From this, the motion pattern of myocardial tissue is reflected by the motion pattern of speckles. By averaging the various regional strains, it is possible to calculate the global longitudinal strain (GLS). Normal reported GLS values, for major echocardiographic systems, are expected to range from 18% to 25% for healthy individuals [48]. One inherent limitation of speckle tracking echocardiography, however, is that it can only be used to calculate strain in 2D. Strain values can only be calculated along the cardiac wall because the speckles cannot be tracked through the image plane. This means that a lot of information about circumferential strain would be lost without separate acquisitions such as the ones shown in Figure 1.3.

Strain is the unitless change in the length of an object during a given period. Strain rate is the rate at which the change in length occurs and has units of $s^{-1}$. Whether the object is shortening or lengthening affects the measured strain value. During systole, when the heart is at its most contracted, the strain and strain rate would be expected to have negative values, but when there is a stretch or lengthening of the myocardium, the strain and strain rate are expected to become positive [9]. In practice, the myocardium increases in thickness during systole, while decreasing in circumference. Recall the directions that radial, circumferential, and longitudinal strain are measured in Figure 1.2. When the myocardium wall increases in thickness, this would mean radial strain is expected to be positive, but because the circumference is decreasing, circumferential strain is expected to be negative. Similarly in systole, the length of the left ventricle decreases, so the longitudinal strain is also expected to be negative. An example of this is shown in Figure 2.5. These charts show each the evaluated strains along the Y axis and time on the X axis, showing the strain that the tissue experiences over time along each axis. In this chart, the normal values for longitudinal strain and circumferential strain all dip to negative values while the radial strain is positive. Note as well in Figure 2.5 that the separate long axis acquisition is required for longitudinal strain whereas the short axis acquisition is required for the circumferential strain.

## 2.3 Registration

Registration refers to techniques that can extract a motion from a pair or sequence of images. In general, registration is the problem of finding some displacement vector, $\phi$, sometimes called the flow field, such that for two images, $I_M$ and $I_F$, $I_F(x) = I_M(x + \phi(x))$. The flow field denotes where each voxel in image $I_M$ must move as closely as possible to create the image $I_F$. $I_M$ is referred to as the moving image, and $I_F$ is referred to as the fixed image. The new image created by applying $\phi$ to the moving image, $I_M$ creates a warped image, $I_W$. The primary goal of registration is to predict a flow field $\phi$ that results in a warped image $I_W$ that resembles the fixed image $I_F$ as closely as possible.

An illustration of the registration process is shown in Figure 2.6. A predicted deformation, $\phi$ is applied to a moving image, Figure $I_M$, creates a new warped image, Figure $I_W$, that matches the fixed image, Figure $I_F$, as closely as possible.

The deformation is applied to the moving image. Generally, the deformation predicted by a registration model will take the shape of a tensor of vectors where each vector represents the displacement that each pixel or voxel must undergo.

Registration methods can be generally grouped into rigid image registration and non-rigid registration. Transformations predicted by rigid image registration are the simplest and in 3D space can be represented by 6 degrees of freedom, 3 translation, and 3 rotation. These degrees of freedom represent the ways the image can be transformed using rigid registration. This means a 3D rigid transformation can only translate the volume in the $X$, $Y$, or $Z$ directions and can only rotate the volume along those same axes. Non-rigid registration is significantly more complex and includes similarity, affine, projective, and deformable transformations. Similarity transformations represent transformations that include translation, rotation, and uniform scaling. An extension of this is the affine transformation, which includes translation, rotation, scaling, and shear. From these definitions, non-rigid and similarity transformations are subsets of affine transformations [44]. Generally, the goal of projective transformations is to show how a given object will change its appearance when the viewpoint of an observer changes. For the domain of medical image registration, projective transformations aren't generally necessary so the two types of registration that are often focused on in this domain are affine and deformable registrations.

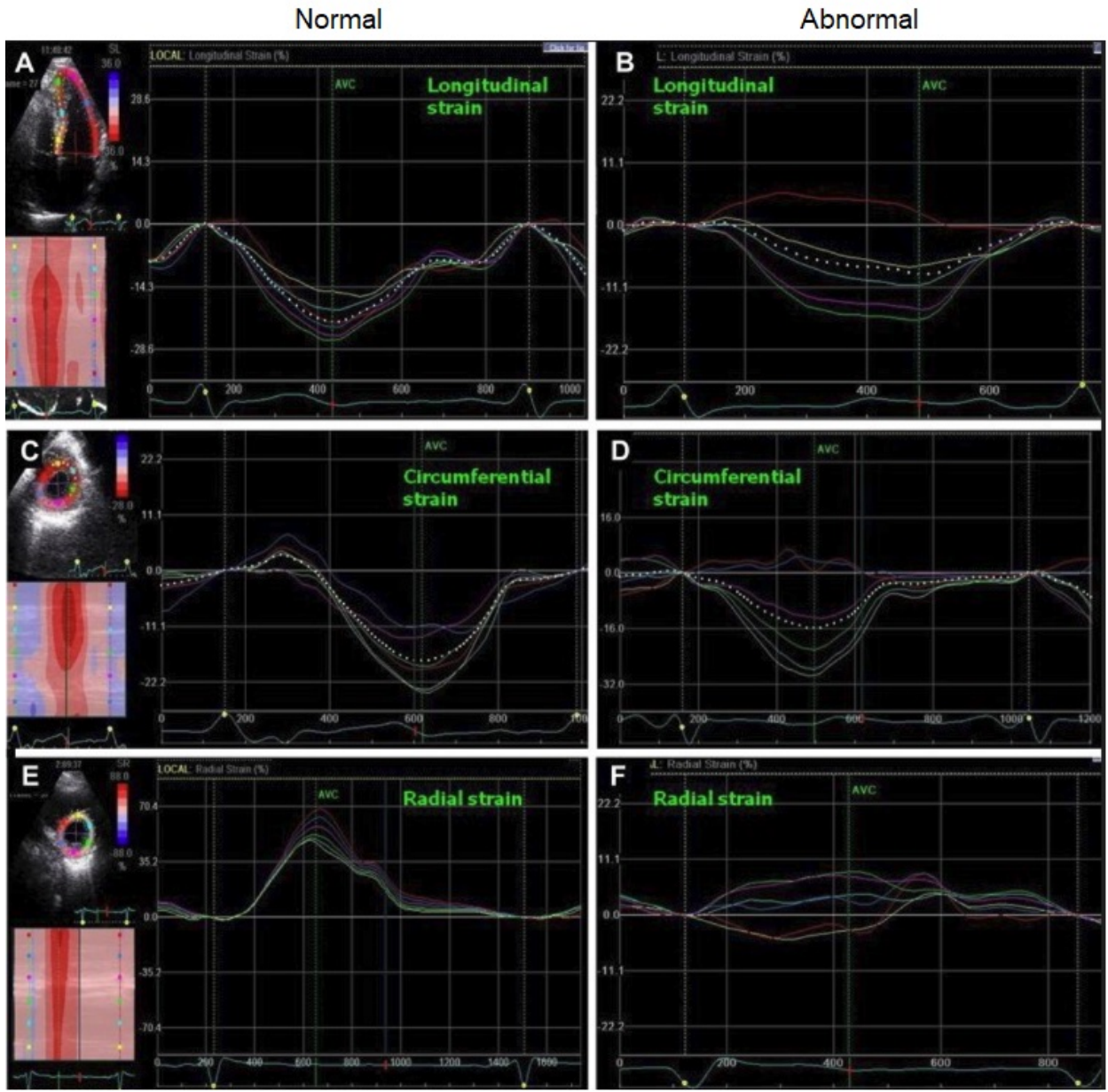

Figure 2.5: Examples of normal longitudinal, circumferential, and radial strain values are shown in charts A, C, and E, respectively. Abnormal longitudinal, circumferential, and radial strain values are shown in charts, B, D, and F, respectively [9].

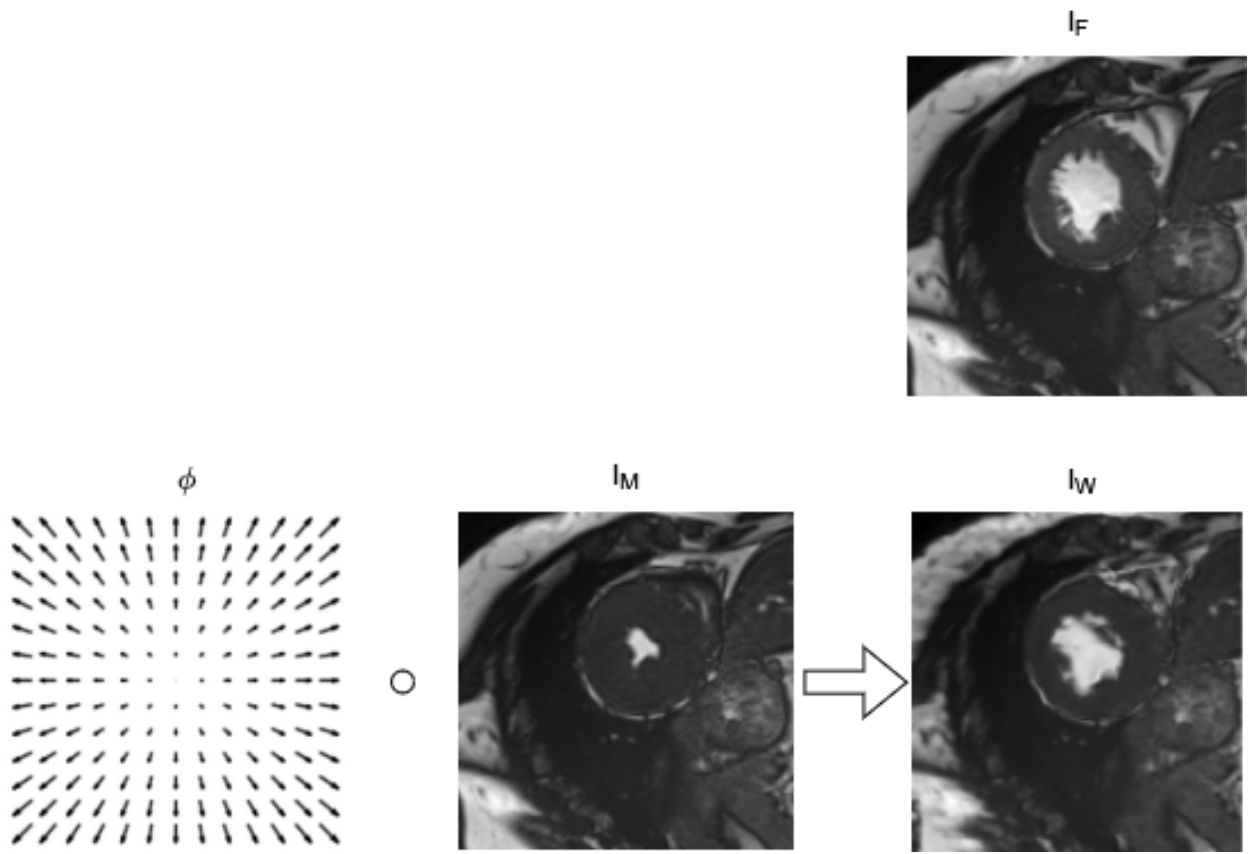

Figure 2.6: Illustration of image registration. For a given fixed image, $I_F$, and a moving image, $I_M$, a deformation $\phi$ is predicted such that when the deformation is applied to $I_M$ it generates a warped image, $I_W$. $I_W$ should look as close to $I_F$ as possible.

Affine registration is often used either for applications where the subject is entirely rigid such as bone, or as a preregistration before a more complex deformable registration. In the following chapters, it is shown that many state-of-the-art models for performing registration will have portions of their architecture dedicated to performing affine registration. The goal of having an affine registration step is to ensure that the subjects in the image are aligned and scaled to each other correctly first so that the portion of the model dedicated to deformable registration will be able to focus solely on predicted non-linear deformations [44].

One common method for medical image registration models is free-form registration, where a model assumes that any transformation is possible [3]. Free-form registration will interpret an image as a grid of control points where each point is moved individually based on displacement vectors as described previously. When moving the control based on this vector it is possible for their final coordinates to not be integer values, which would mean the control point is placed in the space between pixels, or the 3D case, voxels. To resolve this, interpolation methods can be used to reconcile which grayscale values should be set to different pixels. The specific interpolation method used depends on whether the deformation is being applied to a raw image or a segmentation. If the deformation is being applied to the moving image itself then bilinear interpolation needs to be used if it is a 2D image and trilinear interpolation if the deformation is being applied to a 3D volume. If the deformation is being applied to a segmentation of an image, to preserve the segmentation, nearest neighbor interpolation needs to be used. Bilinear interpolation is the method of taking a given value at a point and setting its value to be the weighted average of the 4 nearest adjacent pixels and setting that as the new value. Trilinear interpolation is an extension of bilinear, where bilinear interpolation is used for 2D images, and trilinear interpolation is used in 3D volumes by taking 6 adjacent voxel values. It can be thought of as taking the result of bilinear interpolation at opposite faces and then performing a 1D linear interpolation, or weighted average, between those. This method for interpolation works well when interpolating on a raw image where there are not expected to be rigid boundaries of intensity, but intuitively these methods don't work for segmentations where a rigid boundary with specific voxel intensities is expected. For example, if a voxel with a signal intensity of 1 is deformed so that it is in the space between two voxels, those two voxels will have their values set to 0.5 following bilinear or trilinear interpolation. If the interpolation is being applied to a segmentation, however, where it is expected that each voxel is either 0 or 1, having voxels with decimal values makes the segmentation unusable. Instead, if the interpolation is being applied to a segmentation, the simpler method of nearest neighbor is used. In nearest neighbor interpolation, the value of the closest voxel is used with no consideration to any other nearby voxel values, compared to bilinear or trilinear interpolation where a distance weighted average of all the neighboring voxels is used. This ensures that rigid boundaries of segmentations are preserved.

The common alternative method to free-flow is elastic models, which represent images as elastic solids [3]. The core idea behind elastic models is that there are internal elastic forces that oppose the deformation which work against the external deformation being applied until the system reaches an equilibrium. Other elastic methods operate similarly to finite element modeling where an input image is broken up into smaller cells and each cell is given physical tissue properties [22].

The goal of this thesis is to find a method for performing registration that is both fast and predicts deformations that represent realistic deformations. For this purpose, free-flow methods are investigated because of the significant computation time that would be added by including calculations based on tissue elasticity to the registration model. Intuitively, because the goal of the model described here is to identify areas of myocardium infarction based on tissue deformation, making assumptions about the elasticity of tissue may affect the model's ability to identify areas of scar tissue effectively.

Registration is typically a slow and computationally intensive process. Existing methods for free-flow registration, such as iterative closest point (ICP), compare every point in two-point clouds or meshes and attempt to find the minimal distance that needs to be moved between them [71]. Even on modern computer hardware, this is a very slow process and it also requires that the inputs be meshes and not raw images, which makes it infeasible for this thesis as we are interested in finding a flow field between two images. Additionally, for the model to be clinically valuable, the problem of long processing times for registration methods should be addressed. Inference time can be improved by front-loading a lot of the processing by using a machine learning model to perform the registration process. Machine learning, specifically deep learning, for registration is described later in this chapter.

## 2.4 Principles of Artificial Intelligence

### 2.4.1 Artificial Intelligence

John McCarthy offers the definition that "AI is the science and engineering of making intelligent machines, especially intelligent computer programs". It is related to the similar task of using computers to understand human intelligence, but AI does not have to confine itself to methods that are biologically observable [4]. Artificial intelligence (AI) is the area of knowledge that develops models programmed to learn and identify patterns from "training data" that can be subsequently applied to new datasets without being explicitly programmed to do so.

AI can be used to automate difficult tasks and improve performance by systemically collecting relevant knowledge. Its use is becoming popular in medicine because of its ability to speed up processes and achieve a high degree of accuracy, contributing to improved patient care [39].

Over the past few decades, AI has been developed to address a variety of needs in a wide range of industries. In the context of CMR, AI prediction strategies have been used to automate multiple clinical workflows. For example, manual delineation of image contours by experts is currently the standard clinical practice in CMR [65]. This is a labor-intensive task and prone to variability among clinicians. Winther et al. performed experiments using datasets from four independent sources for training and validation of an AI model. The model proved to be capable of reliably producing high-quality segmentations independent of aspects such as different image acquisition techniques and diverse MRI protocols and vendors. The AI model was able to perform equally or outperform human cardiac experts in measurements of the left ventricle and right ventricle [65]. AI can also be used clinically for a recently applied technique called texture analysis. Texture analysis employs various algorithms and models to quantify spatial heterogeneity and the relationship of adjacent pixels to compute imaging metrics [35]. Texture analysis is effective at differentiating hypertrophic cardiomyopathy from hypertensive heart disease, whereas conventional methods of global T1 analysis typically can't [69]. Mancio et al. have used AI-based texture analysis to analyze routine time series images to identify subtle tissue alterations and identify patients with a low probability of having scar tissue. This could potentially be used to identify patients where LGE imaging would not be necessary because as discussed previously, LGE imaging has been shown to have harmful long-term effects and its use should be limited [42]. AI applications in CMR have also contributed significantly to the acceleration of image acquisition and analysis. AI models have been applied to reconstruct data from rapidly acquired undersampled MRI images across different sequences where an AI-based super-resolution CMR angiography framework has enabled the reconstruction of low-resolution data acquired in 50 seconds scan time [13]. Additionally, by using an AI-based approach, Qi et al. were able to achieve a 9-15x speed up in the acquisition of 3D time series images in a 10-15 second breath hold [51].

### 2.4.2 Machine Learning

Although colloquially the terms "machine learning" (ML) and AI might be used interchangeably, it is more accurate to think of ML as a subset of AI. AI is generally the field of study of creating smart systems, and ML more specifically describes algorithms that can learn from data and make decisions based on patterns observed [25]. ML algorithms can be classified as supervised, unsupervised, semi-supervised, and reinforcement learning [6].

In supervised learning contexts a dataset used for training can be thought of as having $N$ number of samples, where each sample has $x$ features and some label $y$. The goal of supervised learning is generally to be able to predict a label $y$ given some features $x$ that it has not seen before [6]. Supervised learning is classically the family of ML models used for tasks in the medical domain such as classification and image segmentation [16]. As the data is fed into a supervised ML model, the model can adjust its weights until the model has been fit based on parameters defined when designing the model. Unsupervised learning, by comparison, makes use of an unlabelled dataset to create a model that takes as input $N$ samples that only contain features $x$ [6]. The goal of unsupervised learning is often to transform the features $x$ in some way based on how the model is defined [47]. This can include reducing the dimensions of the input data, finding patterns or groups in the input data that had previously not been identified, or identifying large outliers in data [59]. Semi-supervised learning, as the name suggests, contains a mix of labeled and unlabelled data. The motivation for adding unlabelled data is to improve the resulting model by providing a more comprehensive idea of the probability distribution of the labeled data [60]. Lastly, reinforcement learning interprets variables in a system at different states during a given process. Decisions made by the ML model in reinforcement learning are rewarded based on a defined reward function when the model achieves "success" for a given state of the system. Reinforcement learning is meant as an iterative process where repeating this process with different decisions moves the model into different states and allows the machine to determine which decisions should be made to achieve success [58]. As an example, reinforcement learning is often used in games to create models that can solve puzzles or respond to changes in a player's behavior [10].

### 2.4.3 Deep Learning

Deep learning is a subset of machine learning, which focuses on deep learning neural networks (DLNNs), i.e., neural networks with three or more layers: an input layer, at least one "hidden" layer, and an output layer [41]. These neural networks attempt to simulate the behavior of the human brain by allowing it to learn from large amounts of data. While a neural network with a single layer can still make approximate predictions, additional layers can help to optimize and refine for accuracy. Neural networks are so-called because of their interconnected structure of 'neurons' or nodes. Each layer builds upon the previous layer to refine and optimize the prediction. Each feature in a sample being used for either training or testing is fed into the first layer of a neural network. This means that the first layer of a neural network must have the same shape as the sample that is being fed into it. After the first layer can be any number of hidden layers, where each neuron in the hidden layers calculates an output using an activation function that takes as input some or all of the features from the layer before it, each multiplied by some weight. The weights are calculated by a process called backpropagation where an algorithm will calculate the error in a given prediction and then adjusts the weights to each node to train the model to achieve better predictions. If each node in a layer takes as input the output of each node from the layer before it, the layer is said to be "fully connected". The output of the layer of a neural network will be whatever shape the network is designed to give as an output. For example, if the network is designed to perform classification the network could be designed to either have only one node at its output with the value at that node associated with a label, or it could give its output as "one hot encoded" and have a number of nodes equal to the number of possible labels with each node outputting a 0 except for the node associated with the label being predicted.

An example architecture of a deep neural network is shown in Figure 2.7. In this example, the nodes in the visible layers are highlighted in blue and the nodes in the hidden layers are highlighted in green.

### 2.4.4 Convolutional Neural Networks

Convolutional Neural Networks (CNNs) are a type of DLNN that is designed to work particularly well on image processing tasks [33]. CNNs are often built using several types of layers: convolution layers, pooling layers, and fully connected layers. Typically a CNN will use a convolution layer followed by a pooling layer, with that pattern repeating several times [23].

**Convolution Layer**

A fundamental component of CNNs, convolution layers generally consist of a convolution operation and an activation function. Convolution is a linear operation used for feature extraction where an array of values make up a filter, called a kernel [67]. The kernel is applied across the input which should have the same number of dimensions as the kernel' i.e., a 2D input should use a 2D kernel, and a 3D input should use a 3D kernel. Commonly for a 2D input a kernel will be 3x3. The values in the kernel are defined by what filtering operation is intended, for example, a kernel designed to detect edges will have different values than a kernel meant to blur the input. To apply the kernel to the input the kernel is placed at the beginning of the input "on top" of the elements. Geometrically, for a 2D input the beginning can be thought of as the top left. Then for each element of the input where the kernel is, that element is multiplied by the corresponding element of the kernel. Each product from the multiplication is added together because of the result of that kernel operation. The kernel is passed over the entire input, performing its multiplication and then the sum of the product at each location until it has filtered the entire input [67]. The distance the kernel moves between each operation is called its stride, but a stride of 1 is typically used. Intuitively, filtering an input with a kernel will shrink the input, and a stride greater than 1 will shrink the input even more. Shrinking the input can be desirable, and is the purpose of the pooling layers discussed in the next section. For a convolution layer, however, generally, the shape of the input should remain unchanged. To keep the input shape the same the input is generally padded with 0's such that the center of the kernel is over the very first element of the input array and a stride of 1 is used.

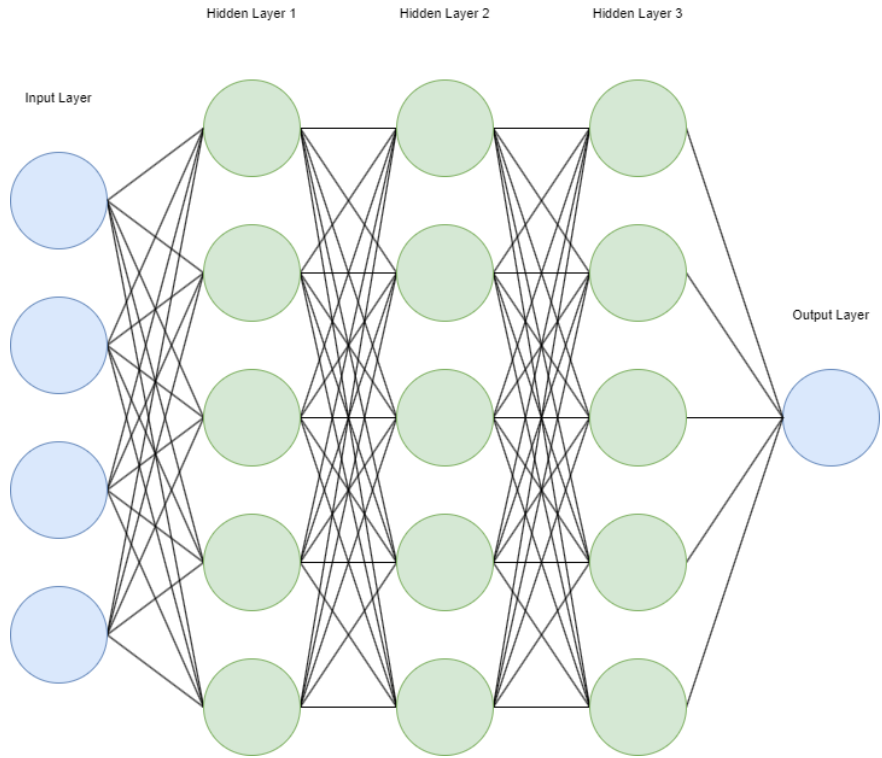

Figure 2.7: A deep neural network. The visible input and output layers are shaded blue with the hidden layers shaded green.

In Figure 2.8 an example of convolution is shown where a 3x3 kernel is passed over a 5x5 input that is zero-padded to retain its original shape.

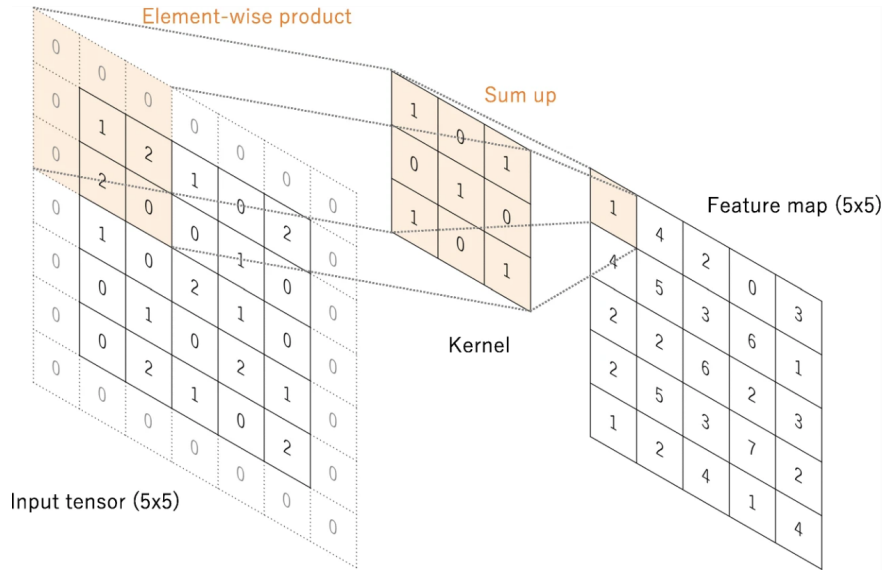

Figure 2.8: An example of convolution, using zero padding to retain the original shape of the input [67].

**ReLU Activation Function**

One of the most common activation functions is the rectified linear unit (ReLU) and is very effective for tasks in computer vision and bioinformatics [8], [24]. ReLU can be described as the function $f(x) = \max(0, x)$. It sets a threshold for values at 0 such that any value less than 0 is returned as 0 and a linear function for any value greater than 0 [1]. The ReLU activation function has been known, however, to create a problem referred to as "Dead Neuron". When the input to the ReLU function is negative the output will always be zero, meaning that the first derivative is also zero and the neuron will be unable to update any parameters. To assist with this limitation, Leaky ReLU is introduced that adds a factor $k$ to negative input. Leaky ReLU is defined as the following function [34]:

$$LeakyReLU = f(x) = \begin{cases} x, & \text{if } x > 0. \\ kx, & \text{if } x \leq 0. \end{cases} \quad (2.1)$$

This adds a small slope to the negative input so the derivative is never zero which means that the neurons are still able to slightly learn even after the function enters the negative interval [66]. Typically 0.01 is used as a value for $k$.

**Pooling Layer**

Pooling layers provide a basic down-sampling operation that reduces the shape of an input to introduce a translation that is invariant to small shifts and distortions, to decrease the number of parameters that a network must learn, and to merge semantically similar features into one [67], [46]. Two noteworthy pooling methods are max pooling and global pooling, with max pooling being the most popular form of pooling [67]. Similar to convolution a filter is passed over the input to the max pooling layer. For max pooling; instead of taking a product with the filter, the pooling filter will take the highest value in the patch being sampled. The filter will have a large enough stride that no elements of the input are sampled multiple times. What this commonly looks like in practice is a 2x2 max pooling filter with a stride of 2, which will sample 4 elements at a time and choose the largest, reducing the size of the input by a factor of 2. A 4x4 input in this example will be down-sampled to a 2x2 input. An illustration of max pooling is shown in Figure 2.9.

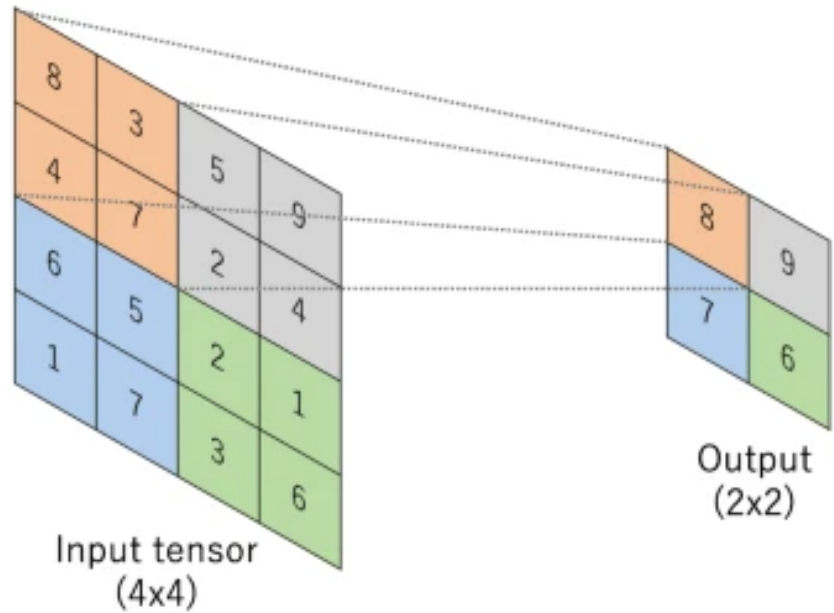

Figure 2.9: An example of max pooling of a 4x4 input tensor that is down sampled into a 2x2 tensor [67].

**Fully Connected Layer**

After a series of convolution and pooling layers, the output is typically transformed into a one-dimensional array and connected to one or more fully connected, or dense layers. Recall that this is the type of layer described in Section 2.4.3 where each node in a layer is connected to each node in a layer before it.

**Final Activation Function**

Typically the activation function of the output layer should be chosen to put the output into a format based on what the output is intended to be. To use the example of classification in Section 2.4.3 where a CNN is designed to classify an input into one of a predetermined number of categories, there should be a node for each category possible to generate an output for. In this case, for example, a softmax layer should be used which will cause the value at each node in the output together to sum up to 1. This will mean that each node will represent the probability that the input is the label that each node represents.

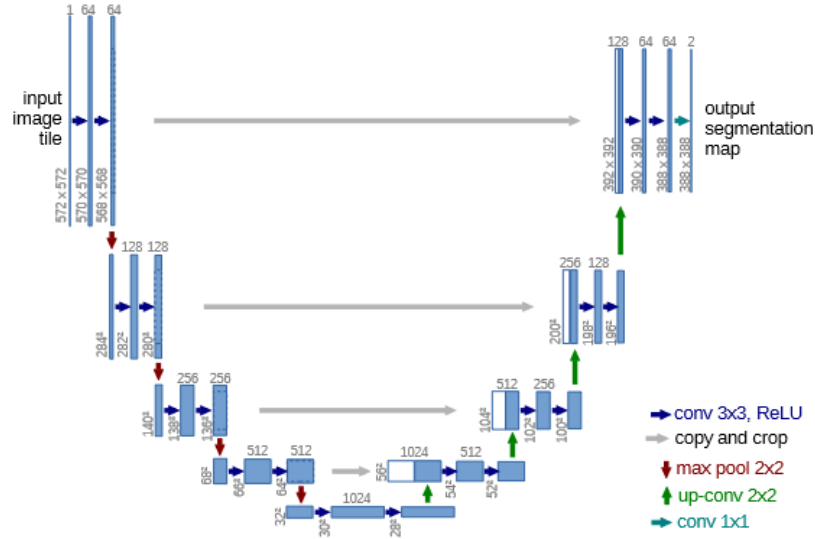

Figure 2.10: U-Net architecture example which downsamples a 572x572 image to 32x32 at its smallest resolution. The number above each box denotes the number of channels at that layer [54].

### 2.4.5 U-Net

A popular deep learning network utilizing an encoder-decoder architecture is the U-Net which has become a core building block for medical image processing models [28]. In domains such as medical image processing, the output should include some localization information, such as a class label assigned to specific pixels or voxels to produce a segmentation [54]. The architecture for the U-Net consists of a contracting series of unpadded convolution layers followed by a ReLU unit and a 2x2 max pooling layer with a stride of 2 to perform downsampling. At each downsampling step, the number of feature channels is doubled. These alternating convolution and downsampling operations comprise the 'encoder' half of the U-Net. After the input has been downsampled a sufficient number of times, the feature map is upsampled followed by a convolution that halves the number of channels and is concatenated with the corresponding convolution layer from the encoder half, and has two 3x3 convolutions followed by another ReLU activation function. Finally, a 1x1 convolution is performed to each feature vector to the desired number of classes. The architecture as it was originally described by Ronneberger et al. [54] consists of 23 convolution layers as shown in Figure 2.10.

## 2.5 Unsupervised Registration with Deep Learning

For the goal of achieving a deformation that will be between two images, we will need to use unsupervised learning. Unsupervised learning is necessary because ground truth deformations are not feasible to create. For an input volume, if each voxel is considered a variable, then to perform supervised learning the exact displacement vector would need to be known for every voxel in a volume and this is just impossible to obtain. Because the purpose of registration is to take the moving image and transform it to look like the fixed image, we can use that in creating a loss function. By using a CNN to predict a transform that creates a warped image that is correlated with the fixed image, we can easily create a model that can find transformations that warp an image such that it visually looks correct. Unfortunately, this doesn't necessarily solve the problem of registration. Registration is considered an ill-posed problem, meaning that multiple trivial solutions might appear to give a warped image that is correlated with the fixed image, however, if the goal is to predict a deformation that closely models how human cardiac tissue moves then many of these solutions are not feasible. To illustrate this, consider Figures 2.11a and 2.11b that show fixed and moving images, respectively. Figure 2.11c and 2.11d are the fixed and moving images, respectively, but with only pixels with signal intensity (SI) of 100. Any of the pixels shown in 2.11d could be moved to the positions shown in 2.11c and this would create an image that visually appears to be correct, but many of the possible solutions that would create this do not realistically model cardiac tissue. A pixel representing tissue in the bottom left of Figure 2.11d would not realistically move to the top right of Figure 2.11c, for example.

This is the core of the problem that deep learning registration models try to solve. The following sections will explore some methods that have been attempted to solve this problem and identify a gap in the research that a new model developed for this thesis attempts to solve.

### 2.5.1 Related Works

Many methods have been proposed that demonstrate effectiveness for performing 3D cardiac image registration. These methods aim to create the most realistic possible deformations using a combination of network architectures designed to let the model mimic a realistic deformation as it learns, and by using loss functions that are tuned to punish complex deformations.

**VoxelMorph**

The VoxelMorph method developed by Balakrishnan et al. is designed for fast learning of deformations designed for use on brain MR images [7]. Despite being designed for brain image use, it is frequently cited and used as a baseline for other medical image registration applications. It is worth analyzing the VoxelMorph model as it effectively lays the groundwork that many other high-performing 3D registration deep learning models are built on.

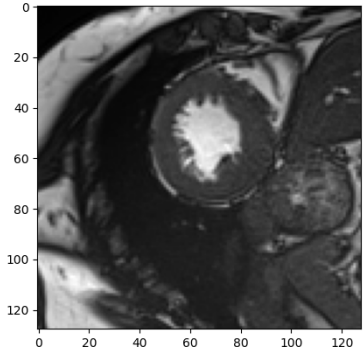

(a) Sample Fixed Image

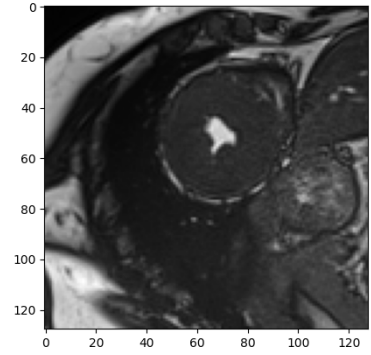

(b) Sample Moving Image

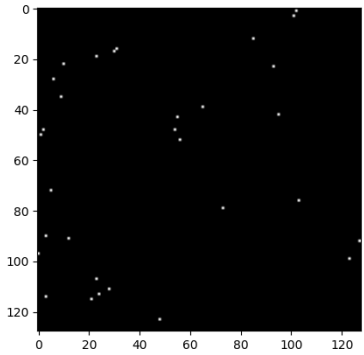

(c) Every pixel with SI 100 from (a)

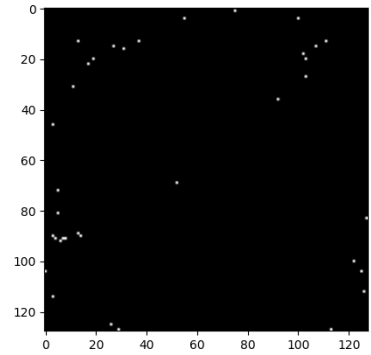

(d) Every pixel with SI 100 from (b)

Figure 2.11: A sample fixed and moving image and each pixel with SI of 100 from each of them. This motivates how simply finding a deformation that moves all pixels into positions that created a seemingly well-correlated warped image might not be realistic for real tissue deformation.

The architecture used for VoxelMorph is based on U-Net, a popular architecture for many deep-learning applications. The exact architecture is shown and described in Figure 2.12. A particularly noteworthy part of the VoxelMorph architecture is the three final convolution layers, two of which have 16 channels. These 16 channel layers are used to refine the final output from the U-Net before the output layer. Balakrishnan et al. use $f$, $m$, and $\phi$ to represent a fixed image, a moving image, and the predicted deformation, respectively in the diagram used for Figure 2.12. For consistency going forward, the fixed, moving, and warped image will continue to be represented as $I_F$, $I_M$, and $I_W$, respectively.

VoxelMorph is designed to use an unsupervised loss function that both penalizes differences in appearance and penalizes local spatial variations in the predicted deformation. By penalizing local spatial variations in the predicted deformation, the goal is to penalize overly complex deformations with the intuition being that a realistic deformation is simpler and will minimize the amount that each voxel should move. The specific loss function to get the loss value $\mathcal{L}_{us}(f, m, \phi)$ is given by

$$\mathcal{L}_{us}(I_F, I_M, \phi) = \mathcal{L}_{sim}(I_F, I_M \circ \phi) + \lambda \mathcal{L}_{smooth}(\phi), \tag{2.2}$$

where $\mathcal{L}_{sim}$ is a similarity metric that penalizes the network for predicting a deformation that generates a warped image that visually doesn't appear correlated with the goal fixed image. The term $\mathcal{L}_{smooth}$ represents the loss function designed to penalize local spatial variations in the deformation which is then multiplied by a regularization factor $\lambda$. Balakrishnan et al. experimented with two different methods for the $\mathcal{L}_{sim}$ term; mean square error and cross-correlation. Experimentally they decided on using cross-correlation which is given by

$$CC(I_F, I_M \circ \phi) = \sum_{p \in \Omega} \frac{(\sum_{p_i} (I_F(p_i) - \hat{I}_F(p) - [\hat{I_M} \circ \phi](p)))^2}{(\sum_{p_i} (I_F(p_i) - \hat{I}_F(p))^2)(\sum_{p_i} ([I_M \circ \phi](p_i) - [\hat{I_M} \circ \phi](p))^2)}, \tag{2.3}$$

where $p$ represents any given voxel, $\hat{I}_F(p)$ and $[\hat{I_M} \circ \phi](p)$ denote local mean intensity images such that $\hat{I}_F(p) = \frac{1}{n^3} \sum_{p_i} I_F(p_i)$ where $p_i$ iterates over a $n^3$ volume around $p$. For the experiments performed by Balakrishnan et al., $n = 9$ was used. Equation 2.3 gives a high value for images that are similar; if the goal is to minimize the similarity loss, negative of the cross-correlation is used which gives the equation for $\mathcal{L}_{sim}$:

$$\mathcal{L}_{sim} = -CC(I_F, I_M \circ \phi). \tag{2.4}$$

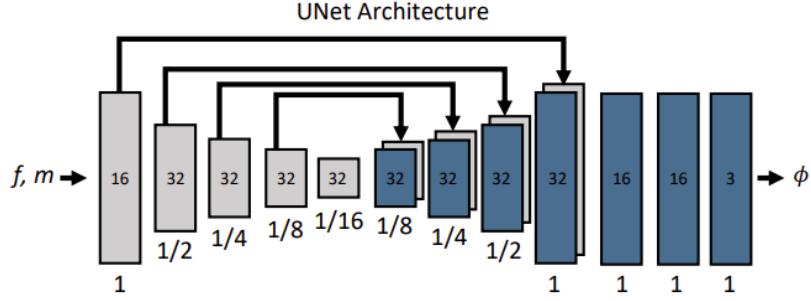

Figure 2.12: Architecture of the VoxelMorph Model [7]. The grey blocks indicate the encoder half of the architecture, using a 3x3x3 kernel with a stride of 2. The stride of 2 results in the resolution being halved in each block, indicated by the fraction under the block. The decoder half is shown by blue blocks which indicate upsampling layers, connected to convolution blocks shown as the grey blocks behind the blue blocks. Each corresponding layer in the encoder and decoder is shown to be connected with skip connections as arrows. After the decoder portion of the U-Net, three convolution layers are shown using 16, 16, and 3 kernels, respectively.

The similarity loss given by Equation 2.4 will encourage the model to learn to give a deformation $\phi$ such that when it is applied to the moving volume $I_M$, it gives a realistic-looking warped image. As previously discussed, this does not guarantee that the $\phi$ represents a realistic deformation. Balakrishnan et al. encourage the model to predict more realistic deformations by adding the smoothing term $\mathcal{L}_{smooth}$. Their implementation of the smoothing term is designed to approximate spatial gradients using the difference between neighboring voxels. The smoothing term is designed to minimize the gradient of each deformation vector in the deformation. If the gradient of one of the displacement vectors $u$ is given as

$$\nabla u(p) = (\frac{\partial u(p)}{\partial x}, \frac{\partial u(p)}{\partial y}, \frac{\partial u(p)}{\partial z}), \tag{2.5}$$

the partial derivatives can be approximated as

$$\frac{\partial u(p)}{\partial x} = u((p_x + 1, p_y, p_z)) - u((p_x, p_y, p_z)), \tag{2.6}$$

$$\frac{\partial u(p)}{\partial y} = u((p_x, p_y + 1, p_z)) - u((p_x, p_y, p_z)), \tag{2.7}$$

$$\frac{\partial u(p)}{\partial z} = u((p_x, p_y, p_z + 1)) - u((p_x, p_y, p_z)). \tag{2.8}$$

**Volume Tweening Network**

Zhao et al. aim to further improve the registration performance achieved by VoxelMorph by modifying the architecture in the development of the Volume Tweening Network (VTN). While the simple U-Net-like architecture used by VoxelMorph is effective for performing registration in a variety of 3D image registration contexts, its architecture is designed to make a single straightforward prediction. For images with significant deformations, it proves to be challenging to predict realistic deformations. Zhao et al. instead propose using a series of cascading U-Nets to perform iterative registrations where each U-Net is only responsible for predicting a small portion of the overall deformation [73]. By having several U-Nets, each predicting small deformations, it is proposed that the overall deformation obtained by combining all of the small predictions should be more realistic and will be less likely to predict large variance, unrealistic deformations.

The VTN model is also designed to perform affine registration as the first part of its cascade. Recall that this is a common preprocessing technique, where the goal of affine registration is to translate and rotate an image without performing any actual deformation to it. The cascading architecture has the benefit of being able to include the affine registration as a part of a single model as opposed to needing to be trained separately.

As its similarity metric, the VTN uses the correlation coefficient, which is given as:

$$\text{CorrCoef}[I_W, I_F] = \frac{\text{Cov}[I_W, I_F]}{\sqrt{\text{Cov}[I_W, I_W] \cdot \text{Cov}[I_F, I_F]}}, \quad (2.9)$$

where the covariance is defined as

$$\text{Cov}[I_W, I_F] = \frac{1}{|\Omega|} \sum_{x \epsilon \Omega} I_W(x) I_F(x) - \frac{1}{|\Omega^2|} \sum_{x \epsilon \Omega} I_W(x) \sum_{y \epsilon \Omega} I_F(y), \quad (2.10)$$

where $\Omega$ represents the input image grid. The correlation coefficient represents how closely two images, $I_W$ and $I_F$, are related. The closer to 1.0 the correlation coefficient is, the more similar the two images are. The correlation coefficient loss is defined as

$$\mathcal{L}_{\text{corrcoef}} = 1 - \text{CorrCoef}[I_W, I_F]. \quad (2.11)$$

Similar to VoxelMorph, a smoothness term is added to the loss function. Zhao et al. use a total variation loss to punish discontinuity in the predicted flow field. The total variation loss is given by:

$$\mathcal{L}_{TV} = \frac{1}{3|\Omega|} \sum_{x} \sum_{i=1}^{3} (\phi(x + e_i) - \phi(x))^2, \quad (2.12)$$

where $e_{1,2,3}$ form the natural basis of $\mathbb{R}^3$.

The architecture of the VTN is shown in Figure 2.13. It features convolution layers that start at a full resolution and have 16 channels and as the resolution decreases, the number of filters double until the model has 128 filters per layer, where the model then has two convolution layers with 128, 256, and 512 filters on the encoder half of the U-Net. This thesis will show that this particular architecture is very cache efficient and contributes to faster inference times, but the VTN model also includes a cascade dedicated to affine registration, which as discussed, is not necessary for the domain of cardiac registration and so can be discarded to even further improve inference speed.

**RC-Net**

Iterating on their previous work with the VTN, Zhao et al. attempt to further improve the cascading model architecture with RC-Net [72]. The goal with RC-Net is to increase the number of cascades that can be effectively used, with the motivation that a more realistic deformation can be predicted if it is comprised of as many small iterative deformations as possible. Where the VTN is designed to compare a similarity loss after each cascade against the fixed image, the RC-Net instead only compares the similarity at the very end after all registrations have been performed. It is argued that by comparing a similarity loss after each cascade, the individual cascades are learning their own objectives separately from each other. Because the sub-networks are not truly cooperating with each other, Zhao et al. suggest that this reduces the benefit of adding more cascades and that further improvement is difficult to achieve simply by adding more cascades. The motivation of the RC-Net then is to train a model that can be shown to be more effective as more cascades are added. Instead of comparing a similarity loss after each cascade, with RC-Net similarity is only measured on the final warped image, which they pose will enable each cascade to learn more cooperatively.

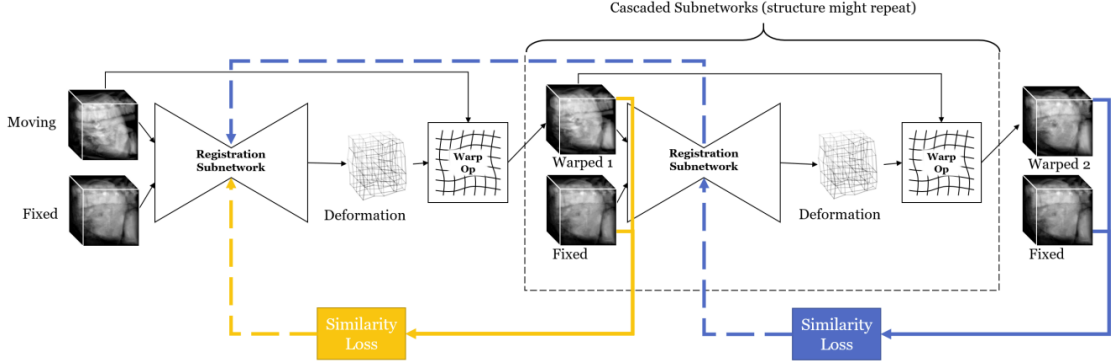

Figure 2.13: Illustration of the cascading structure of the VTN model. End-to-end learning is achieved by back-propagating the similarity loss to all the previous cascades.

Unlike previous works, such as VoxelMorph, RC-Net does not use a specialized smoothness term in its loss function. Zhao et al. argue that a smoothness loss term is not necessary, as the cascading nature of the model itself solves the same problem that the smoothness loss term aims to solve. Instead, the model is trained end-to-end using an image similarity metric that is calculated at the end of all the cascades and an L2 variation loss on each of the intermediate deformations, $\phi$.

The pipeline for this method is shown in Figure 2.14. Each cascade, $k$, takes the warped image predicted by the previous cascade, $k-1$, as $I_m^{k-1}$, and reconstructs a new warped image $I_m^k$ using the intermediate predicted deformation, $\phi_k$. Note the key difference here is that the intermediate deformations are predicted but no similarity loss is calculated after each intermediate deformation. The similarity loss is only back-propagated after the final deformation is calculated.

RC-Net is similar to VTN in that it follows a similar operations-efficient architecture with more low-resolution layers, but has a similar limitation of having an extra sub-U-Net that must be a dedicated affine transformation which, as discussed previously, will add extra operations that are not necessary for the work of this thesis.

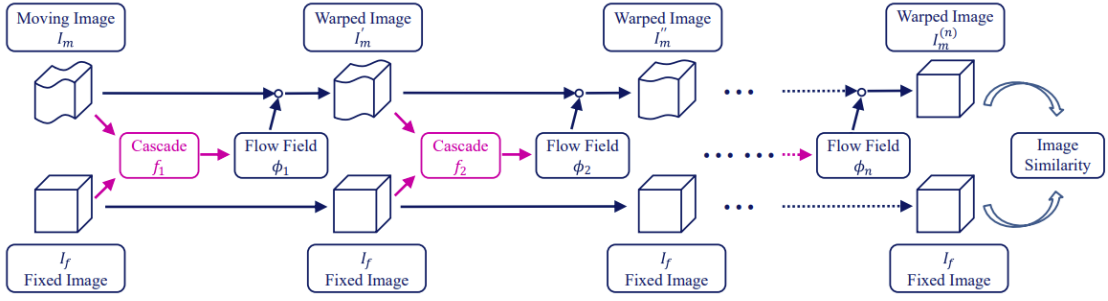

Figure 2.14: Illustration of cascade architecture. For each cascade, $k$, the previous warped image, $I_m^{k-1}$, is reconstructed using predicted deformation $\phi_k$, giving the warped image $I_m^k$ [72].

**VR-Net**

A model developed by Jia et al. proposes even further improvements to an architecture for 3D cardiac image registration [30]. They propose three specially designed layers that, when cascaded together similar to the VTN, and RC-Net, should yield an effective 3D cardiac deformation. The layers they have designed, are referred to as the warping layer, the intensity consistency layer, and the generalized denoising layer. The warping layer uses bilinear interpolation, originally designed for 2D images, extended to 3D. The purpose of this layer is to take as input a deformation $\phi$, moving image $I_m$, and output the warped image $I_w$.

The intensity consistency layer imposes internal consistency between the warped image and the fixed image such that the loss function can be minimized. In other words, this layer is what calculates a metric similar to the similarity loss that VoxelMorph, VTN, and RC-Net use. Specifically, the intensity consistency layer tries to minimize the following function.

$$min_u \frac{1}{s} \int_\Omega |I_m(x + u(x)) - I_f(x)|^s dx + \lambda \mathcal{R}(u(x)). \tag{2.13}$$

Lastly, a generalized denoising layer is described. The input and output of both the internal consistency layer and the generalized denoising layer is a flow field, first predicted by the warping layer. Because these two layers have the same input and output, a residual connection can be used between the layers, which eliminates the need to manually tune parameters for both layers. The model assumes that there is some Gaussian noise in the flow field, with the intention that the generalized denoising layer being to clean the flow field to some extent. An illustration of the network architecture is given in Figure 2.15.

The loss function for this model is given by

$$\mathcal{L} = min_\Phi \frac{1}{N} \sum_{i=1}^{N} ||I_M^i(x + u_i(\Phi)) - I_F^i(x)||_1 + \frac{\alpha}{N} \sum_{i=1}^{N} ||\nabla u_i(\Phi)||_2^2. \tag{2.14}$$

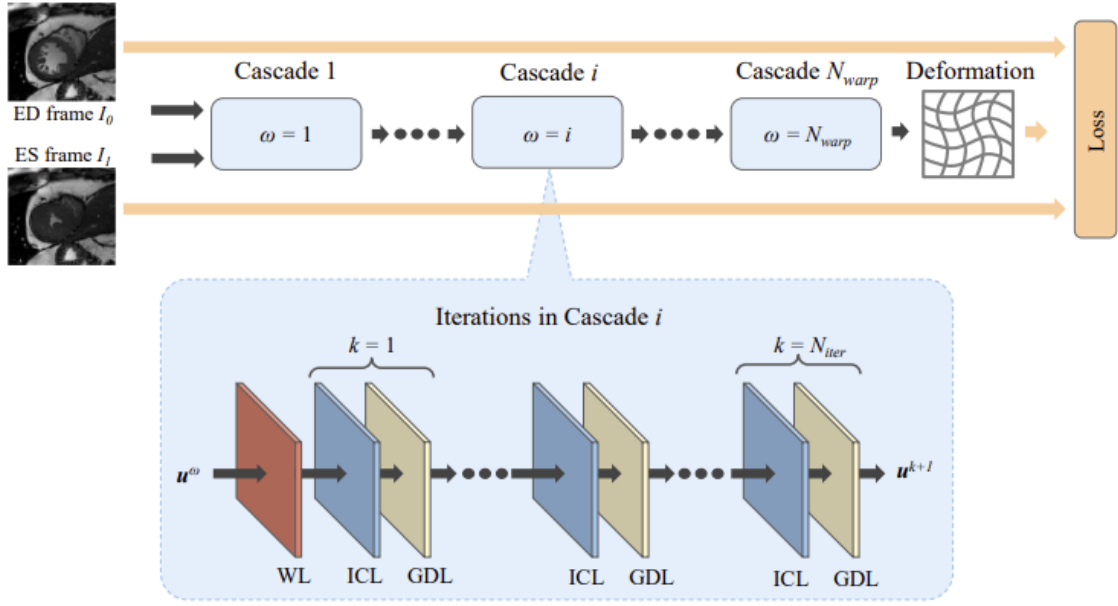

Figure 2.15: Illustration of the architecture of the VR-Net model. In this illustration WL is the warping layer, ICL is the internal consistency layer, and GDL is the general denoising layer [30].

Here, $N$ is the number of training image pairs, $\Phi$ are the network parameters to be learned and $\alpha$ is a manually tuned hyper-parameter. This loss function follows a very similar pattern to the previous 3D registration models shown, where the first term is a similarity metric between the fixed and warped image and the second term enforces smoothness by trying to minimize the gradient of the predicted deformation.

Jia et al. performed analysis against the RC-Net and VoxelMorph methods on two different datasets: the UK BioBank (UKBB) and a dataset of 3D CMR volumes they created for their own experiments which they call 3DCMR. Table 2.2 shows the improvement in dice scores of the VR-Net method, denoted as R-L2-2x1, and the RC-Net model with 2 and 3 cascades compared to VoxelMorph. This table compares the performance of the models performance separately for their dice score performance on the left ventricle (LV), myocardium (myo), and right ventricle (RV). The dice score is a measure of overlap between two segmentations and is given by

$$DSC = \frac{2|X \cap Y|}{|X| + |Y|}, \tag{2.15}$$

where $X$ and $Y$ are two segmentations. This is the metric that each of the shown methods has used for assessing performance so far. The methodology for comparing performance to calculate the dice score is to take segmentations of each of the listed regions of interest, the LV, myo, and RV, and apply the predicted flow field $\phi$ to those segmentations on the moving image to create warped segmentations. Calculating the dice score of the warped segmentations compared to the segmentations for the fixed image gives a consistent metric that can capture how well registration can capture the motion of specific regions of interest and isn't affected by regions of an image such as the background where it may be less important to predict exact deformations.

The warping layer and internal consistency layer implementation details do not mention any downsampling or cropping, so these layers can be assumed to do their operations on the full resolution volumes. The generalized denoising layer features a U-Net but exact details on the number of channels and size of each layer in the U-Net are not provided. However, from the architecture shown in Figure 2.15, we can see that each cascade features the internal consistency layer followed by a generalized denoising U-Net. This means that for every cascaded U-Net used in the VRNet, there is an equivalent number of extra full-resolution layers used which this thesis shows will result in a significant slowdown in inference time.

Table 2.2: Performance results for the VR-Net (R-L2-2x1), RC-Net(RC-Net 2x and RC-Net 3x), and VoxelMorph on the UKBB and 3DCMR datasets. Performance evaluation is done by Jia et al. [30].

| Network | LV Dice | Myo Dice | RV Dice |
| --- | --- | --- | --- |
| VoxelMorph UKBB | 0.931±0.029 | 0.717±0.072 | 0.685±0.102 |
| RC-Net 2x UKBB | 0.942±0.022 | 0.737±0.066 | 0.703±0.099 |
| RC-Net 3x UKBB | 0.944±0.036 | 0.736±0.068 | 0.705±0.105 |
| VoxelMorph 3DCMR | 0.817±0.028 | 0.676±0.051 | 0.634±0.046 |
| RC-Net 2x 3DCMR | 0.820±0.030 | 0.701±0.051 | 0.657±0.047 |
| RC-Net 3x 3DCMR | 0.824±0.027 | 0.692±0.050 | 0.647±0.048 |
| R-L2-2x1 3DCMR | 0.825±0.026 | 0.695±0.050 | 0.649±0.047 |

In Table 2.2 it is shown that RC-Net and VR-Net can achieve dice score improvements over VoxelMorph with VR-Net slightly outperforming RC-Net as well. But for clinical adoption of any of these methods for strain quantification, inference time also needs to be considered. Jia et al. include an evaluation of the inference times of the RC-Net, VR-Net, and VoxelMorph running on both CPU and GPU. The recorded run times are shown in Table 2.3, which shows that there is a dramatic increase in run time for each of the suggested methods. The VR-Net method, which achieves a dice score improvement of 0.007 over VoxelMorph also takes over 3 times as long to run on both CPU and GPU.

VTN, RC-Net, and VR-Net are all shown to be effective for performing medical image registration. For the specific domain of cardiac magnetic resonance image registration, however, some improvements can be made to these architectures. VTN, RC-Net, and VR-Net have portions of their architecture dedicated to performing affine registrations before deformable image registration. Because patient motion during an MR acquisition would result in motion artifacts, making the entire acquisition unviable for analysis, an assumption can be made that affine registration is not necessary. The VR-Net shows the highest registration fidelity of these state-of-the-art models, but it does this at a significant increase to inference time which can be attributed to its complicated architecture and the number of layers that process the volumes at their full resolution. The model shown in this thesis shows that by assuming no affine transformation is necessary and by using an architecture designed to minimize the number of operations performed on the volume full resolution, a similar level of fidelity can be achieved while reducing inference time.

| Method | CPU (S) | GPU (S) |
|---|---|---|
| VoxelMorph | 5.97 | 0.10 |
| RC-Net 2x | 11.95 | 0.21 |
| RC-Net 3x | 17.85 | 0.34 |
| R-L2-2x1 | 18.25 | 0.33 |

Table 2.3: Average inference times over 100 test runs.

## 2.6 Calculating Strain

Performing registration quickly is beneficial for the adoption of strain analysis for cardiac magnetic resonance images. Using the intuition that infarcted tissue moves less than healthy tissue, registration can be used to trace the motion of cardiac tissue as the heart beats. The deformations predicted from transformation can be difficult to visualize and are not a quantitative measurement. To perform analysis, the deformations can be converted into finite strain values which give a quantitative measurement of the tissue movement and can be easily mapped onto 2D and 3D images.

### 2.6.1 Uniform Strain

Strain is a unitless measure of how much an object gets bigger or smaller from an applied load. Normal strain, $\epsilon$, occurs when a force acts perpendicular to an object and is given by

$$\epsilon = \frac{\delta}{L} \tag{2.16}$$

where $\delta$ is the change in length, and $L$ is the original length. An example of normal strain is shown in Figure 2.16a. Shear strain, $\gamma$ occurs when a force acts parallel to an object and is given by

$$\gamma = \frac{\Delta L}{h} \tag{2.17}$$

where $\Delta L$ is a change in the length of the object and $h$ is the height. A visual for this equation is shown in Figure 2.16b.

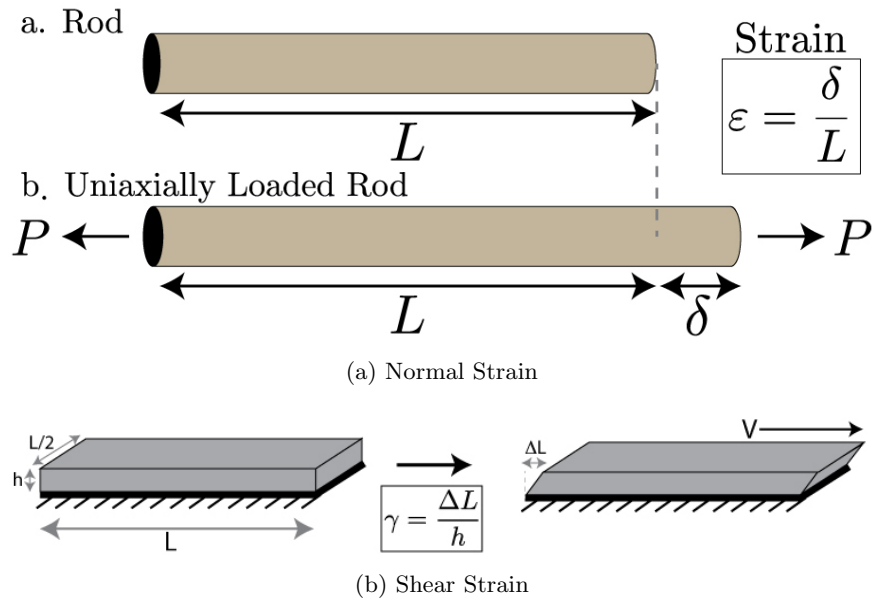

(a) Normal Strain

(b) Shear Strain

Figure 2.16: A visual representation of normal and shear strain[27]

### 2.6.2 Finite Strain

To calculate nonuniform strain, a different method is required. If the object can be modeled with a finite number of points and if the motion of each point is used, then the strain experienced by each point can, in theory, be calculated. This introduces a new challenge though. Using Equation 2.16, $\delta$ can be found easily from the displacement of each point, but $L$ can't be found for individual points in an object which makes calculating normal strain unfeasible.

Finding the strain of an object using a mesh of points also introduces the challenge of how to handle translation. If, for example, an object experiences rotation but no deformation, each point in the object will have a different displacement vector, but the object as a whole will experience no total strain. This means that to calculate the strain experienced by each point, each point's motion will need to take into account the motion of all points relative to it, while also ignoring any rotation the object experiences. To illustrate this problem, an example 1x1 mesh is shown in Figure 2.17a which is stretched to a 2x1 mesh in Figure 2.17b, and translated but not stretched in Figure 2.17c. In this example, the points on the right of the mesh cannot be used alone to calculate the strain experienced by the mesh. The calculation of the strain must include all points on the left of the mesh as well to differentiate between the case where the mesh is deformed as in Figure 2.17b, and if the whole mesh is being translated as in Figure 2.17c.

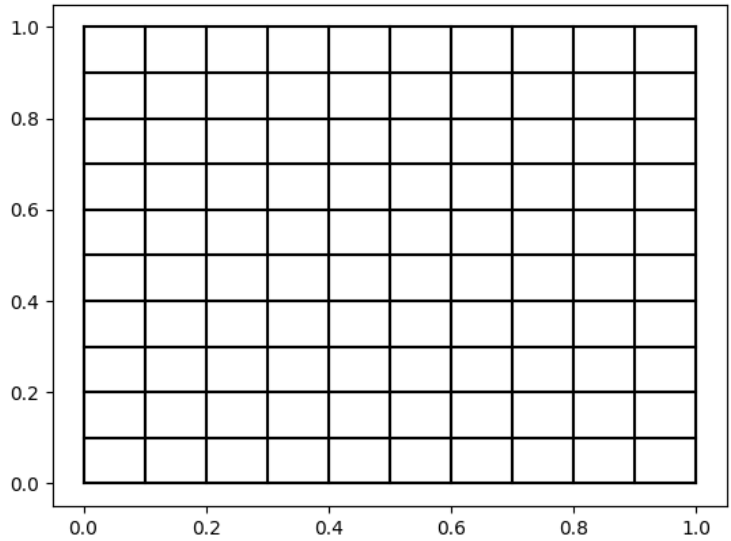

(a) An example undeformed mesh

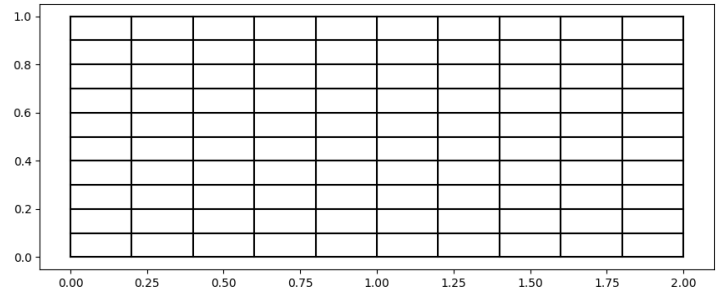

(b) An example deformed mesh

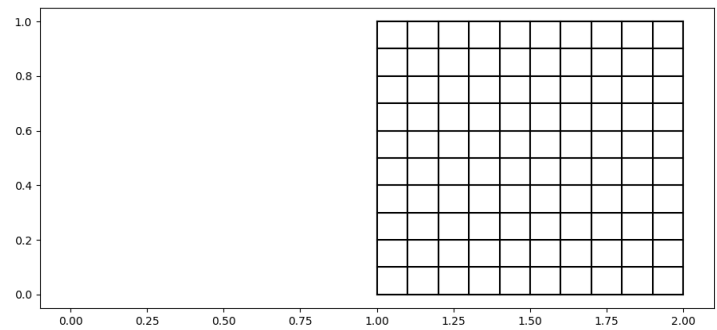

(c) An example translated mesh

Figure 2.17: An example of an undeformed and a deformed mesh to illustrate how each point in the mesh must be tracked relative to each other point to calculate finite strain.

To do this we can use the established method of finite strain quantification, which will be explained in this chapter and shown how it can be used to determine finite strain.

**Deformation Gradient**

For an undeformed object, $x$, and a deformed object $X$, we can find a deformation gradient $F$ by taking the derivative of each component of $x$ with respect to each component of $X$. This gives us:

$$F = \frac{\delta x}{\delta X}. \tag{2.18}$$

An important quality of the deformation gradient is that it can be written as the product of two matrices, $R$ and $U$. $R$ represents a rotation matrix, and $U$ is the right stretch tensor. By multiplying the rotation matrix by the stretch tensor we are representing first stretching the object and then rotating it. However, the overall deformation given by $F$ is transitive, so the rotation could come first followed by the stretch. Doing this gives us a left stretch tensor $V$. From this, we can write the deformation gradient as:

$$F = R \cdot U = V \cdot R \tag{2.19}$$

We can make use of the relationship in Equation 2.19 to eliminate the rotation component when we are calculating strain. By multiplying $F$ by its own transpose, we can do the following:

$$F^T \cdot F = R^T \cdot U^T \cdot R \cdot U = V^T \cdot R^T \cdot R \cdot U. \tag{2.20}$$

Now we can make use of the fact that for a rotation tensor, $R^T = R^{-1}$, so this gives us

$$F^T \cdot F = V^T \cdot R^{-1} \cdot R \cdot U = V^T \cdot U. \tag{2.21}$$

**Cartesian Finite Strain**

Equation 2.21 gives us a way of calculating strain based only on the stretch tensors. The $F^T \cdot F$ term is often called the right Cauchy-Green deformation tensor and is expressed as:

$$C = F^T \cdot F. \tag{2.22}$$

The Cauchy-Green deformation tensor needs to be normalized to give realistic strain values due to the nature of it being the product of two stretch tensors. This is done by subtracting the identity matrix from the right

Cauchy-Green deformation tensor and then dividing the difference by 2. This gives us the following final equation for calculating strain using the displacement gradient tensor $F$:

$$E = \frac{1}{2}(C - I). \tag{2.23}$$

Equation 2.23 gives the strain in the Lagrangian frame of reference which refers to points in the original undeformed material coordinates. This is relevant when visualizing the strain on the object because it will mean that the strain values are projected onto the undeformed object and the deformations themselves will not be present in the visualization at all. To map the strain onto the deformed object instead, we need to use the Eulerian frame of reference for calculating the strain. The equation to find strain in the Eulerian frame of reference is given by:

$$e = \frac{1}{2}(I - c), \tag{2.24}$$

where c is the left Cauchy-Green deformation tensor defined as:

$$c = F^{-T} \cdot F^{-1}. \tag{2.25}$$

**Cylindrical Finite Strain**

Assuming that the deformation gradient tensor is using Cartesian coordinates, Equations 2.23 and 2.24 will give strain values in terms of the principal Cartesian coordinates, namely, the $x$, $y$, and $z$ directions. Commonly when calculating cardiac strain, strain values are instead given in terms of radial, circumferential, and longitudinal values. Radial strain can be thought of as strain in the direction normal to the myocardium wall, directed towards the center of the chamber. If a plane is assumed at the base of the chamber, and another plane at the apex, circumferential strain is the strain that is in the direction of the wall parallel to these planes. Conversely, longitudinal strain can be thought of as the strain along the myocardium wall perpendicular to these planes. These directions are illustrated in Figure 1.2.

To calculate strain in these directions, the strain values must be converted to the cylindrical coordinate space. The cylindrical coordinate system is used for 3D as opposed to the spherical coordinate system because we are trying to map strain values onto the left ventricle of the heart, which can be much closer approximated to be the shape of a cylinder as opposed to a sphere. The cylindrical coordinates are easy to obtain using a rotation matrix $R$ given by

$$R = \begin{bmatrix} cos(\theta) & -sin(\theta) & 0 \\ sin(\theta) & cos(\theta) & 0 \\ 0 & 0 & 1 \end{bmatrix}, \tag{2.26}$$

with

$$\theta = arctan(x/y). \tag{2.27}$$

Here $x$ and $y$ are the x and y components of the strain tensor found in the Cartesian coordinate system.

## 2.7 Chapter 2 Summary

In this chapter, an overview of magnetic resonance imaging was given to illustrate how, owing to the nature of magnetic resonance imaging, scar tissue generally appears the same as healthy tissue during the common imaging acquisitions. Typically to identify scar tissue in MR imaging a heavy metal contrast agent, gadolinium is required. Strain quantification, a method already established and used in echocardiography, can be adapted as a possible alternative for identifying scar tissue. This method involves using registration to predict the deformation of the heart as it beats which can be done efficiently with deep learning. Deep learning describes a subset of machine learning models that focuses on deep learning neural networks, neural networks that feature at least one hidden layer. Convolutional neural networks, a type of deep learning neural network, have been used effectively in state-of-the-art models to achieve a high degree of registration fidelity. Three state-of-the-art registration models namely VTN, VR-Net, and RC-Net were described. These models use the VoxelMorph model as a common baseline for comparison. A comparison of these models shows that they can achieve a high level of fidelity but at a cost of inference time caused by their complicated architectures. These state-of-the-art models all have components designed to align the images before deforming them and also feature a cascading architecture. In cardiac magnetic resonance images where images from the same acquisition are being processed, any misalignment would also cause artifacts specific to patient motion to appear which would make the images unviable for analysis. This assumption allows the rigid alignment portions of these models to be removed to improve performance time. Using this assumption and building a model with a cascading architecture, which the state-of-the-art models have shown to be effective, a model can be tuned to give registration fidelity that is on a similar level to these start-of-the-art models but performs inference much faster.

# Chapter 3

# Proposed FLIR Model

This chapter provides an overview of the proposed FLIR deep learning model. First, the motivation of the deep learning model is described based on an analysis of existing work discussed in Chapter 2, followed by a description of the architecture and implementation of the model, and lastly details of the implementation of the strain calculator that is used to convert the flow fields predicted by the FLIR model into strain quantified strain values.

## 3.1 Deep Learning Registration Model

The previously shown methods for performing 3D registration with deep learning show a trend; trying to improve the accuracy or fidelity of a deep learning model increases the complexity and number of parameters of the model. As shown in Table 2.3, the added complexity and parameters come at a significant cost to run time. If these models were to be used to process a large dataset such as UKBB, which contains over 100,000 samples, these models would become too unwieldy to use even with advanced GPUs. Some 3D models have been proposed for brain image registration that focus on inference time, such as Quicksilver and unnamed deep learning models developed by Dalca et al [68, 17]. These methods, however, perform patch-based registration as opposed to calculating the flow field for the image in one network pass. In these approaches, a neural network is used to estimate deformation at sparse locations with a small field of view, which is then transformed into a dense flow field through complex interpolation techniques, such as the Large Deformation Diffeomorphic Metric Mapping (LDDMM) used in Quicksilver. By using a patch-based method large deformations become hard to model because of points or voxels moving over the boundary of multiple patches. For a domain such as brain imaging, where deformations are expected to mostly be small, for a highly dynamic structure such as the heart, patch-based methods might struggle to fully capture deformations. Instead, a new model called the Fast and Lightweight Image Registration method (FLIR) is proposed here. FLIR can compute a flow field for an entire image volume with one network pass with a fast run-time.

Cascading sub-networks are effective for producing realistic deformations. A similar architecture is used for FLIR. All of the models discussed in Sec. 2.5.1 also have a portion of their architecture designed to perform affine transformations. While performing an affine transformation is typically an important preprocessing step for domains where the images being registered are from different acquisitions and so are probably not aligned, for cases where the images are from the same acquisition, such as volumes taken from the same SSFP sequence, the assumption can be made that for any clinically usable SSFP sequence, there should be no affine deformation. Any affine deformation between two images of an SSFP sequence would mean that the patient physically moved during the acquisition. Because of how MRI images are captured, this wouldn't just add affine motion to the image, it would result in the presence of significant motion artifacts in the image that would make it unusable for any clinical analysis [70]. Recall that in the implementations of VTN, RC-Net, and VR-Net, the first cascade is restricted such that it can only perform affine, rigid transformations to align the input volumes to the model. By assuming that there will be no affine transformations in these data being registered, any part of VTN, RC-Net, and VR-Net that are designed to perform affine registration can be ignored which should significantly reduce the number of parameters in the model and improve inference time. Looking at the architecture of VoxelMorph in Figure 2.12, for example, we can see that 5 total layers perform operations on the full resolution volume, with 3 of these full resolution layers added at the end to refine the output. Instead of adding more full-resolution layers, the FLIR model is designed to make use of cascades that are effective in achieving good registration performance. These cascades are inspired by similar models such as VTN [73], RC-Net [72], and VR-Net [30]. By performing more operations at a lower resolution, there are fewer convolution operations required, which should result in a significant improvement in inference time. Intuitively this will decrease registration fidelity slightly when designing a model with only a single iteration of a U-Net architecture, so a cascading architecture is utilized that uses multiple faster U-Nets to achieve high-fidelity results. From these observations, the FLIR model is designed to iterate on the improvements proposed by the reviewed models by first removing the cascades designed to perform affine registration, and then by optimizing the architecture to use smaller layers to reduce the number of convolution operations being performed.

As illustrated in Figure 3.1a, the FLIR model uses a cascading U-Net architecture where each U-Net features eight layers in the encoder half and six layers in the decoder half. This architecture is similar to the one developed for the VTN. The model applies a 3x3x3 kernel with a stride of 2 for the first three layers and then alternates between a stride of 1 and a stride of 2 for the next four layers to create blocks with two convolution layers each. The first layer starts with 16 channels which is then doubled in each block of the encoder, whenever the resolution halves, and then doubled again each time the resolution increases in the decoder. The output of each sub-U-Net is a flow field. To get this flow field, a final layer with 3 channels is required where each channel represents the displacement of each voxel in the x, y, and z directions. Any number of these U-Nets can be cascaded together; an example with two cascades is shown in Figure 3.1b. Figure 3.1b also shows that the FLIR model is using the learning format of the VTN, as opposed to the RC-Net. In this setup, the similarity loss is calculated between the fixed image and a partially warped image after each cascade and then back propagated to each previous cascade.

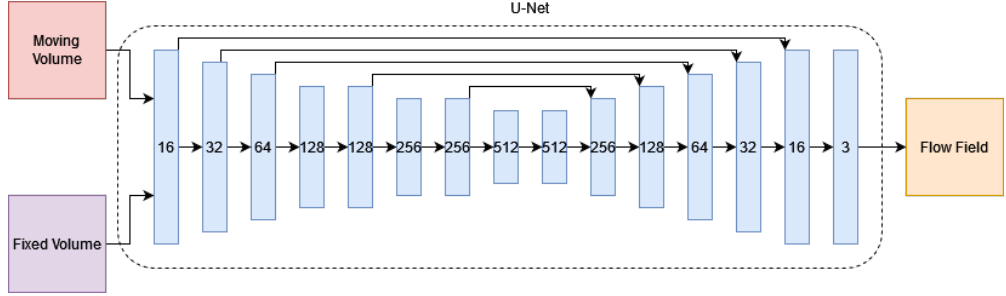

(a) U-Net architecture of each sub-network

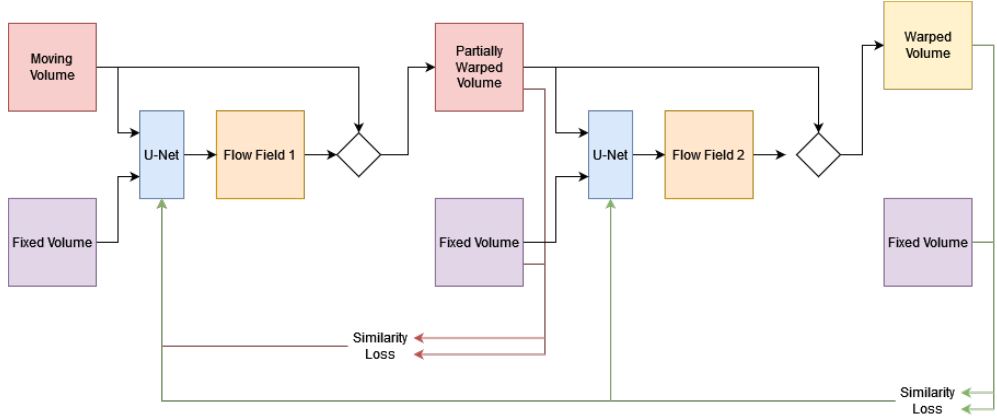

(b) Cascading architecture with 2 cascades

Figure 3.1: Implementation of the U-Net used as the sub-networks (top) and an example of the end-to-end pipeline of the proposed FLIR model (bottom).

Like the loss functions shown for the VoxelMorph, VTN, RC-Net, and VR-Net methods, the loss function for FLIR consists of a similarity loss term and a smoothing term. The similarity loss is calculated as

$$\mathcal{L}_{\text{corrcoef}} = 1 - \text{CorrCoef}[I_W, I_F], \tag{3.1}$$

and the smoothing term is calculated as

$$\mathcal{L}_{TV} = \frac{1}{3|\Omega|} \sum_x \sum_{i=1}^{3} (\phi(x + e_i) - \phi(x))^2.  \tag{3.2}$$

The VTN paper also suggests using its inevitability loss. While ground truth landmarks may have been available for use on the liver CT data that the VTN was trained on, landmarks like this are not available on the data that will be used for training the FLIR model. It would take significant effort and a high degree of expertise to create good landmarks to use for a loss function like this for CMR data because the z resolution is typically very low for CMR SSFP data, such that each voxel in a CMR SSFP acquisition potentially represents several centimeters in the original patient. This means that if the model is trying to learn registration based on how the landmarks move, a landmark being off by even a single voxel in the z-axis represents an error of potentially several centimeters of real tissue deformation. Adding the landmark loss also means that the model is not truly unsupervised, and is instead semi-supervised. For these reasons, the landmark loss is not implemented for the FLIR method. To further fine-tune the model performance, manually tuned weights are used to give a final loss function for the FLIR model as

$$L_{total} = \omega_{corrcoef} \cdot L_{corrcoef} + \omega_{TV} \cdot L_{TV}, \tag{3.3}$$

where $\omega_{corrcoef}$ and $\omega_{TV}$ are manually tuned weights.

In Chapter 4, an experiment will be designed for training the FLIR model with tuned weights and evaluating its performance. The goal of these experiments is to demonstrate that a level of registration fidelity comparable to the state-of-the-art methods shown can be achieved without adding to the inference time.

## 3.2 Strain Calculator

The strain calculator can be expressed as the function $E = f(\phi)$ which takes as input the deformation predicted by the registration model, $\phi$ and outputs a strain map $E$ in either the Lagrangian or Eulerian frame of reference. Recall that the Lagrangian and Eulerian strain can be calculated using Equations 2.23 and 2.24, respectively. The registration model generates a flow field of the shape (16, 128, 128, 3) where the first three dimensions represent the size of the volume that is fed into the network, and the last dimension represents the motion in the x, y, and z directions. The variable with the 3D tensor representing the motion of each of the $x$, $y$, and $z$ directions will be referred to as $dx$, $dy$, and $dz$, respectively. The strain calculator as it is written only cares that the shape has four dimensions. The last dimension is three to ensure that there is a motion being measured in all three dimensions; it is agnostic to the actual shape of the volume being measured, i.e. the first three dimensions.

The strain calculator takes as input the deformation, the center of the subject being measured, and a parameter to define if the strain values returned should be in the Lagrangian or Eulerian frame of reference. The gradient of each of $u_x$, $u_y$, and $u_z$ is taken, which gives partial derivatives in each other direction. For $u_x$ this gives the three partial derivatives $\frac{\partial u_x}{\partial x}$, $\frac{\partial u_x}{\partial y}$, and $\frac{\partial u_x}{\partial z}$. The gradient of $u_y$ gives the partial derivatives $\frac{\partial u_y}{\partial x}$, $\frac{\partial u_y}{\partial y}$, and $\frac{\partial u_y}{\partial z}$, and lastly, the gradient of $u_z$ gives $\frac{\partial u_z}{\partial x}$, $\frac{\partial u_z}{\partial y}$, and $\frac{\partial u_z}{\partial z}$. All of these partial derivatives are stacked into a gradient tensor as follows:

$$\nabla u = \begin{bmatrix} \frac{\partial u_x}{\partial x} & \frac{\partial u_x}{\partial y} & \frac{\partial u_x}{\partial z} \\ \frac{\partial u_y}{\partial x} & \frac{\partial u_y}{\partial y} & \frac{\partial u_y}{\partial z} \\ \frac{\partial u_z}{\partial x} & \frac{\partial u_z}{\partial y} & \frac{\partial u_z}{\partial z} \end{bmatrix} \qquad (3.4)$$

Next, an identity matrix is defined to be the same shape as $\nabla u$ and added to $\nabla u$ to get the deformation gradient $F$. Here, $F$ is a 5D tensor where the first 3 dimensions represent the shape of the input volume. For the calculation of strain in the Eulerian frame of reference, recall that Eulerian strain is given as

$$e = \frac{1}{2}(I - c) \qquad (3.5)$$

and

$$c = F^{-T} \cdot F^{-1}. \qquad (3.6)$$

Here $F^{-1}$ is found by taking the pseudo inverse of $F$. The pseudo-inverse of $F$ is defined as

$$F^{-1} = (F^T F)^{-1} F^T. \qquad (3.7)$$

As mentioned previously this will give the strain values in the Cartesian coordinate space. Using the transformation defined in Equation 2.26, the Cartesian strain can be converted to a cylindrical coordinate system which is what will convert the principle strains from being in terms of $x$, $y$, and $z$ to being in terms of radial, circumferential, and longitudinal, which is the standard frame of reference for quantifying cardiac strain. The conversion is illustrated in Figure 3.2. In this illustration, radial strain, circumferential strain, and longitudinal strain are along the axis of $r$, $\theta$, and $z$, respectively. For this thesis, the assumption is made that the center of the LV is at the center of the entire volume.

To validate the strain calculator a comprehensive set of unit tests is written where a deformation in every possible combination of directions is evaluated, including shear and translations. The results of the unit tests are given in Appendix A.

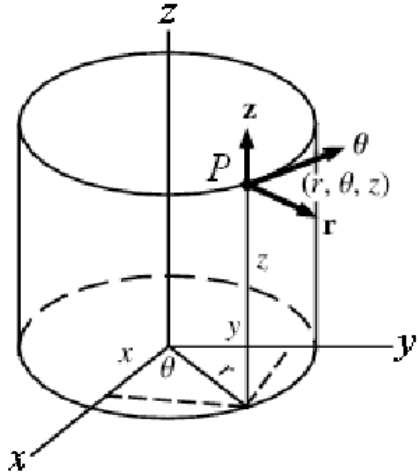

Figure 3.2: Illustration of the relationship between cartesian and cylindrical coordinate spaces [52].

## 3.3 Chapter 3 Summary

In this chapter, the architecture of the proposed model is described. State-of-the-art models show that cascading architectures are highly effective for breaking down complicated registration problems into small iterative transformations. This architecture has been implemented in the FLIR model. Similar to the state-of-the-art models, the loss function is given as a weighted sum of the correlation coefficient between the $I_W$ and $I_F$ and a smoothing term that constrains the predicted deformations to be as simple as possible. Unlike the state-of-the-art models, however, the FLIR model is designed to be entirely the cascading U-Nets, without any portion of the architecture designed for affine transformations or rigid alignment. The implementation of a strain calculator is also given which takes the flow field, $\phi$, predicted by the FLIR model and converts it to quantified strain values in cylindrical coordinates.

# Chapter 4

# Experiments

## 4.1 VoxelMorph-Lite Registration Experiment

In this chapter, the parameters for training and evaluating the FLIR model are described followed by analysis of the results. In addition, a preliminary experiment is described where a slightly modified version of VoxelMorph is implemented that excludes the full resolution layers which fine-tune the output after the U-Net. This experiment illustrates the effect that the full-resolution layers have on inference time to further motivate the efficiency improvements that can be achieved by cascading layers that use more layers at a low resolution.

### 4.1.1 Data

The data used for training and testing the FLIR model and the version of VoxelMorph that is to be used as a baseline was taken from two datasets of cardiac MR cine SSFP images. The first dataset was the publicly available ACDC challenge dataset [11] and the publicly available UK Biobank project [57]. A sample slice from the moving and fixed image are shown in Figures 4.1 and 4.2, respectively. Each patient study contains a pair of 3D volumes at the ED and ES phases, considered to be the fixed and moving images, respectively. 70 cases from ACDC and 500 cases from UKBB were used for training. A validation set was prepared using 10 cases from ACDC and 50 cases from UKBB. The models were evaluated using 20 cases from ACDC and 30 cases from UKBB. All of the volumes used for training include segmentations of the left ventricle (LV), right ventricle (RV), and LV myocardium (myo). Ethics approval was not necessary for this data.

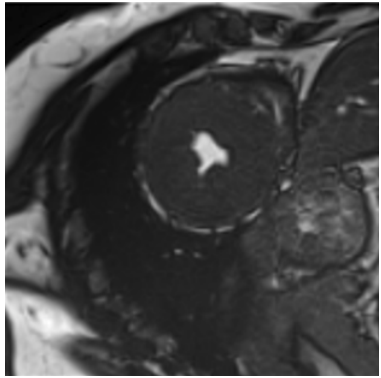

Figure 4.1: Sample slice from a moving volume in the ACDC dataset

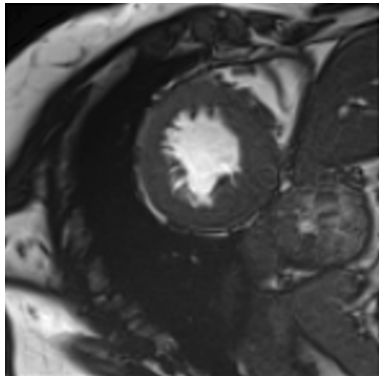

Figure 4.2: Sample slice from a fixed volume in the ACDC dataset

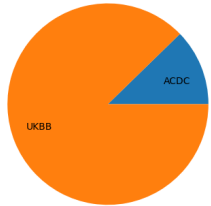

(a) Train Data Split

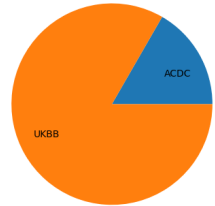

(b) Val Data Split

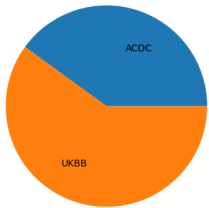

(c) Test Data Split

Figure 4.3: Distribution of train, validation, and test data over datasets.

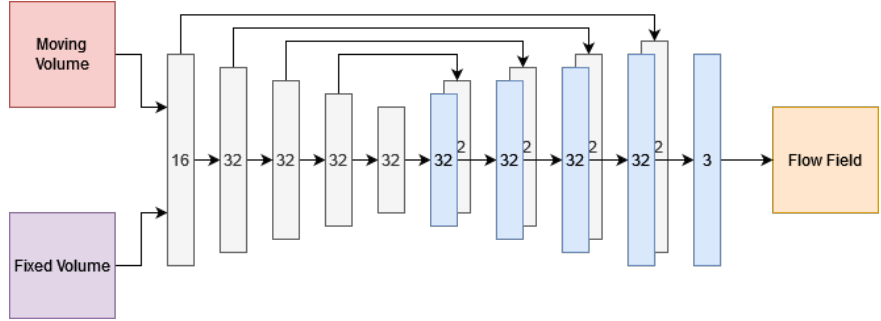

Figure 4.4: VoxelMorph Lite architecture. It is an exact copy of the VoxelMorph architecture but without the additional fine-tuning layers after the U-Net.

### 4.1.2 VoxelMorph-Lite Methodology

To evaluate the hypothesis that full-resolution layers contribute the most to inference, a more lightweight version of VoxelMorph was created that will be called VoxelMorph-Lite. This implementation of VoxelMorph is the same as the model proposed by Balakrishnan et al. [7] except with two of the final full-resolution layers removed. In this model, the last layer of the decoder still has 32 channels which is then followed by a full resolution. The last 32-channel full-resolution layer of the decoder remains unchanged so that there are as few changes to the initial model as possible. The last layer with 3 channels is required to give an output that represents a 3-dimensional displacement tensor. An illustration of the architecture is shown in Figure 4.4.

This model was only trained for 1 epoch because the goal of this experiment is not to demonstrate the effect that removing these layers would have on registration fidelity, but to instead demonstrate the relationship between the number of full-resolution layers in a model and its inference time. To compare the methods, the test set was fed into the network one patient study at a time. Because the initialization of graph variables takes a significant amount of time, results recorded for the first test study were dropped and the number of floating point operations performed, inference time, and all the CPU instructions performed were recorded and averaged across each test run.

The inference time and the amount of time to perform convolution were measured using the Pytorch profiler library which computes the number of milliseconds that each function in a model is run during inference. The results for the VoxelMorph-Lite model are tabulated and shown in Section 4.2.2.

## 4.2 FLIR Model Registration Experiment

**Training**

The training datasets were combined and used to train the network with a batch size of 1 over 400 epochs with a learning rate of $10^{-4}$. Different FLIR models were trained using 1, 2, 3, and 5 cascades. The model was trained using PyTorch 2.1.1 using an Nvidia RTX A6000 and an AMD Ryzen 9 7950x 16-Core processor.

### 4.2.1 FLIR Model Methodology

The models will be evaluated by comparing the dice scores on each of the LV, myocardium, and RV, the average time spent performing convolution operations, and the average total inference time. The convolution times and inference times were evaluated by using PyTorch 2.1.1 which includes a profiler that tracks the amount of time spent in different function calls while running the model. PyTorch version 2.1.1 also includes functionality to compile a model as a preprocessing step to further optimize runtime on Linux systems. Each of the models was evaluated on an AMD Ryzen 9 7950x 16-Core Processor (CPU), an NVidia RTX A6000 (GPU 1), and an NVidia GTX 1080 (GPU 2). The system used for evaluation with GPU 2 was using Windows 10. As of PyTorch 2.1.1 model compilation is not supported on Windows; the evaluation of the compiled models could not be performed using GPU 2.

Because each of the state-of-the-art models was trained on a different data set, to keep the comparison fair the main performance evaluation performed will be done by measuring the improvement of each metric compared to a version of VoxelMorph trained on the same dataset. A performance evaluation was performed by Jia et al. for their development of the VR-Net model which compared the VR-Net and RC-Net against VoxelMorph – all trained using consistent datasets. The improvement of each of these models against their implementation of VoxelMorph will be compared against a version of VoxelMorph trained using the same data that is used to train and test the FLIR model. Results in Table 4.1 show the performance of the FLIR model compared to an implementation of VoxelMorph trained and tested on the same data. Performance is measured using the dice score, defined in Eq. 2.15 and is a measurement of the overlap between the segmentation of the goal image, $I_F$, and the segmentation of the warped image, $I_W$, where a dice score closer to 1.0 signifies higher performance. The dice score is calculated separately for the left ventricle (LV), myocardium (Myo), and right ventricle (RV). To compare the improvement of the FLIR model over an implementation of VoxelMorph, two tables from a performance evaluation performed by Jia et al. [30] are shown in Table 4.4 and Table 4.5. Table 4.4 shows the RC-Net with 2 and 3 cascades and VoxelMorph trained on the UKBB model. Table 4.5 shows the RC-Net with 2 and 3 cascades, the VR Net with 2 cascades and L2 loss, and VoxelMorph trained on the 3DCMR dataset. The purpose of including Table 4.4 and Table 4.5 is to evaluate the improvement of the FLIR model over the implementation of VoxelMorph trained on the same data as the FLIR model, as shown in Table 4.1, compared to the improvement of the state-of-the-art RC-Net and VR-Net models over implementations of VoxelMorph trained for their respective papers. This should provide a fair comparison for showing how the FLIR model can perform compared to other state-of-the-art methods.

### 4.2.2 FLIR Results

| Network | LV Dice | Myo Dice | RV Dice |
|---|---|---|---|
| VoxelMorph | 0.835±0.063 | 0.683±0.070 | 0.683±0.113 |
| 1xFLIR | 0.801±0.083 | 0.680±0.082 | 0.674±0.107 |
| 2xFLIR | **0.839±0.062** | **0.693±0.069** | **0.689±0.099** |
| 3xFLIR | **0.864±0.062** | **0.737±0.060** | **0.692±0.109** |
| 5xFLIR | **0.882±0.050** | **0.749±0.056** | **0.734±0.099** |

Table 4.1: Summary of dice results for the left ventricle (LV), myocardium (Myo), and right ventricle (RV) segmentations for VoxelMorph and the FLIR model with 1, 2, 3, and 5 cascades.

| Network | CPU Conv. Time (ms) | CPU Total Time (ms) | GPU 1 Conv. Time (ms) | GPU 1 Total Time (ms) | GPU 2 Conv. Time (ms) | GPU 2 Total Time (ms) |
|---|---|---|---|---|---|---|
| VoxelMorph | 58.057±1.102 | 76.182±1.141 | 1.422±0.003 | 2.193±0.004 | 10.425±0.580 | 14.219±0.726 |
| VoxelMorph Lite | 52.067±0.998 | 67.879±1.710 | 1.355±0.003 | 2.122±0.004 | 7.351±4.526 | 10.499±4.518 |
| 1xFLIR | 33.407±0.592 | 47.211±2.880 | 0.927±0.002 | 1.282±0.004 | 6.654±0.536 | 8.819±0.605 |
| 2xFLIR | 79.926±1.911 | 98.891±2.315 | 1.873±0.004 | 2.568±0.006 | 13.548±1.546 | 17.956±1.929 |
| 3xFLIR | 96.060±2.412 | 133.075±3.308 | 2.824±0.024 | 3.856±0.027 | 19.666±0.857 | 26.561±0.971 |
| 5xFLIR | 173.023±1.402 | 237.974±3.181 | 4.709±0.035 | 6.452±0.039 | 72.735±13.824 | 95.561±17.233 |

Table 4.2: Summary of timing data for VoxelMorph, VoxelMorph-Lite, and the FLIR model with 1, 2, 3, and 5 cascades. All results were collected without using the PyTorch compile optimizations.

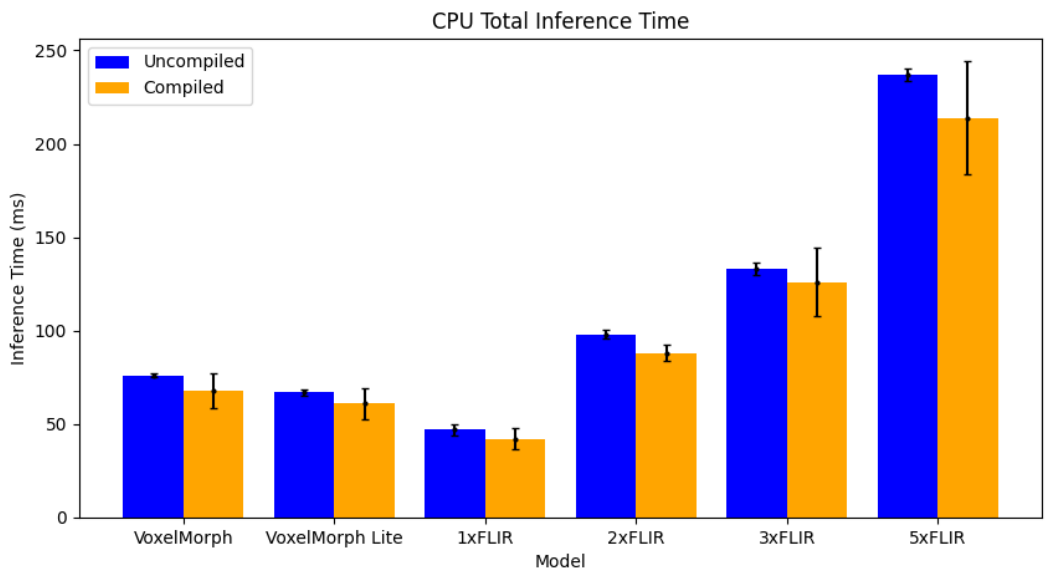

Figure 4.5: Inference times of FLIR models with various cascades compared to VoxelMorph and VoxelMorph Lite models on CPU.

| Network | CPU Conv. Time (ms) | CPU Total Time (ms) | GPU 1 Conv. Time (ms) | GPU 1 Total Time (ms) |
| --- | --- | --- | --- | --- |
| VoxelMorph | 59.135±8.703 | 68.552±9.499 | 0.00145±0.000900 | 0.0269±0.00216 |
| VoxelMorph Lite | 53.609±8.366 | 61.486±8.415 | 0.00146±0.00119 | 0.0267±0.00256 |
| 1xFLIR | 37.927±12.471 | 42.886±5.591 | 0.00147±0.000911 | 0.0330±0.00271 |
| 2xFLIR | 73.062±4.175 | 88.615±4.390 | 0.00153±0.000821 | 0.0637±0.00452 |
| 3xFLIR | 109.602±30.203 | 126.604±18.221 | 0.00155±0.000830 | 0.0941±0.00713 |
| 5xFLIR | 182.676±43.286 | 214.188±30.112 | 0.00157±0.000799 | 0.155±0.0117 |

Table 4.3: Summary of timing data for VoxelMorph, VoxelMorph-Lite, and the FLIR model with 1, 2, 3, and 5 cascades. All results were collected using the PyTorch compile optimizations.

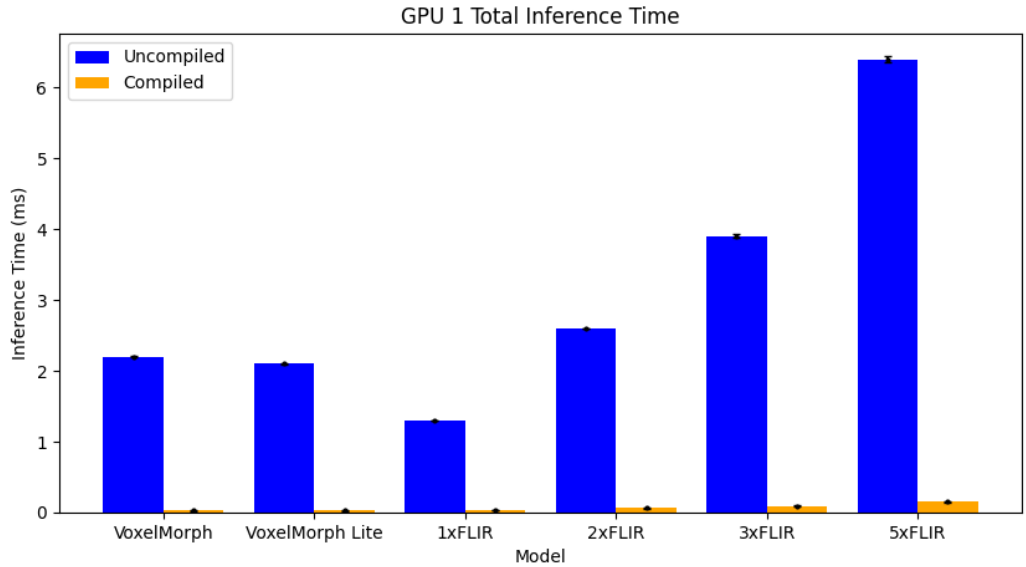

Figure 4.6: Inference times of FLIR models with various cascades compared to VoxelMorph and VoxelMorph Lite models on GPU 1.

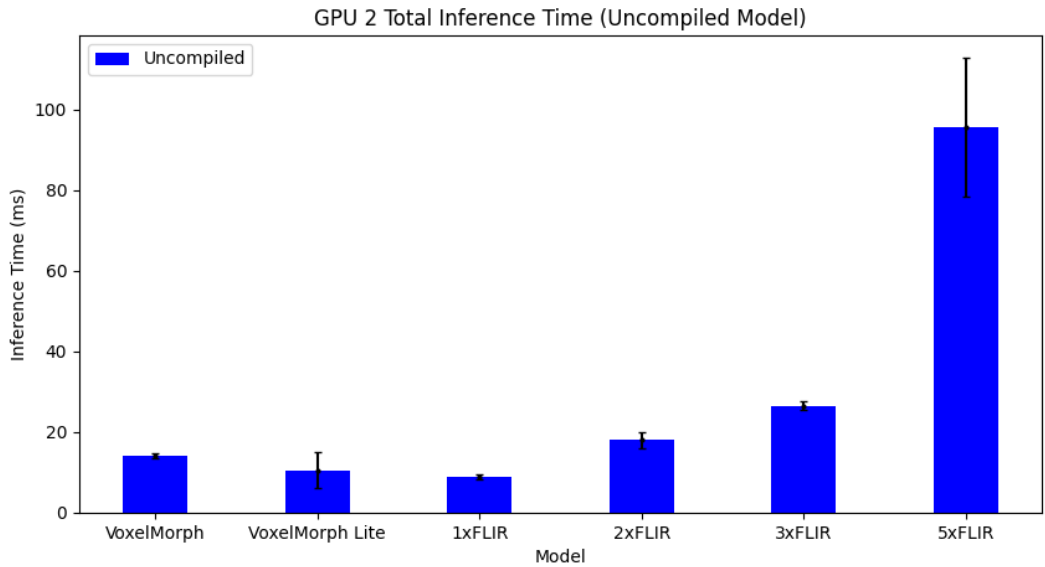

Figure 4.7: Inference times of FLIR models with various cascades compared to VoxelMorph and VoxelMorph Lite models on GPU 2.

| Network | LV Dice | Myo Dice | RV Dice |
|---|---|---|---|
| VoxelMorph UKBB | 0.931±0.029 | 0.717±0.072 | 0.685±0.102 |
| RC-Net 2x UKBB | 0.942±0.022 | 0.737±0.066 | 0.703±0.099 |
| RC-Net 3x UKBB | 0.944±0.036 | 0.736±0.068 | 0.705±0.105 |

Table 4.4: Performance summary of the VoxelMorph, RC-Net 2x and, RC-Net 3x trained on the UKBB dataset. Values are obtained from a performance evaluation done by Jia et al. [30]. Inference time is reported in seconds.

| Network | LV Dice | Myo Dice | RV Dice |
|---|---|---|---|
| VoxelMorph 3DCMR | 0.817±0.028 | 0.676±0.051 | 0.634±0.046 |
| RC-Net 2x 3DCMR | 0.820±0.030 | 0.701±0.051 | 0.657±0.047 |
| RC-Net 3x 3DCMR | 0.824±0.027 | 0.692±0.050 | 0.647±0.048 |
| R-L2-2x1 3DCMR | 0.825±0.026 | 0.695±0.050 | 0.649±0.047 |

Table 4.5: Performance summary of the VoxelMorph, RC-Net 2x, RC-Net 3x, and R-L2-2x1 trained on the 3DCMR dataset. Values are obtained from a performance evaluation done by Jia et al. [30]. Inference time is reported in seconds.

| Network | Inference Time (CPU) (ms) | Inference Time (GPU) (ms) |
|---|---|---|
| VoxelMorph | 5.97 | 0.10 |
| RC-Net 2x | 11.95 | 0.21 |
| RC-Net 3x | 17.85 | 0.34 |
| R-L2-2x1 | 18.25 | 0.33 |

Table 4.6: State-of-the-art model inference times [30].

Table 4.1 shows that when comparing the FLIR model with a different number of cascades against VoxelMorph, the FLIR model starts showing gradual improvements over VoxelMorph starting with two cascades.

The difference in inference time of the FLIR model compared to VoxelMorph is summarized in Table 4.2 without using the PyTorch model compiler, and results using compiled models are shown in Table 4.3.

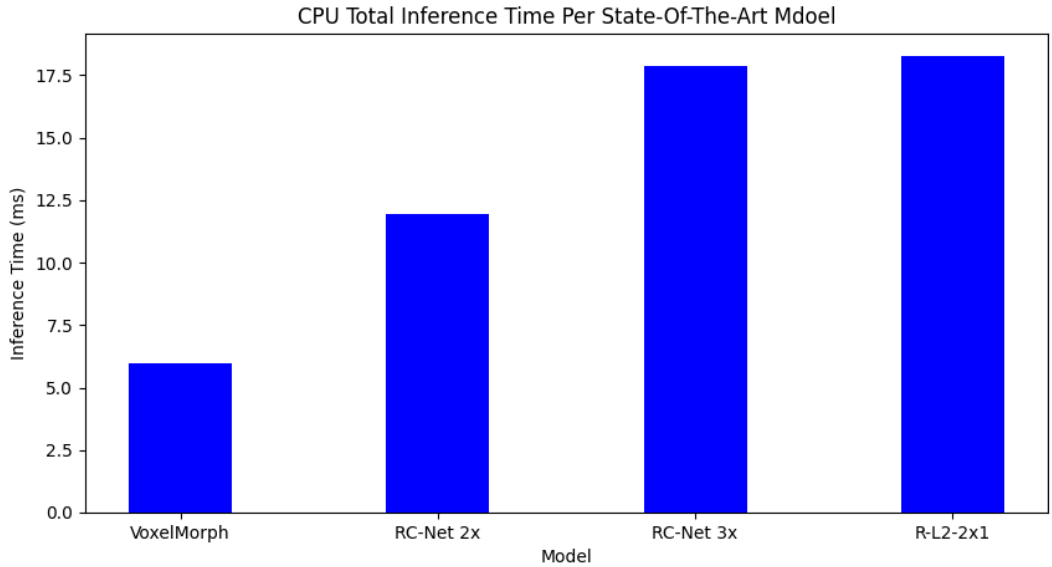

Figure 4.8: Inference times of state-of-the-art models with various cascades compared to VoxelMorph models on CPU.

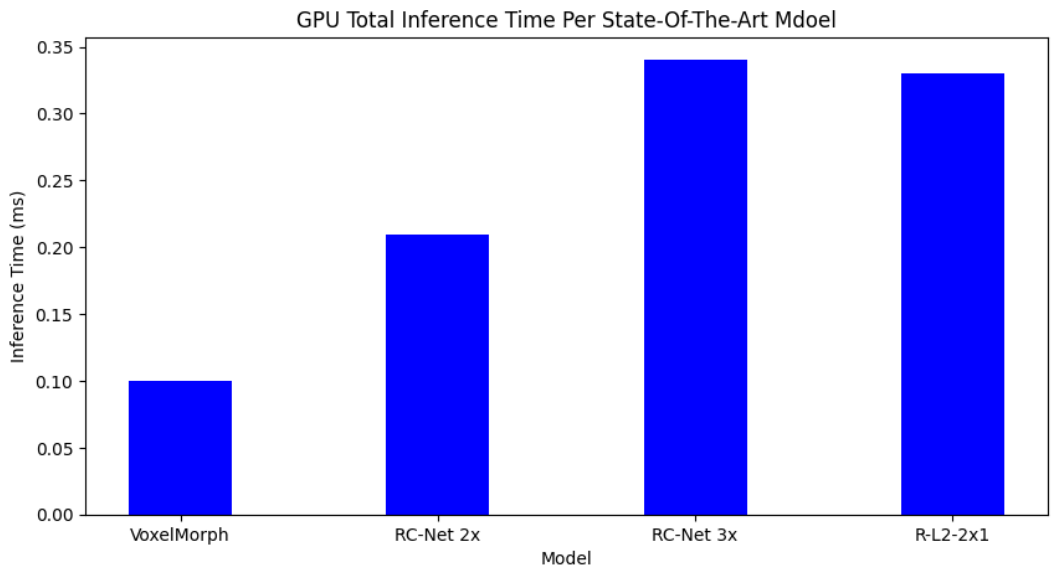

Figure 4.9: Inference times of state-of-the-art models with various cascades compared to VoxelMorph models on GPU.

In both of these tables, the VoxelMorph Lite model is shown to be faster than the original VoxelMorph implementation, with the 1xFLIR model inference times being on par with VoxelMorph Lite, illustrating the effect of the efficient U-Net design. These tables also illustrate an almost linear relationship with the number of FLIR cascades and inference times on all the different hardware configurations used. A graphical illustration of the inference time measurements in Table 4.2 and Table 4.3 is given for CPU, GPU 1, and GPU 2 inference times in Figures 4.5, 4.6, and 4.7, respectively.

To compare the evaluation of the FLIR model to state-of-the-art models, the dice scores of the state-of-the-art models are shown in Tables 4.4 and 4.5, with their inference times given in Table 4.6.

### 4.2.3 FLIR Analysis

The FLIR model with only a single cascade shows a lower dice score than VoxelMorph in Table 4.1 but it performs inference faster than both the VoxelMorph and VoxelMorph-Lite models as shown in Table 4.2 and Table 4.3. This is likely because the final full-resolution layer of the FLIR model is designed to have 16 channels whereas VoxelMorph has 32 channels in its final full-resolution layers as shown in Figure 3.1 and Figure 2.12, respectively. By comparing the inference times of the VoxelMorph model to VoxelMorph Lite in Table 4.2 and Table 4.3, the full resolution layers are shown to contribute significantly to the overall runtime by adding to the amount of time spent performing convolution. Starting with 2 cascades the FLIR model is shown to give better dice score results than VoxelMorph which gradually improves as more cascades are added. While the variations of the FLIR model with 2 and more cascades are slightly slower than the VoxelMorph model, the relative slowdowns should be compared to the slowdowns reported by the other state-of-the-art models. Analysis of the RC-Net shown in Table 4.4 shows that the RC-Net model with 2 cascades and 3 cascades are respectively 2.0 times slower and 3.0 times slower than VoxelMorph on CPU, and 2.1 times slower and 3.4 times slower on GPU. Similarly, the VR-Net model with 2 cascades shown in Table 4.5 also runs 3.1 times slower than VoxelMorph on CPU, and 3.3 times slower on GPU. On CPU, the 2xFLIR, 3xFLIR, and 5xFLIR models show a slowdown of 1.3, 1.7, and 3.1. On GPU1, the 2xFLIR, 3xFLIR, and 5xFLIR models show a slowdown of 1.3 times, 1.7 times, and 3.1 times, respectively on GPU 1. Using the compiled models, the slowdowns of the reported models are 1.3, 1.8, and 3.1 for the 2x, 3x, and 5x models on GPU 1. on GPU 2, the 2xFLIR, 3xFLIR, and 5xFLIR models show a slowdown of 1.3, 1.9, and 7.0 respectively. The relative slowdowns and change in LV dice score of each model over their relative versions of VoxelMorph are shown in Figures 4.10 and 4.11 for CPU and GPU, respectively.

This comparison shows that the FLIR model, while not being strictly faster than VoxelMorph, can achieve an increase in dice score similar to other state-of-the-art models while adding less to inference times shown by the RC-Net and VR-Net models. The 5xFLIR model shows that it is actually able to achieve an improvement over VoxelMorph greater than the state of the art, but shows the largest slowdown in inference time on CPU. Visualization of the changes in inference times are shown in Figures 4.5, 4.6, 4.7, 4.8, and 4.9 for the FLIR on CPU, GPU 1, GPU 2, state-of-the-art models on CPU and GPU, respectively. Analysis of the charts verifies that on CPU and GPU 1 there is a linear increase in inference times as cascades are added to the FLIR model. GPU 2 shows a significant increase in the 5xFLIR model which is likely because GPU 2 is lower end than GPU 1 with less memory to load the entire model with five cascades. Comparing the charts for the FLIR model against the state of the art models on both CPU and GPU, it is shown that even the most lightweight of the compared state-of-the-art models shows double the inference time of VoxelMorph when run on the same hardware.

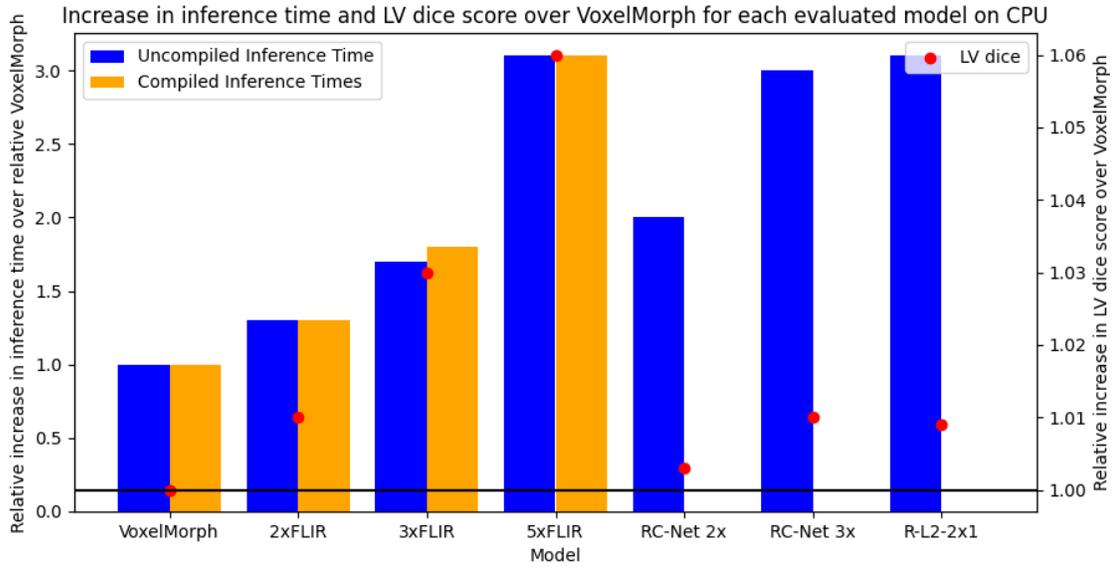

Figure 4.10: Relative slowdown of each model over VoxelMorph on CPU

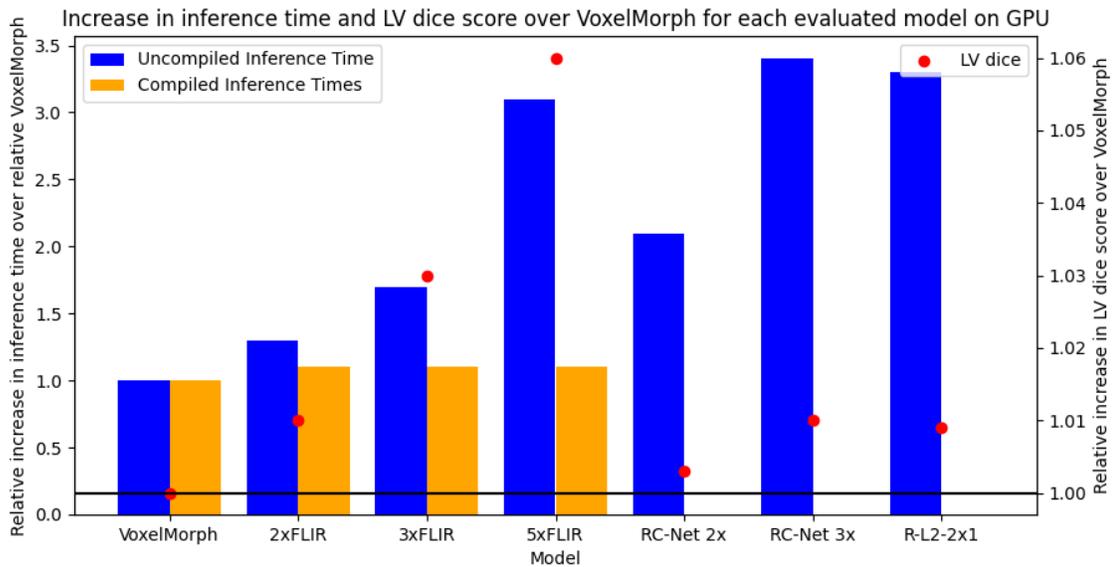

Figure 4.11: Relative slowdown of each model over VoxelMorph on GPU

A visualization of the improvement is shown in Figures 4.12, 4.13, and 4.14 where a slice of a sample volume from a patient study from the test data set is processed using VoxelMorph, 2xFLIR, and the 3xFLIR models. To create this visualization, the flow field was predicted by feeding the moving and fixed image into each of the FLIR and VoxelMorph models. Comparing the fixed image in Figures 4.12, 4.13, and 4.14, we can see that the warped image generated by the FLIR model visually looks more similar to the fixed one than the warped image generated by VoxelMorph. The predicted flow fields from each of these models are also applied to a segmentation of the LV, myocardium, and RV for the fixed image to get warped segmentations. The warped segmentations show a very significant improvement for the FLIR model. The RV (white segmentation) shows that the VoxelMorph model struggles to create a contiguous right ventricle segmentation. This suggests that the deformation predicted by the VoxelMorph model is taking voxels representing tissue outside of the heart and moving them into the heart to represent cardiac tissue. This does not represent a realistic deformation. Another issue with the deformation predicted by VoxelMorph in this example is the rough boundary shown between the myocardium (light grey) and LV (dark grey) segmentations. Looking at the fixed image, the boundary between LV and myocardium does not appear to be smooth but these rough edges likely represent portions of papillary muscles and should be considered to be separate from the myocardium. Moving voxels representing myocardium tissue into the LV to represent papillary muscles also does not represent a realistic deformation. Comparing the issues present in the flow field predicted by VoxelMorph with the deformation predicted by the 2xFLIR and 3xFLIR models, there are still some minor issues but the FLIR model shows significant improvements in the predicted flow field. Here we can see the RV segmentation is much more contiguous than the one predicted by VoxelMorph. There are still a couple of disconnected regions in the segmentation and there appears to be a portion of voxels representing myocardium tissue that has been moved into the RV. The study chosen for the evaluation did show a large portion of myocardium that appears to have been moved to the inside of the LV blood pool, but the rest of the myocardium segmentation appears to be very high quality. Overall though the region of the flow field that predicted the deformation of the RV is far more realistic for the flow field predicted by the FLIR model than the one predicted by VoxelMorph. The boundary between the LV and myocardium is also much smoother for both the 2xFLIR and 3xFLIR models.

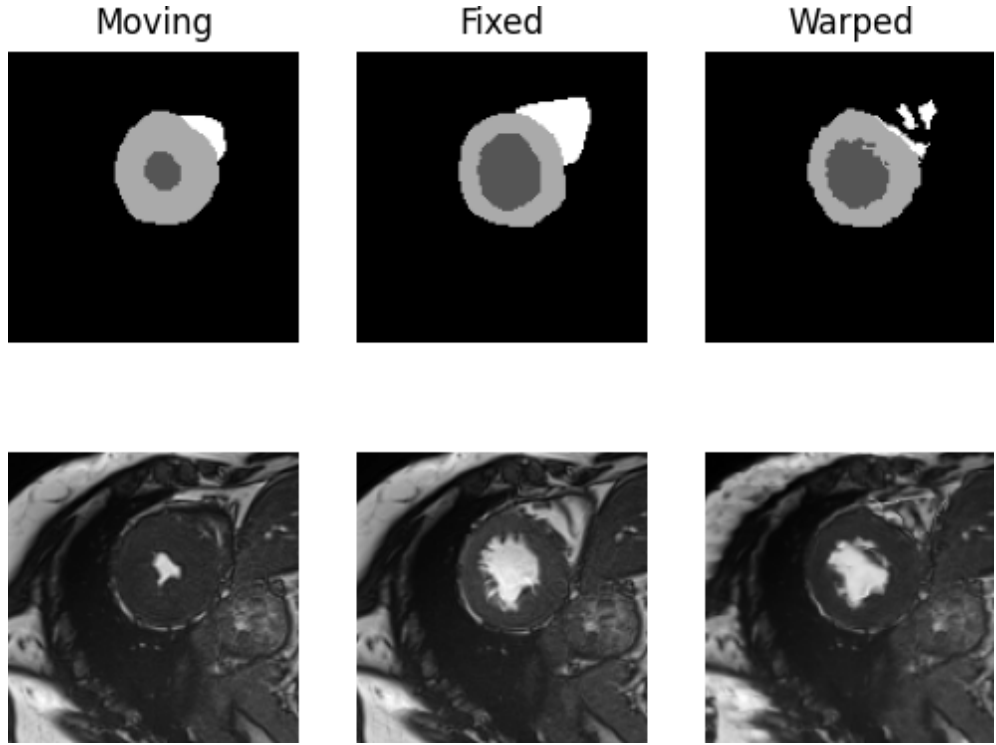

Figure 4.12: Visual comparison of results of the VoxelMorph registration.

## 4.3 Strain Comparison

### 4.3.1 Strain Comparison Methodology

FLIR is shown to be more effective than VoxelMorph for performing registration. Next, the FLIR model will be evaluated on how well its registration can be used for evaluating strain. To perform this comparison, a test-retest dataset from Johannes Gutenburg University Mainz (UMainz) is used where each patient study contains two SSFP acquisitions. Because there are two acquisitions from each patient, it is expected that the strain calculated in each of the acquisitions should be very close. Both acquisitions for each patient study will be fed into the FLIR model. The computed flow fields will be used to calculate the difference in the peak radial, circumferential, and longitudinal strain between both acquisitions in each patient study. For this comparison, the 5xFLIR model was chosen to show its best-case scenario.

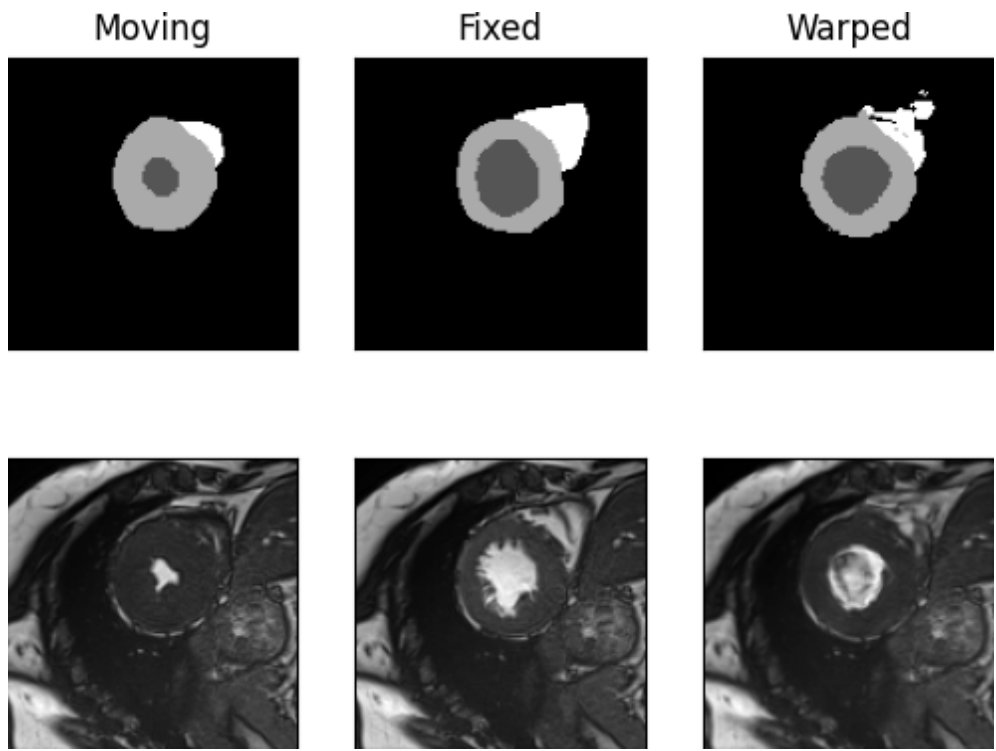

Figure 4.13: Visual comparison of results of the 2xFLIR registration.

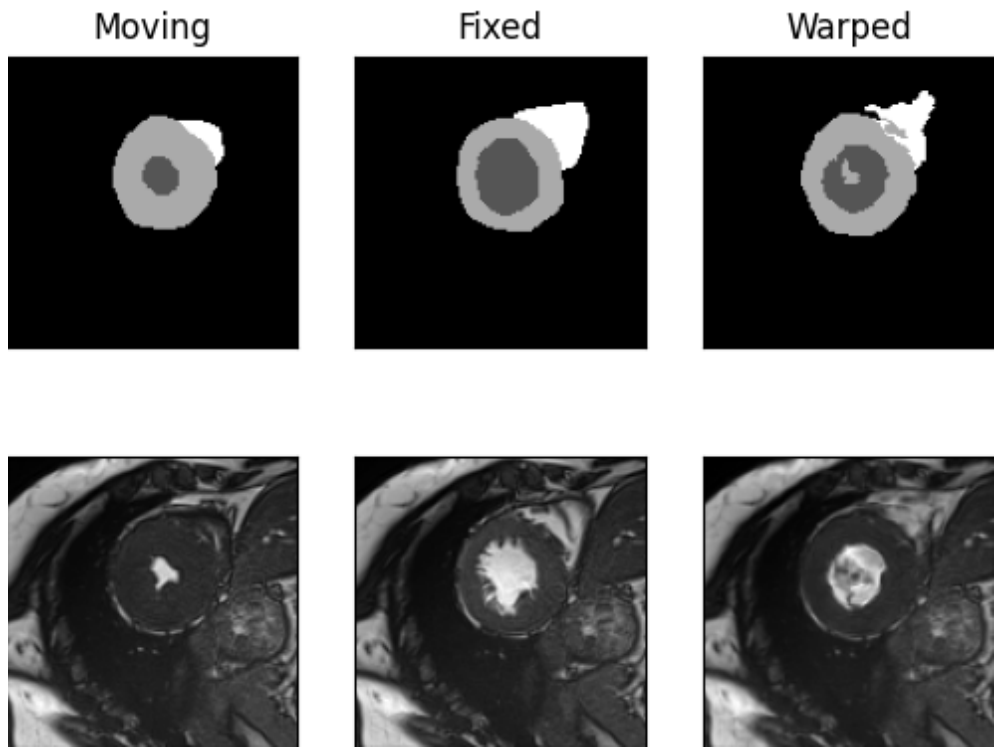

Figure 4.14: Visual comparison of results of the 3xFLIR registration.

For this experiment, the first and last 2 slices of the volumes were removed before calculating the strain values with the FLIR model. This was done because the volumes in the test-retest contain several slices above the most basal point of the myocardium and below the most apical slice of the myocardium. The data used to train the VoxelMorph and FLIR models evaluated here were not trained on volumes with data above basal and below apical, and were found to give poor predicted deformations on these regions. Because the myocardium is the primary region of interest for evaluating strain, the slices above basal and below apical were removed to ensure that the peak strain values measured were coming from the myocardium.

| Patient Study | Peak Radial Strain | Peak Circumferential Strain | Peak Longitudinal Strain |
|---|---|---|---|
| Patient 1 Test | 0.888 | 1.78 | 0.778 |
| Patient 1 Retest | 0.799 | 1.24 | 0.883 |
| Patient 2 Test | 1.79 | 1.70 | 0.802 |
| Patient 2 Retest | 1.34 | 1.85 | 0.804 |
| Patient 3 Test | 1.92 | 1.26 | 1.14 |
| Patient 3 Retest | 1.45 | 1.40 | 0.998 |
| Patient 4 Test | 1.97 | 1.63 | 0.753 |
| Patient 4 Retest | 1.30 | 1.82 | 0.737 |
| Patient 5 Test | 2.53 | 1.81 | 1.15 |
| Patient 5 Retest | 3.06 | 3.04 | 1.10 |
| Patient 6 Test | 1.07 | 0.987 | 1.01 |
| Patient 6 Retest | 1.35 | 1.22 | 0.939 |
| Patient 7 Test | 1.97 | 2.43 | 1.01 |
| Patient 7 Retest | 1.87 | 2.01 | 1.01 |
| Patient 8 Test | 1.74 | 2.75 | 0.957 |
| Patient 8 Retest | 2.26 | 3.20 | 0.949 |
| Patient 9 Test | 1.30 | 1.27 | 0.673 |
| Patient 9 Retest | 1.16 | 1.36 | 0.700 |
| Patient 10 Test | 0.763 | 0.932 | 0.818 |
| Patient 10 Retest | 0.763 | 0.932 | 0.818 |

Table 4.7: 5xFLIR Strain Results

| Patient Study Study | Peak Radial Strain Difference | Peak Circumferential Strain Difference | Peak Longitudinal Strain Difference |
|---|---|---|---|
| Patient 1 | 0.089 | 0.540 | 0.005 |
| Patient 2 | 0.450 | 0.150 | 0.002 |
| Patient 3 | 0.530 | 0.140 | 0.142 |
| Patient 4 | 0.670 | 0.190 | 0.016 |
| Patient 5 | 0.530 | 1.23 | 0.050 |
| Patient 6 | 0.280 | 0.233 | 0.071 |
| Patient 7 | 0.100 | 0.420 | 0.000 |
| Patient 8 | 0.520 | 0.450 | 0.008 |
| Patient 9 | 0.140 | 0.090 | 0.027 |
| Patient 10 | 0.000 | 0.000 | 0.000 |

Table 4.8: 5xFLIR Strain Differences

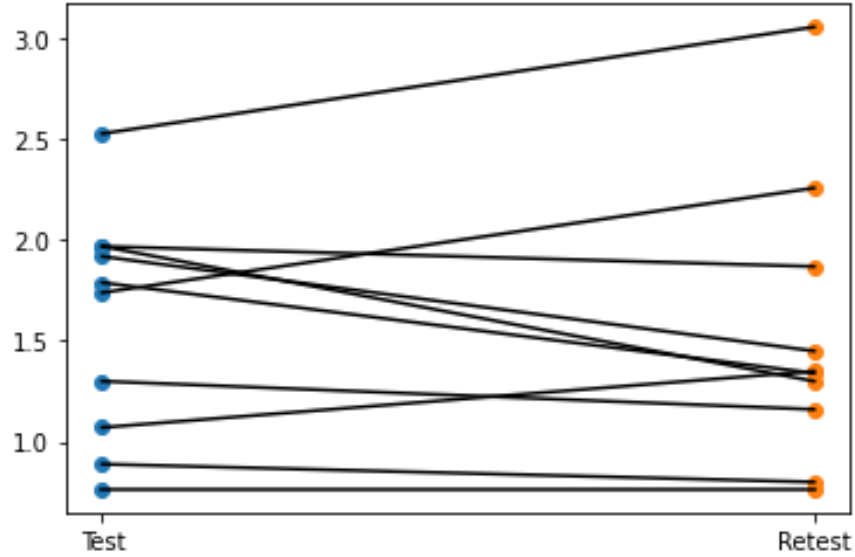

Figure 4.15: Quantified radial strain values paired between test and retest for each patient.

### 4.3.2 Strain Analysis

Recorded strain results for the 5xFLIR model are shown in Table 4.7. These tables show the raw results recorded for the test and retest acquisitions for each patient. Visualizations for the paired test and retest data for radial, circumferential, and longitudinal strain are shown in Figures 4.15, 4.16, 4.17, respectively. The agreement between each of the radial, circumferential, and longitudinal strains between the datasets is illustrated as Bland-Altman plots in Figures 4.18, 4.19, and 4.20, respectively.

In these illustrations, the raw test and retest strain values are plotted where the paired test and retest values from each patient study are connected with a black line. In these figures, a smaller slope in the black line signifies better registration performance because there's less difference between the test and retest values. The model is validated based on its ability to consistently quantify strain between the test and retest acquisitions. Because it is the same patient in each acquisition and the acquisitions are taken shortly after each other, the difference between the test and retest images should be small.

The measured strain values are all fairly consistent between the test and retest acquisitions. The average differences with standard deviations are given in Table 4.9. Here it is shown that the average difference is very small for all of the radial, circumferential, and longitudinal strains. The longitudinal strains show the most consistent values, with the lowest average difference. Circumferential strain shows the largest average difference. The large difference can be attributed to Patient 5 who shows a difference in peak circumferential strain that is much larger than the other patients; Patient 5 may have just been a uniquely challenging image.

| model | Average Radial Strain Difference | Average Circumferential Strain Difference | Average Longitudinal Strain Difference |
|---|---|---|---|
| FLIR | $0.332 \pm 0.236$ | $0.344 \pm 0.356$ | $0.032 \pm 0.045$ |

Table 4.9: Average difference in radial, circumferential, and longitudinal strain between each of the evaluated methods.

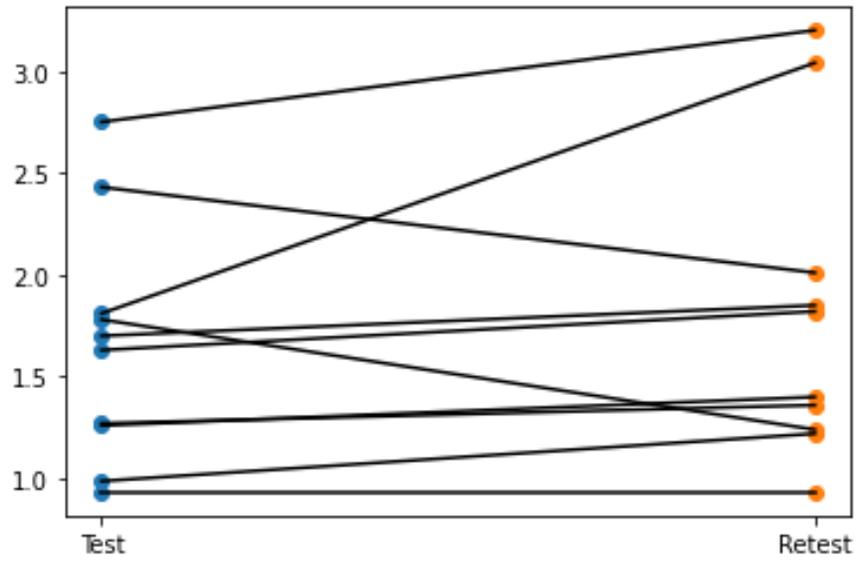

Figure 4.16: Quantified circumferential strain values paired between test and retest for each patient.

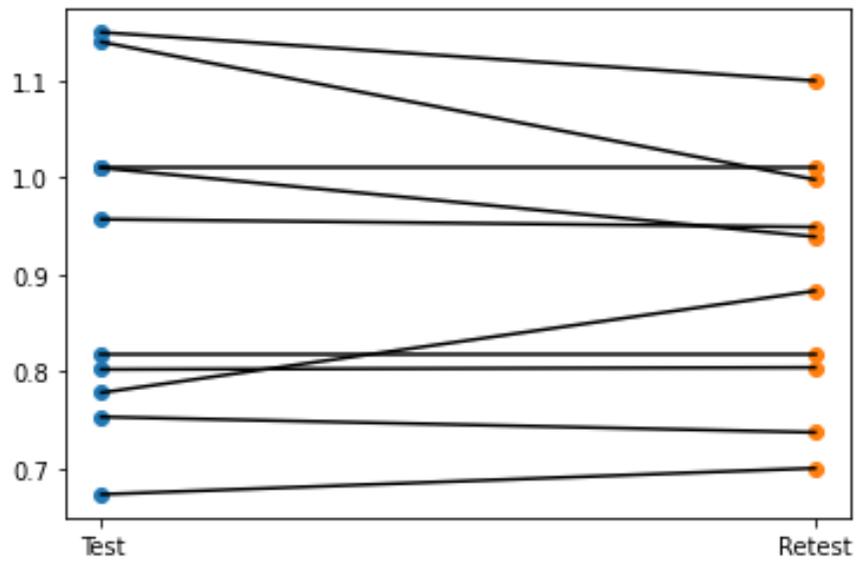

Figure 4.17: Quantified longitudinal strain values paired between test and retest for each patient.

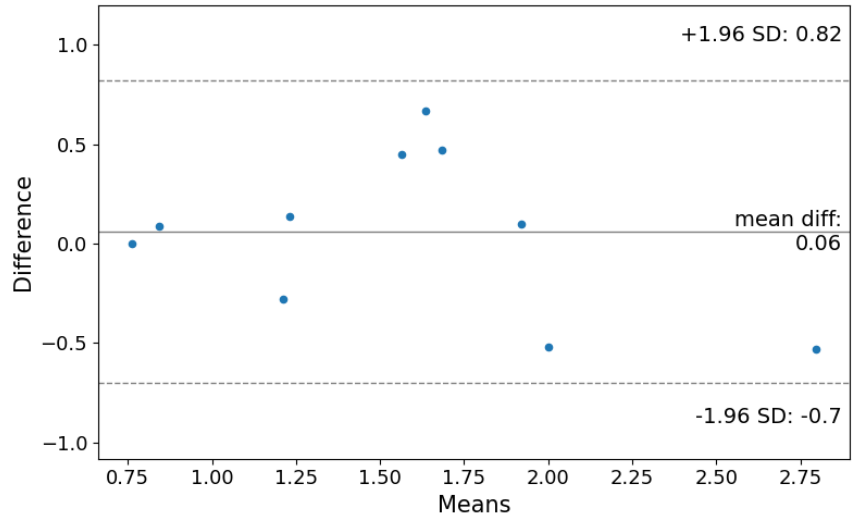

Figure 4.18: Bland-Altman plot of the agreement between radial strains

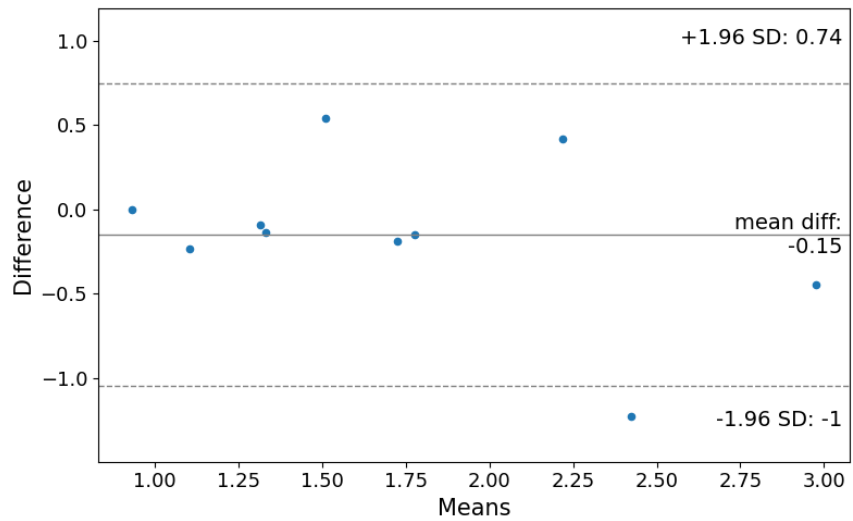

Figure 4.19: Bland-Altman plot of the agreement between circumferential strains

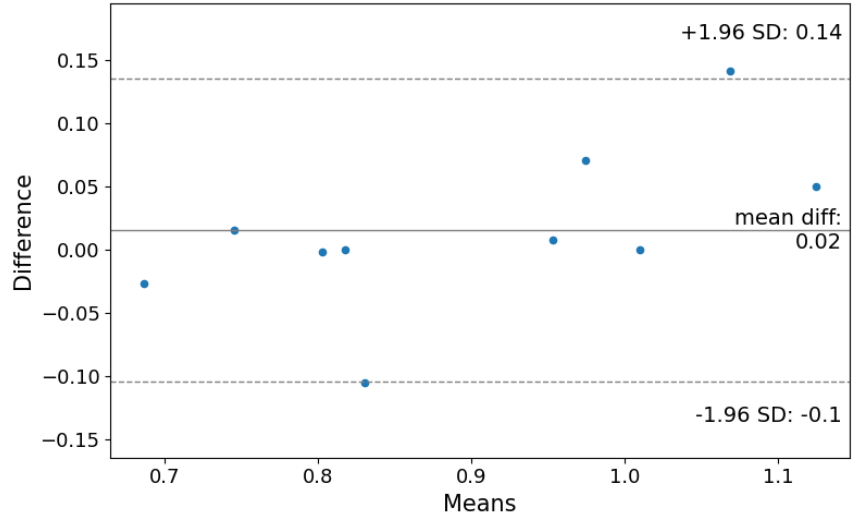

Figure 4.20: Bland-Altman plot of the agreement between longitudinal strains

Recall that when the heart goes from systole to diastole, the circumference and length of the myocardium increase, but the myocardium thickness decreases which would mean that we expect the circumferential and longitudinal strain to be positive and the radial strain should be negative. With the values obtained in this experiment shown in Table 4.7, the peak radial strain values are measured as positive. To assess this discrepancy, the entire strain maps generated with the 5xFLIR model are visualized. Visualizations of the strain maps for radial, circumferential, and longitudinal strain and the slice location, where the peak strain was measured, are shown in Appendix B. These radial, circumferential, and longitudinal strain maps for all of the test and retest images for each patient were visualized to assess why the sign of the measured peak radial strain was different than what was originally expected. In all of the strain maps, large positive strain values are shown in red and negative strains are shown in dark blue according to the color scale shown with each strain map. For the radial strain of all 10 patients, there are large positive strain values in the blood pool of the left ventricle while the strain measured in the myocardium is mostly negative, which is what would be expected. In some cases such as Patient 3 as shown in Figure B.3, where there is some positive strain measured in the myocardium, the myocardium is still shown to have the largest component being negative. The absolute values of the negative strains measured in the myocardium are slightly less than the large positive strain values measured in the blood pool which is why the peak values recorded in Table 4.7 are positive. To more accurately assess the radial strain values for the myocardium, the comparison of strain values between the test and retest acquisitions was performed again, but only recording the peak negative strain instead of the absolute peak; results are shown in Table 4.10, with the absolute differences between these values shown in Table 4.11.

The results in Table 4.7 show positive values for the circumferential and longitudinal strain which would be expected for a healthy patient since when the heart goes from systole to diastole, the circumference increases and the length from the base to the apex of the left ventricle increases. However, some of these cases may not be healthy so the strain maps should still be analyzed visually. Visual validation of the circumferential and longitudinal strain maps for each of the patient test and retest acquisitions in the test data set was performed with the outputs for circumferential strain shown in Figures B.11, B.12, B.13, B.14, B.15, B.16, B.17, B.18, B.19, B.20 and the longitudinal strain values shown in Figures B.21, B.22, B.23, B.24, B.25, B.26, B.27, B.28, B.29, B.30. Similar to the radial strain maps, all of these show very high strain values inside the blood pool, but in the myocardium itself, we can see mostly positive strain values for each patient study shown. Positive circumferential and longitudinal strain values in the myocardium with this method are normal and identify a patient as healthy. Take note that for most cases the range of strain values recorded starts at approximately 0, so most cases show longitudinal strain values that are generally positive. In contrast to most other patients, Patient 3 and Patient 6 shown in Figures B.23 and B.26, show much larger negative values, approximately -0.4 for both cases, recorded in the myocardium which signifies unhealthy tissue. Recall as well that because of the alignment of the myocardium fibers, scar and dead tissue will tend to create abnormal longitudinal strain values before abnormal circumferential and radial values, which is consistent with the assessment of the strain maps.

As shown in this analysis, the peak values for recorded strain given with the FLIR model are affected by very large positive values recorded in the myocardium, which does limit the viability of the proposed FLIR method if no segmentations are used to mask out the myocardium as a region of interest and if only the peak values are used for analysis. Realistically, however, in a clinical or research environment the entire strain map would be evaluated, where a user would see that the large positive strain values are coming from the myocardium. Hence, they would limit their analysis to just the values in the myocardium. If a user desired to evaluate strain maps by only recording the peak strain values, then it would be recommended to generate or apply a segmentation of the myocardium after performing registration and then only measure strain values from within the segmentation.

| Patient Study | Peak Negative Radial Strain |
|---|---|
| Patient 1 Test | -0.427 |
| Patient 1 Retest | -0.408 |
| Patient 2 Test | -0.496 |
| Patient 2 Retest | -0.471 |
| Patient 3 Test | -0.492 |
| Patient 3 Retest | 0.441 |
| Patient 4 Test | -0.396 |
| Patient 4 Retest | -0.470 |
| Patient 5 Test | -0.494 |
| Patient 5 Retest | -0.497 |
| Patient 6 Test | -0.423 |
| Patient 6 Retest | -0.462 |
| Patient 7 Test | -0.487 |
| Patient 7 Retest | -0.486 |
| Patient 8 Test | -0.449 |
| Patient 8 Retest | -0.466 |
| Patient 9 Test | -0.438 |
| Patient 9 Retest | -0.480 |
| Patient 10 Test | -0.492 |
| Patient 10 Retest | -0.492 |

Table 4.10: FLIR radial strain results measuring only the most negative values

| Patient Study | Peak Negative Radial Strain |
|---|---|
| Patient 1 Test | 0.019 |
| Patient 2 Test | 0.025 |
| Patient 3 Test | 0.051 |
| Patient 4 Test | 0.074 |
| Patient 5 Test | 0.003 |
| Patient 6 Test | 0.039 |
| Patient 7 Test | 0.001 |
| Patient 8 Test | 0.017 |
| Patient 9 Test | 0.042 |
| Patient 10 Test | 0.000 |
| Average | $0.027 \pm 0.024$ |

Table 4.11: 5xFLIR radial strain differences using only the most negative values

## 4.4 Chapter 4 Summary

In this chapter, the parameters of the experiment for evaluating the FLIR model were described. In addition to this a lightweight VoxelMorph implementation, called VoxelMorph-Lite, was designed to further illustrate the effect full resolution layers have on inference time. In its performance evaluation, the FLIR model was found to run faster than VoxelMorph when only 1 cascade is used, achieving similar inference times to the VoxelMorph-Lite model. Starting with 2 cascades of the FLIR model inference time starts to get gradually slower than VoxelMorph, but registration fidelity measured with the dice score of the left ventricle, myocardium, and right ventricle starts to improve. The performance evaluation shows that it scales very efficiently with additional cascades. The FLIR model with 3 cascades shows a slowdown of 1.7 times on CPU, 1.8 times on GPU 1, and 1.9 times on GPU 2. The 3xFLIR, while being slightly slower than VoxelMorph, achieves a level of inference fidelity very close to the state-of-the-art 2xRC-Net and VR-Net models which are 2.0 times slower and 3.1 times slower on CPU, respectively and 2.1 times slower and 3.3 times slower on GPU respectively.

## Chapter 5

# Conclusion and Future Work

## 5.1 Conclusion

This thesis presents a novel deep learning-based method for 3D CMR registration that can be effectively used for strain quantification. This method is shown to perform registration with a similar level of fidelity to other state-of-the-art methods while taking a fraction of the time to run on a high-end CPU. The contributions of this thesis are summarized as the following:

1. A 3D registration deep learning model called Fast and Lightweight Image Registration (FLIR) is proposed that is optimized for the task of 3D CMR image registration to register images from a single steady-state free precession (SSFP) sequence, from its most compressed phase to its most dilated. With this method the motion of the heart on a per voxel basis can be extracted, capturing the unique motion of different sections of the heart as it beats. This information can be used with principles of finite strain analysis to quantify non-uniform strain throughout the myocardium tissue which can be used to infer unhealthy tissue. Finding unhealthy tissue with this method would make strain quantification an effective alternative to gadolinium-enhanced imaging.

2. Experimental analysis shows a strong relationship between the number and size of full-resolution layers in a deep learning model and its inference time. By designing a model to use fewer blocks that process the image at a full resolution and instead use lower resolution blocks at the center of a U-Net, the model can be efficiently scaled by cascading several U-Nets together, which improves the fidelity of the model while not significantly adding to the inference time of the model compared to other state-of-the-art models. The effect of cascading smaller, more efficient models together was evaluated by comparing the proposed method to VoxelMorph and evaluating the change in inference time compared to the changes in inference time measured by three state-of-the-art models: the RC-Net with 2 and 3 cascades, and the VR-Net with 2 cascades.

3. The proposed FLIR deep learning method was validated for strain evaluation using a test-retest dataset. This dataset consists of a selection of patients, some with necrosis of cardiac tissue, where two acquisitions were taken a short time apart which implies strain values should be consistent between the test and retest acquisitions. In this evaluation, the proposed FLIR model is shown to quantify strain values consistently.

4. Deformations generated by the proposed model are effective for the quantification of strain. The proposed deep learning model determines patients with abnormal cardiac strain to show abnormal strain primarily in the longitudinal direction, which is consistent with physiological assessments that determine cardiac muscle fibers in the longitudinal direction are generally the first to be affected by necrosis and will show abnormal longitudinal strain values before showing abnormal radial and circumferential strain.

## 5.2 Limitations and Future Works

In this thesis, a deep learning model is developed to quickly predict tissue motion for the quantification of cardiac strain. Because the images being registered are taken from the same acquisition, the assumption is made that there should be negligible movement of the patient between each of the input images to the model. This should be a safe assumption as generally significant motion causes artifacts in MRI images that affect analysis. Because of this assumption, the model was trained and evaluated on data with no significant artifacts or motion present so the performance evaluation might not accurately reflect performance on low-quality data where motion artifacts are present. Additionally, as part of converting quantified strain values from the Cartesian to cylindrical coordinate space, the center of the cylinder needs to be known. In this thesis, it is assumed that the center of the cylinder is at the exact center of the volume for the conversion of coordinate spaces. Another limitation of this method is that using deformations predicted by the FLIR model for strain quantification will give strain values throughout the entire volume, including the background and in the left ventricle blood pool. This thesis showed that strain values calculated this way will be very high in the left ventricle blood pool which will affect the analysis of peak strain values that are taken alone without visual assessment of the strain maps.

From these limitations, future work could be done to improve and further refine the model. Areas for future work are summarized as follows:

1. Inclusion of a segmentation model into the analysis pipeline before performing registration would give an exact center of the left ventricle that could be used to better convert from Cartesian to cylindrical coordinate systems. Adding a segmentation model to the analysis pipeline would increase processing time a bit, which reinforces the need for the registration portion of the pipeline to be as fast as possible to reduce the compounding run times for other stages that might be added to the pipeline.

2. As shown in the comparison of strain values calculated using the FLIR model, the peak strain values are affected by large displacements in the blood pool. Because the myocardium specifically is the region of interest, using a segmentation here as well would allow for the myocardium to be masked so that only the strain values quantified for the myocardium can be analyzed. The CVI42 model shows that using the segmentation as part of the registration model is prone to compounding errors from the segmentation so the best way to integrate segmentations into the analysis pipeline would be to generate a myocardium segmentation before performing registration and then apply the deformation predicted from the FLIR model to the segmentation. From this, the center of the left ventricle can be determined for a more accurate calculation of the cylindrical coordinate space and the mask can be applied to the warped image created from the deformation to isolate strain values from just the myocardium.

In summary, this thesis proposes a method for quantifying strain using 3D CMR registration that generates effective strain maps on a 3D volume on its own and has high potential to be easily integrated into larger analysis pipelines thanks to the efficient design of the FLIR model allowing for high registration fidelity while not posing a bottleneck for a deep learning based pipeline.

# Appendix A

# 3D Strain Calculator Unit Tests

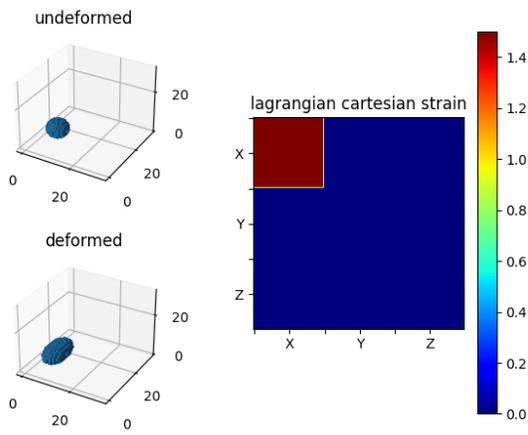

Figure A.1: X uni-axial dilation unit test

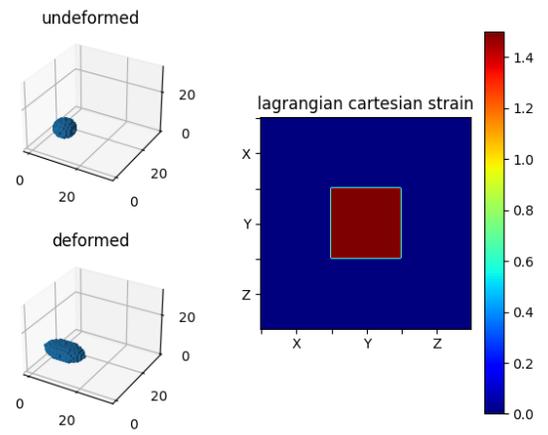

Figure A.2: Y uni-axial dilation unit test

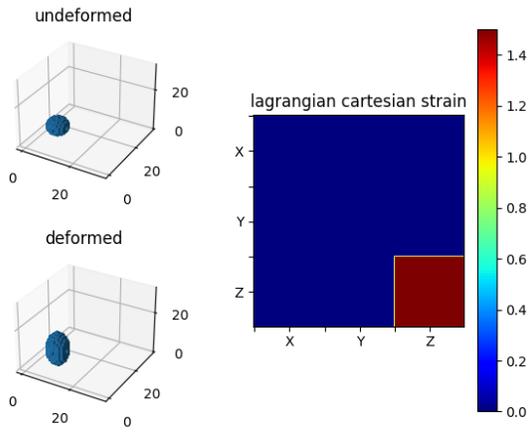

Figure A.3: Z uni-axial dilation unit test

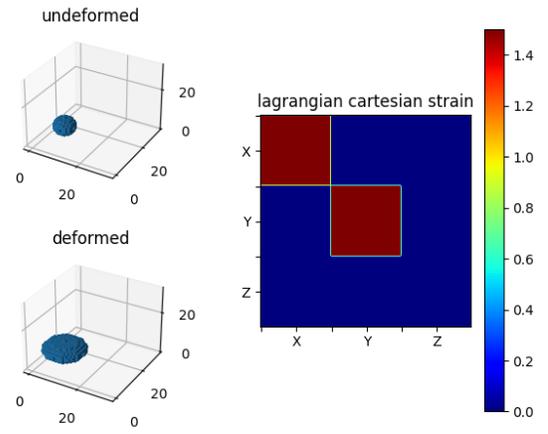

Figure A.4: XY bi-axial dilation unit test

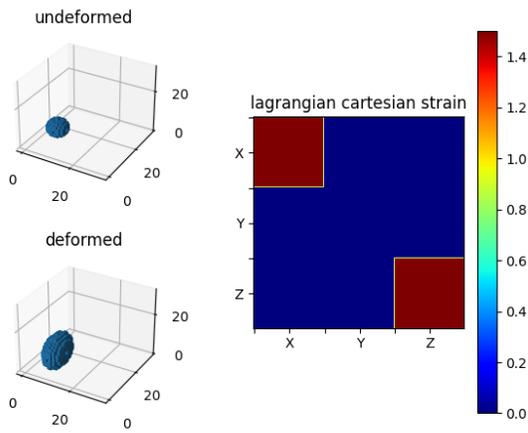

Figure A.5: XZ bi-axial dilation unit test

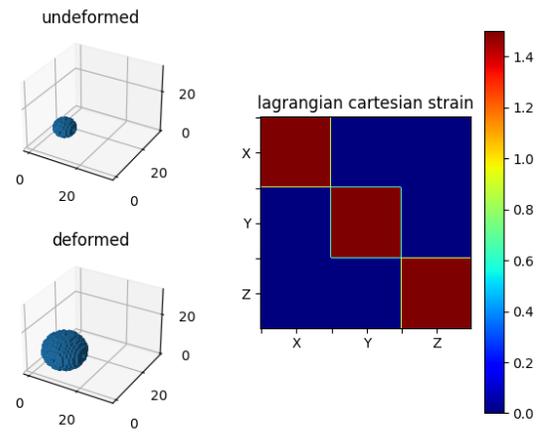

Figure A.6: Tri-axial dilation unit test

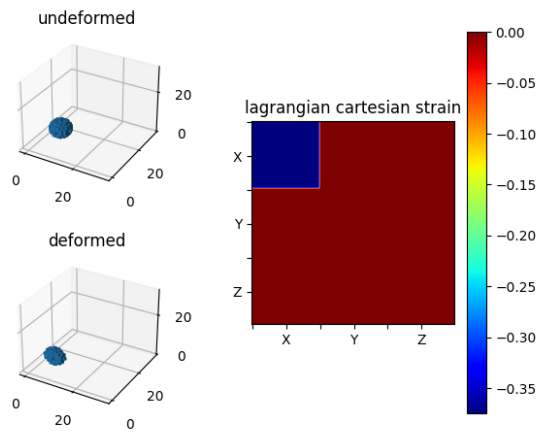

Figure A.7: X uni-axial shortening unit test

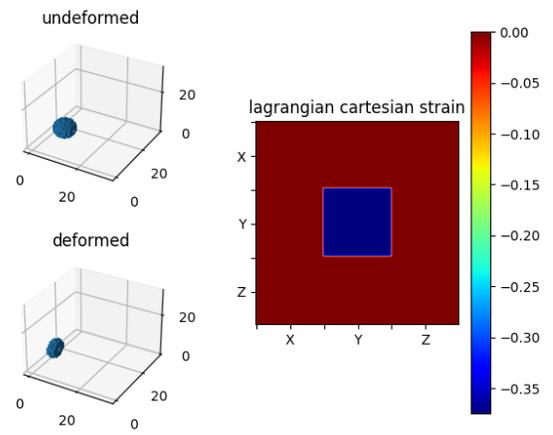

Figure A.8: Y uni-axial shortening unit test

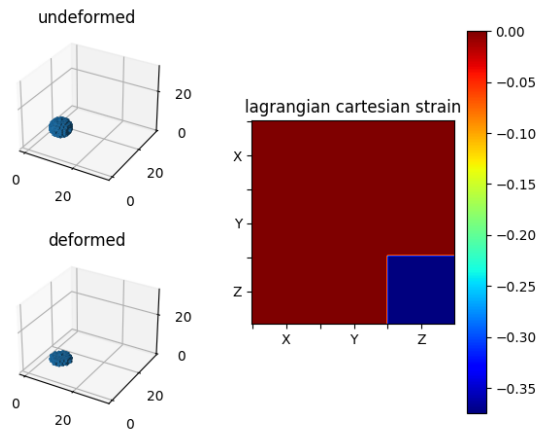

Figure A.9: Z uni-axial shortening unit test

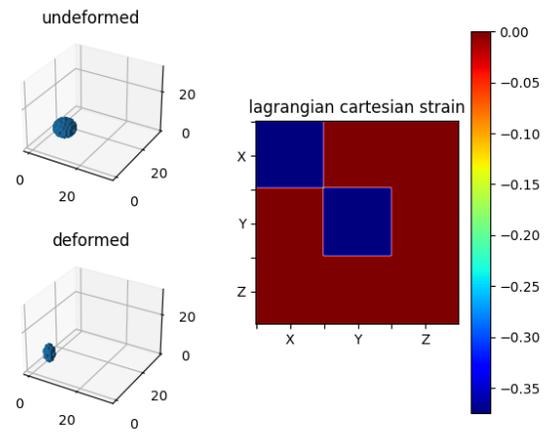

Figure A.10: XY bi-axial shortening unit test

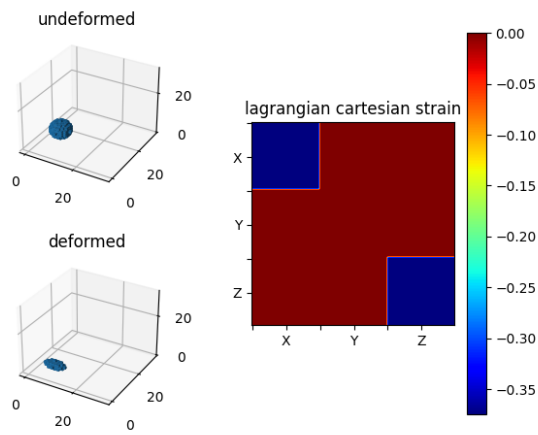

Figure A.11: XZ bi-axial shortening unit test

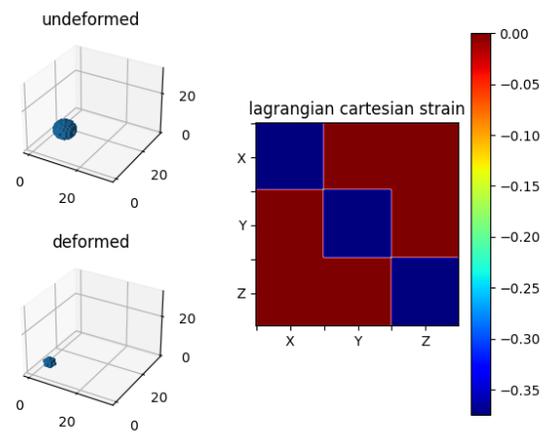

Figure A.12: Tri-axial shortening unit test

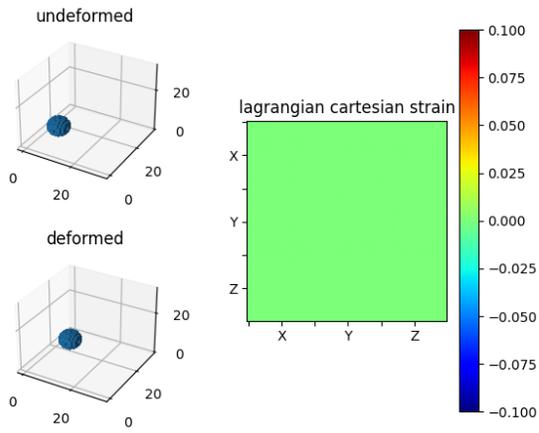

Figure A.13: X translation unit test

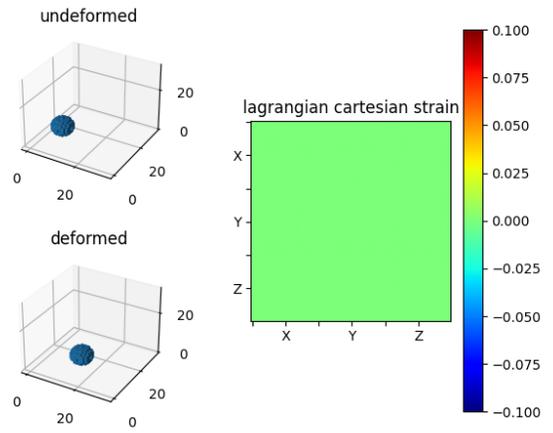

Figure A.14: Y translation unit test

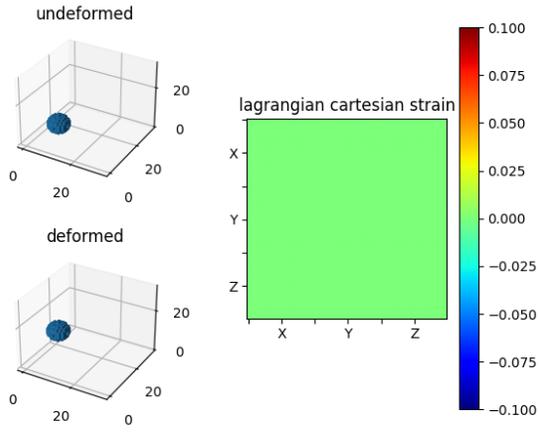

Figure A.15: Z translation unit test

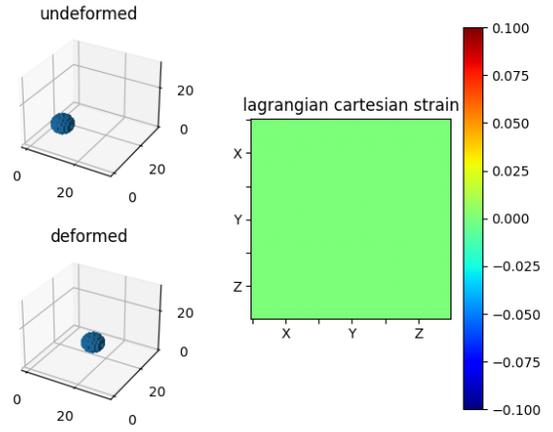

Figure A.16: XY translation unit test

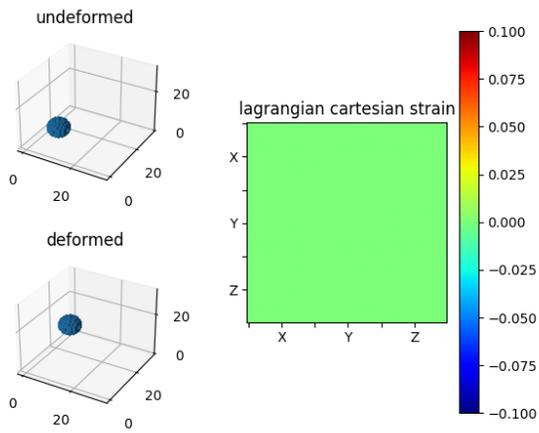

Figure A.17: XZ translation unit test

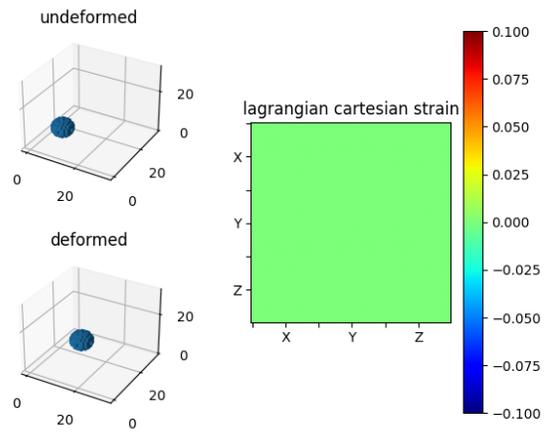

Figure A.18: YZ translation unit test

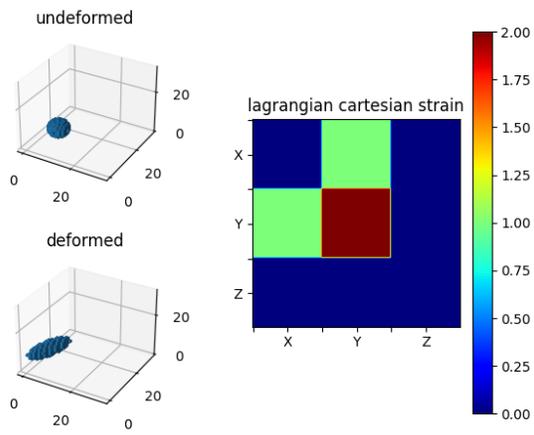

Figure A.19: X shear Y unit test

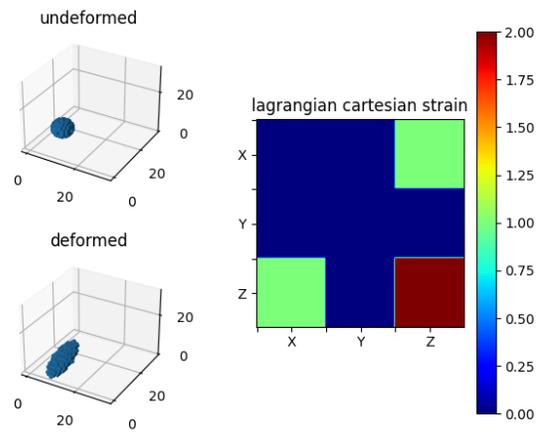

Figure A.20: X shear Z unit test

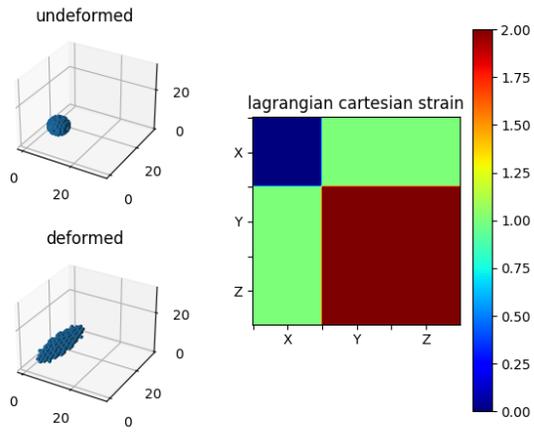

Figure A.21: X shear Y and Z unit test

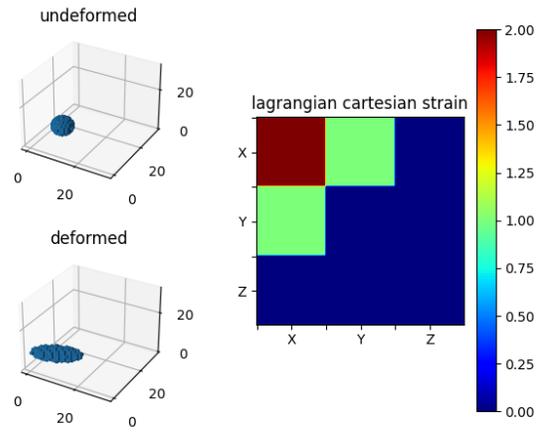

Figure A.22: Y shear X unit test

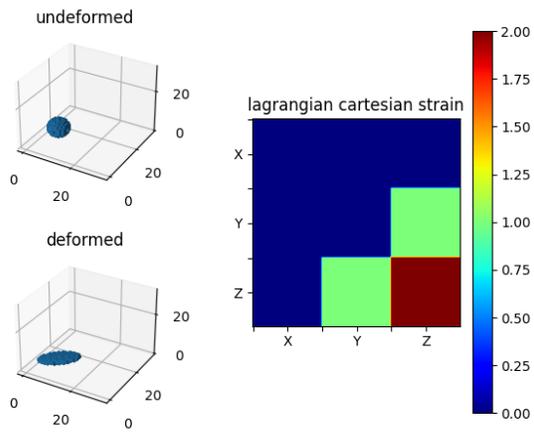

Figure A.23: Y shear Z unit test

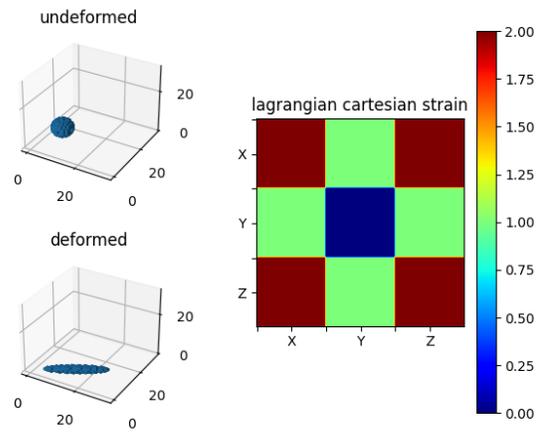

Figure A.24: Y shear X and Z unit test

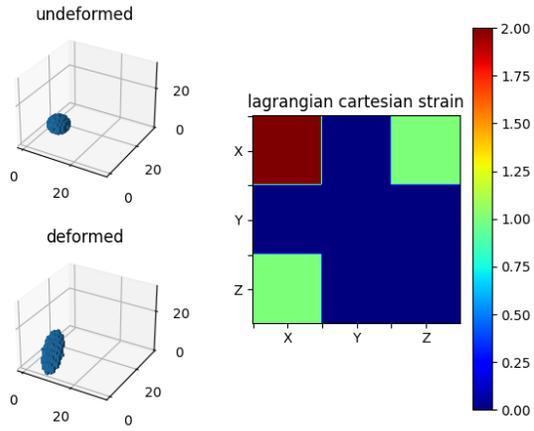
Figure A.25: Z shear X unit test

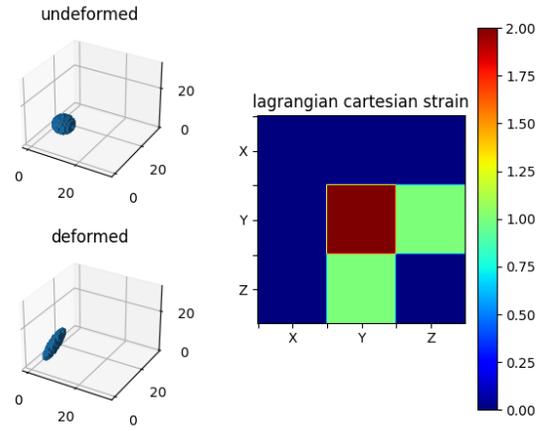
Figure A.26: Z shear Y unit test

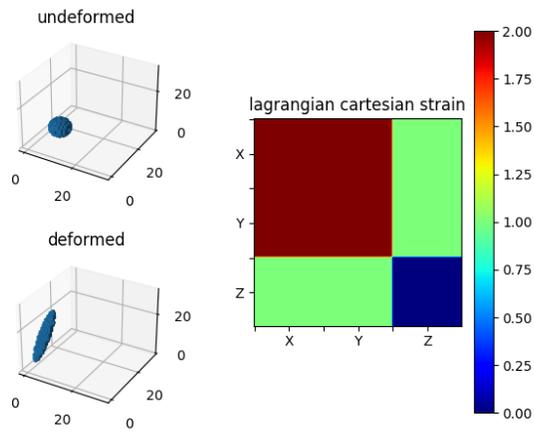
Figure A.27: Z shear X and Y unit test

# Appendix B

# Strain Maps

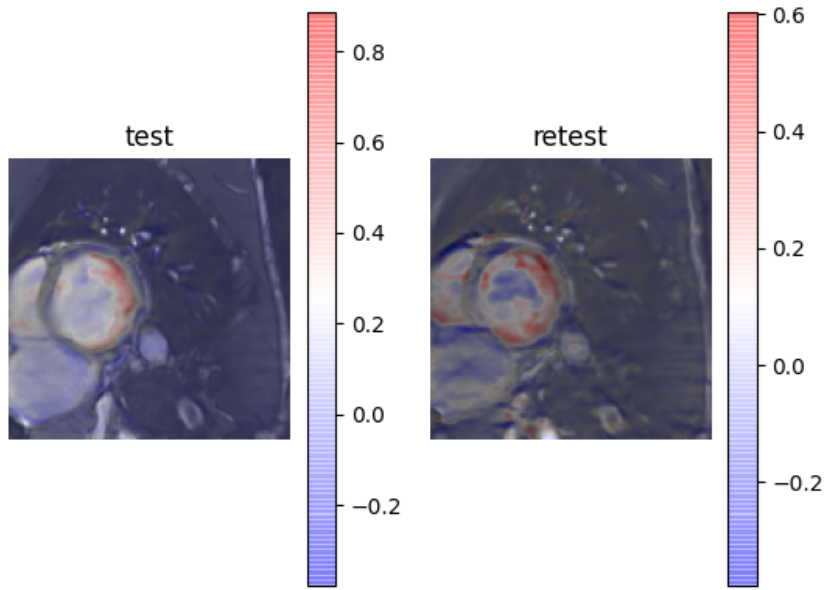

Figure B.1: Patient 1 test and retest acquisition radial strain map

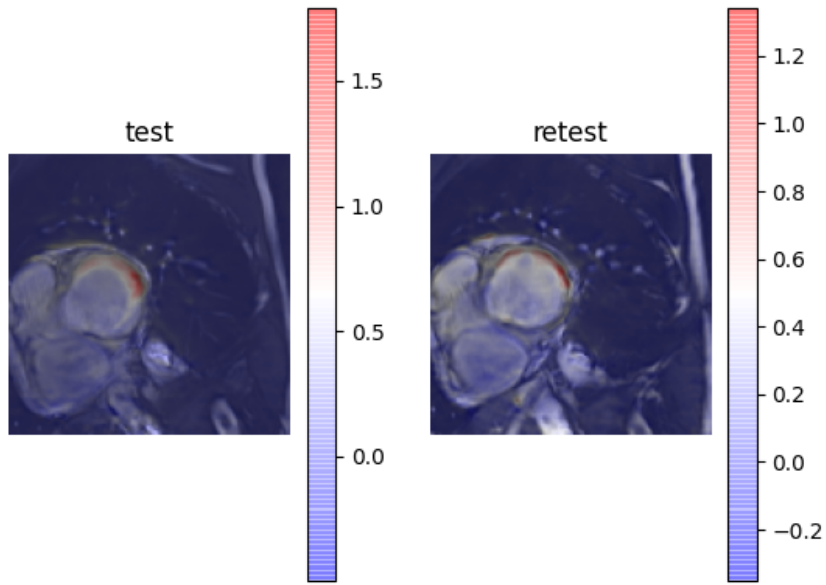

Figure B.2: Patient 2 test and retest acquisition radial strain map

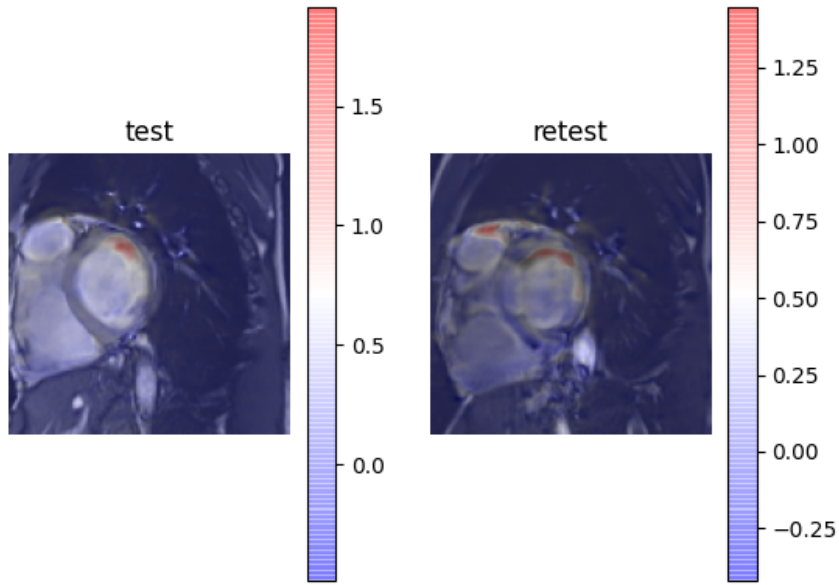

Figure B.3: Patient 3 test and retest acquisition radial strain map

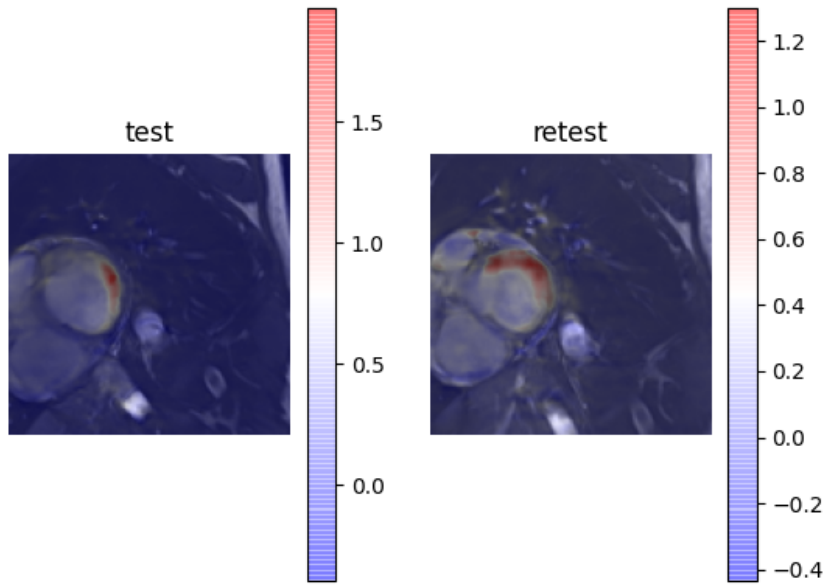

Figure B.4: Patient 4 test and retest acquisition radial strain map

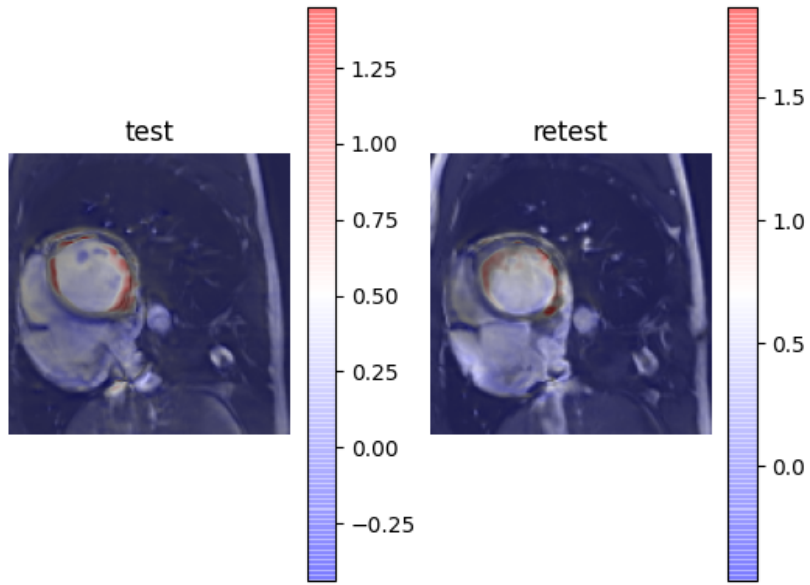

Figure B.5: Patient 5 test and retest acquisition radial strain map

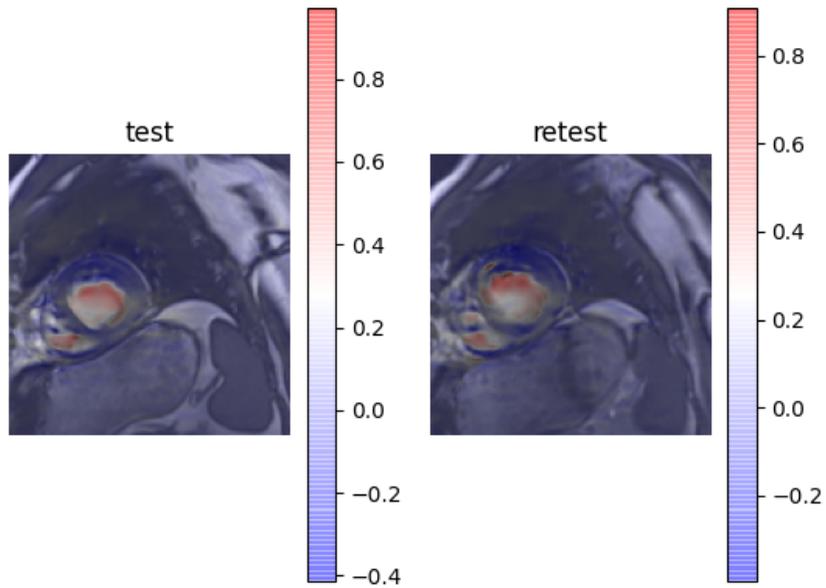

Figure B.6: Patient 6 test and retest acquisition radial strain map

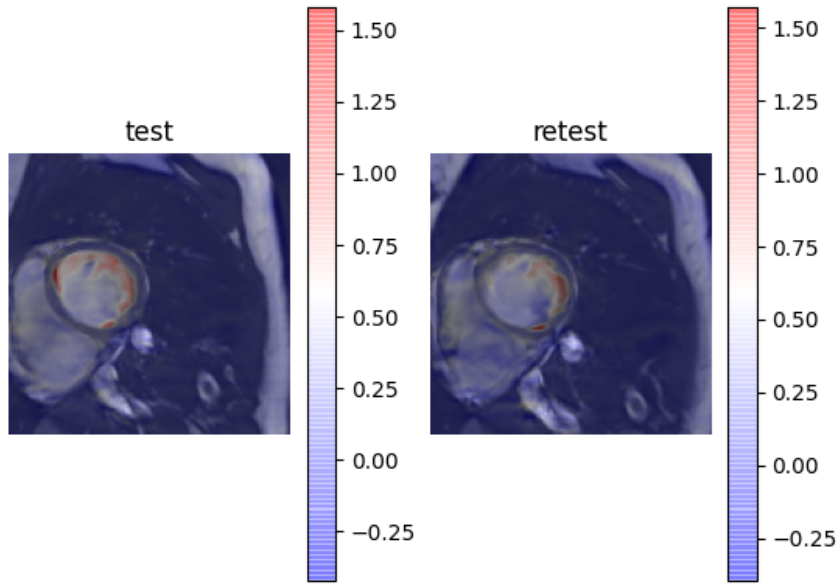

Figure B.7: Patient 7 test and retest acquisition radial strain map

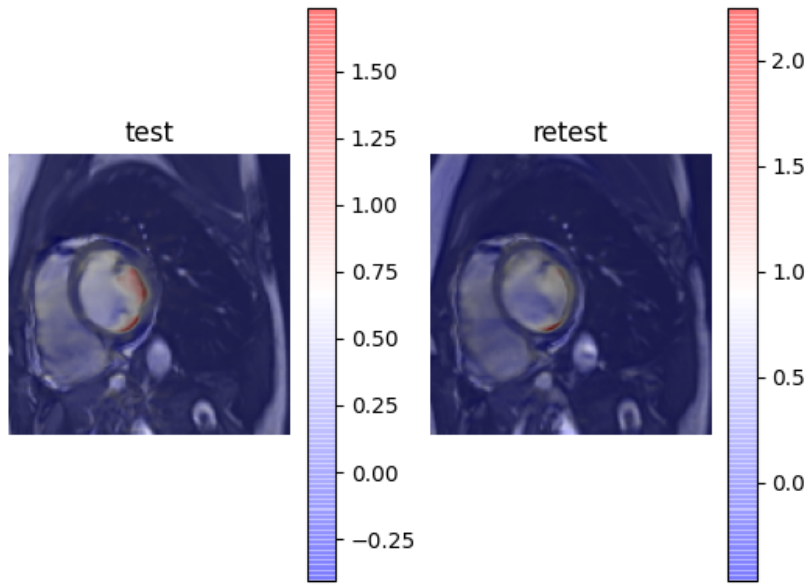

Figure B.8: Patient 8 test and retest acquisition radial strain map

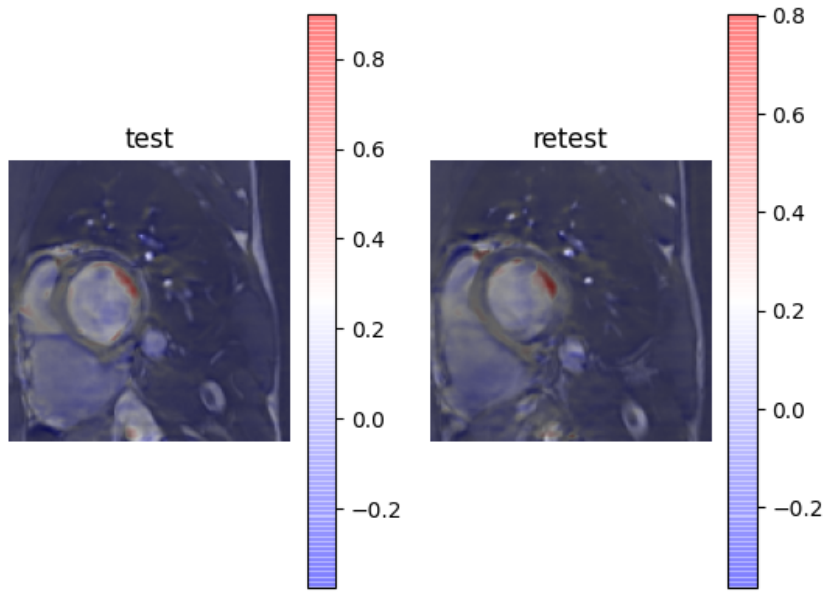

Figure B.9: Patient 9 test and retest acquisition radial strain map

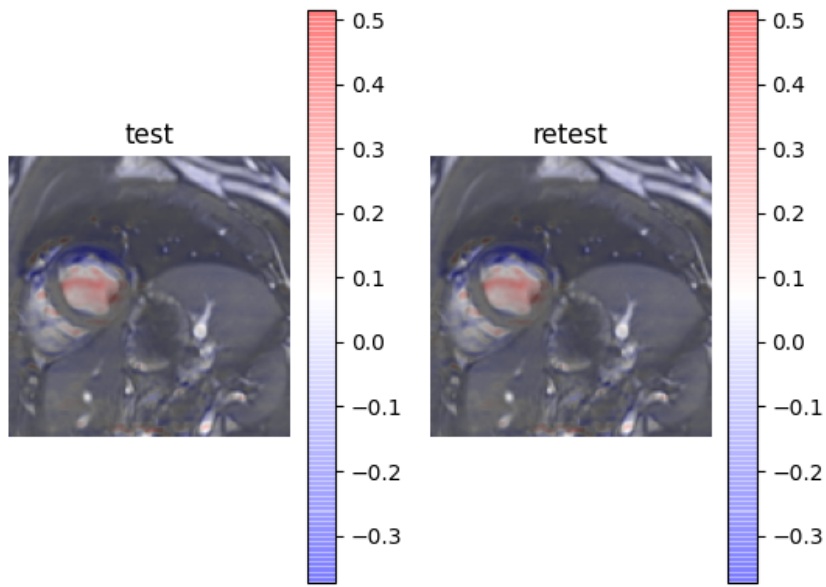

Figure B.10: Patient 10 test and retest acquisition radial strain map

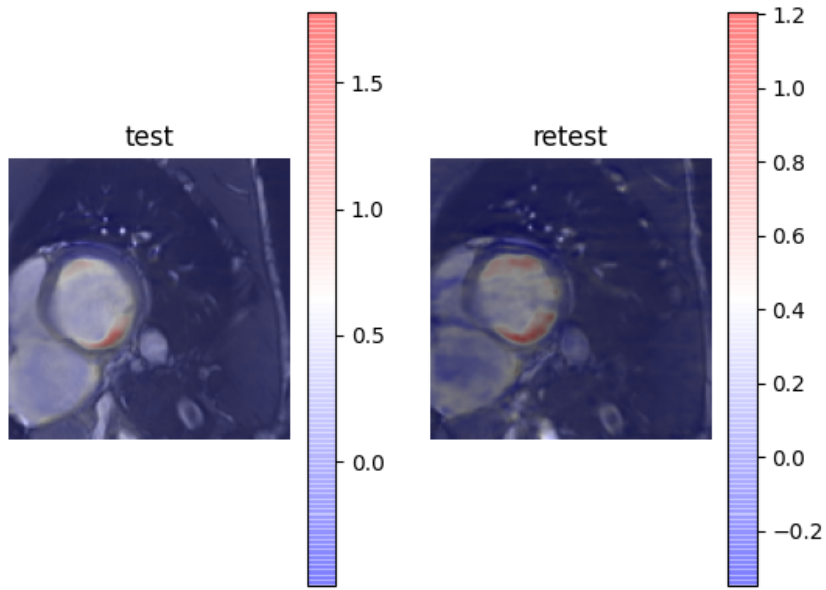

Figure B.11: Patient 1 test and retest acquisition circumferential strain map

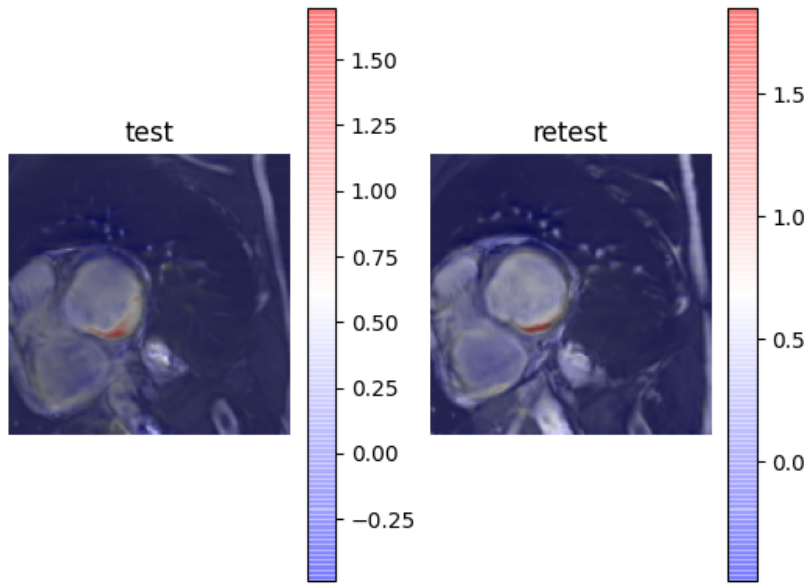

Figure B.12: Patient 2 test and retest acquisition circumferential strain map

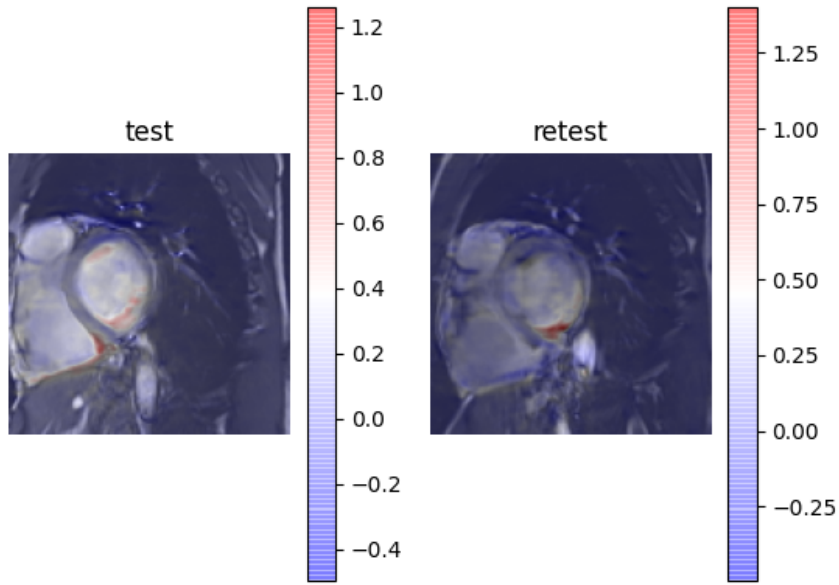

Figure B.13: Patient 3 test and retest acquisition circumferential strain map

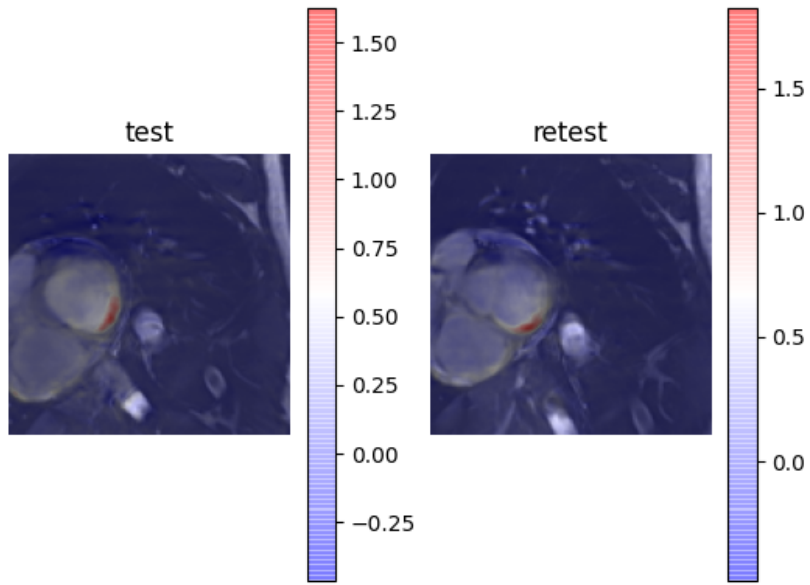

Figure B.14: Patient 4 test and retest acquisition circumferential strain map

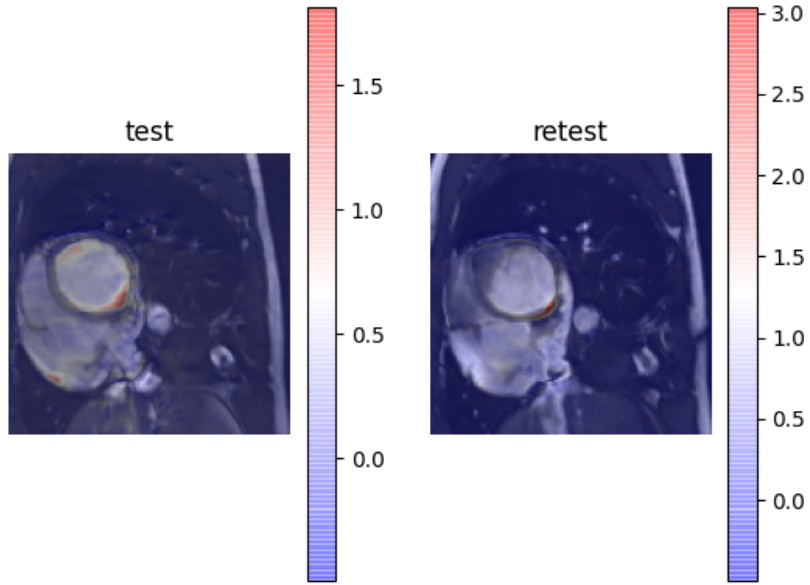

Figure B.15: Patient 5 test and retest acquisition circumferential strain map

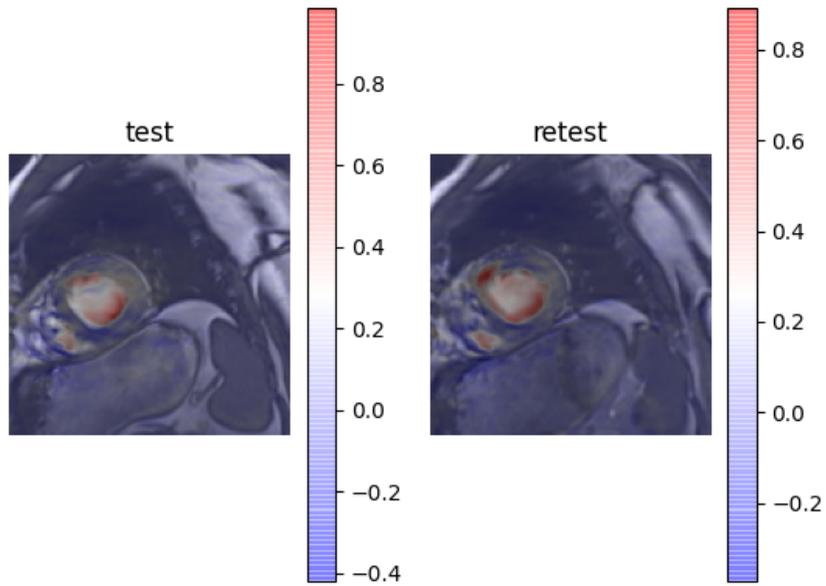

Figure B.16: Patient 6 test and retest acquisition circumferential strain map

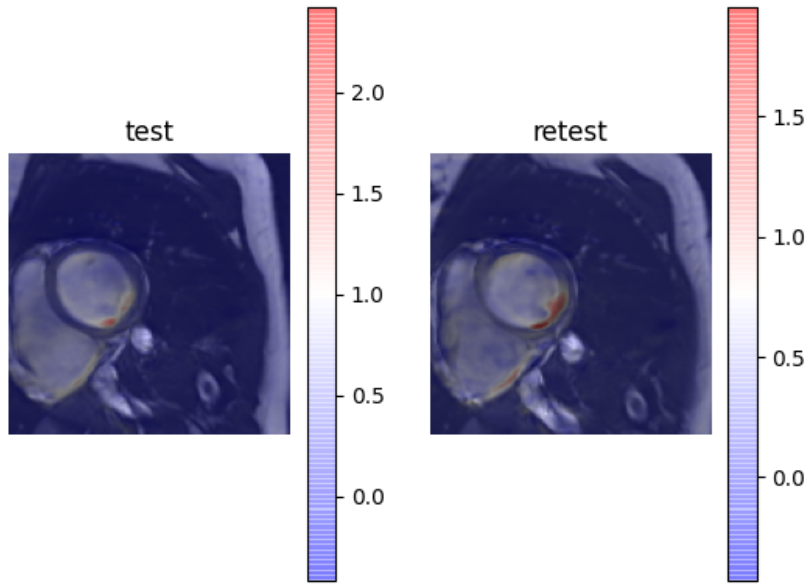

Figure B.17: Patient 7 test and retest acquisition circumferential strain map

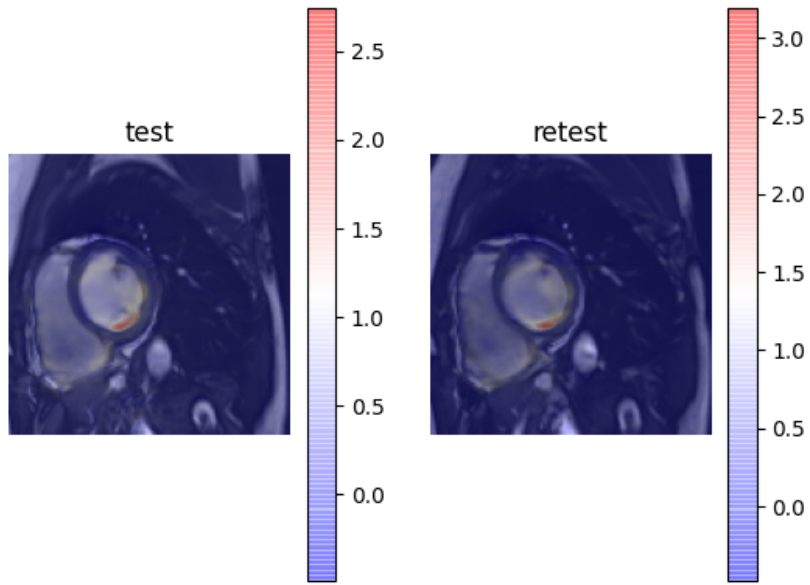

Figure B.18: Patient 8 test and retest acquisition circumferential strain map

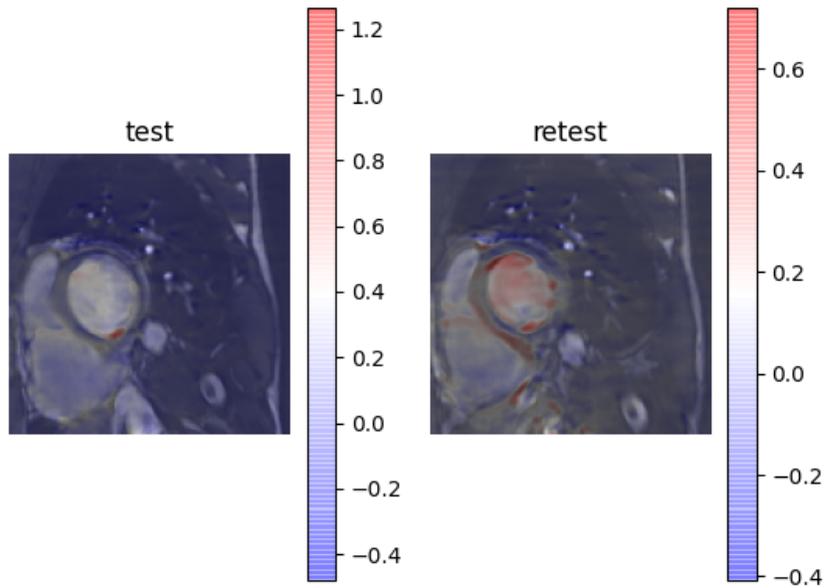

Figure B.19: Patient 9 test and retest acquisition circumferential strain map

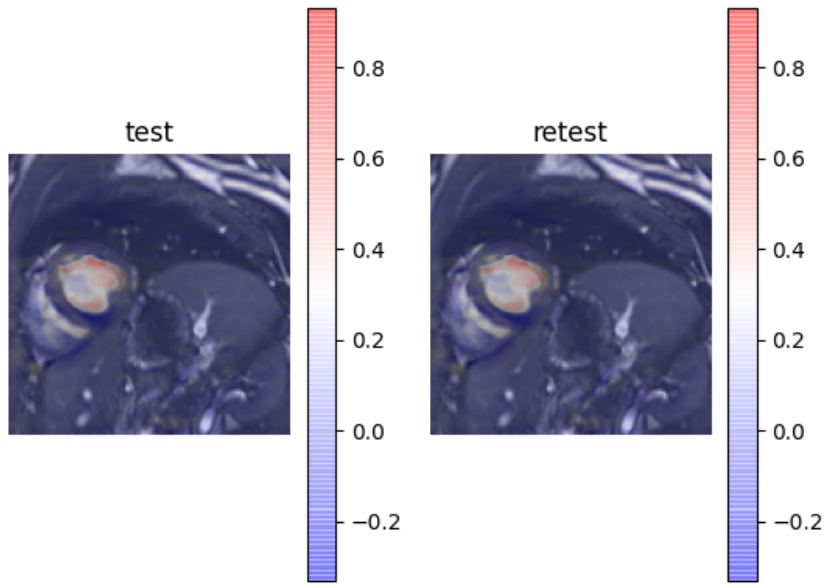

Figure B.20: Patient 10 test and retest acquisition circumferential strain map

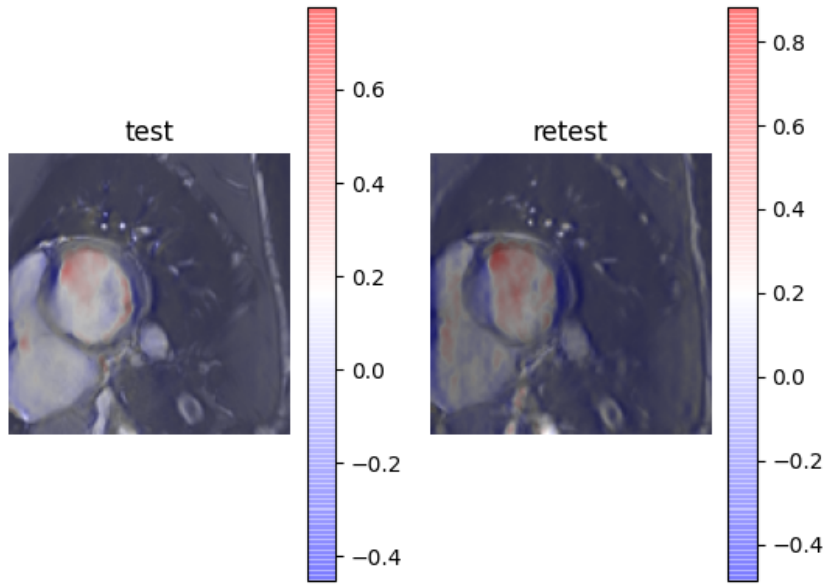

Figure B.21: Patient 1 test and retest acquisition longitudinal strain map

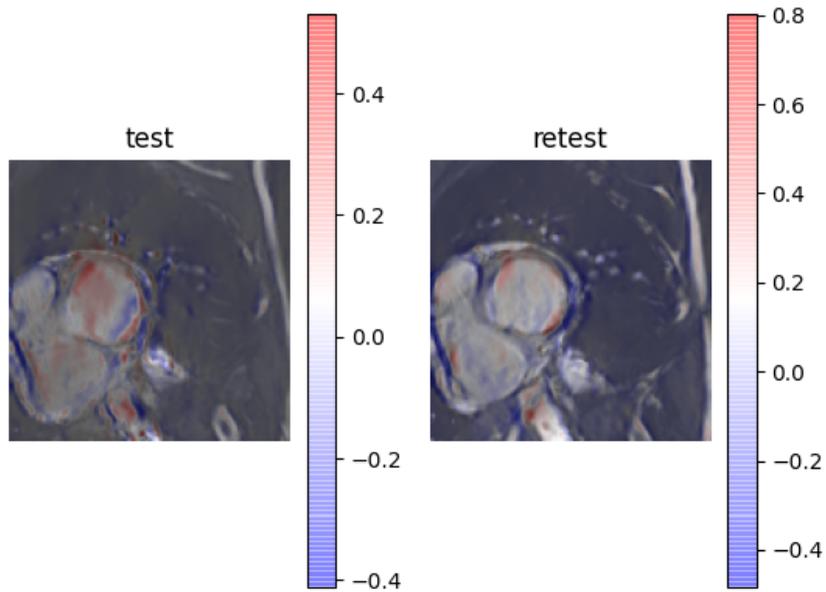

Figure B.22: Patient 2 test and retest acquisition longitudinal strain map

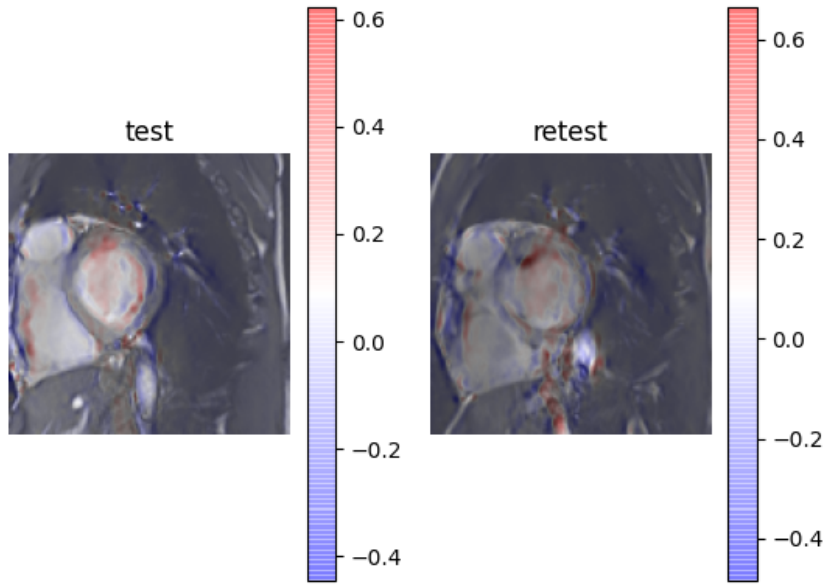

Figure B.23: Patient 3 test and retest acquisition longitudinal strain map

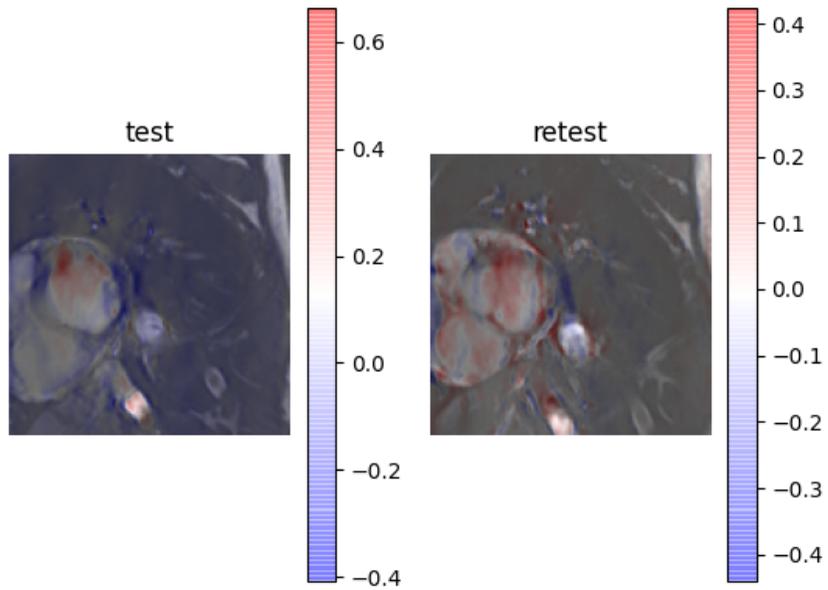

Figure B.24: Patient 4 test and retest acquisition longitudinal strain map

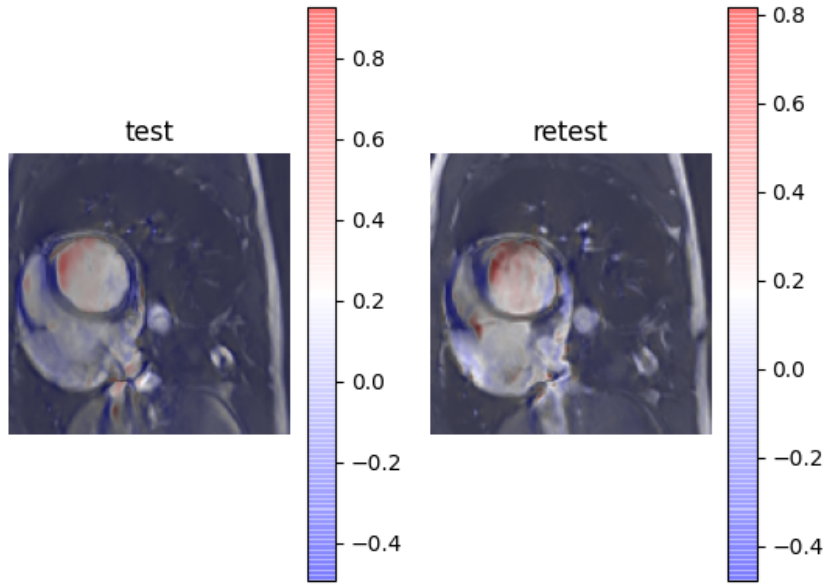

Figure B.25: Patient 5 test and retest acquisition longitudinal strain map

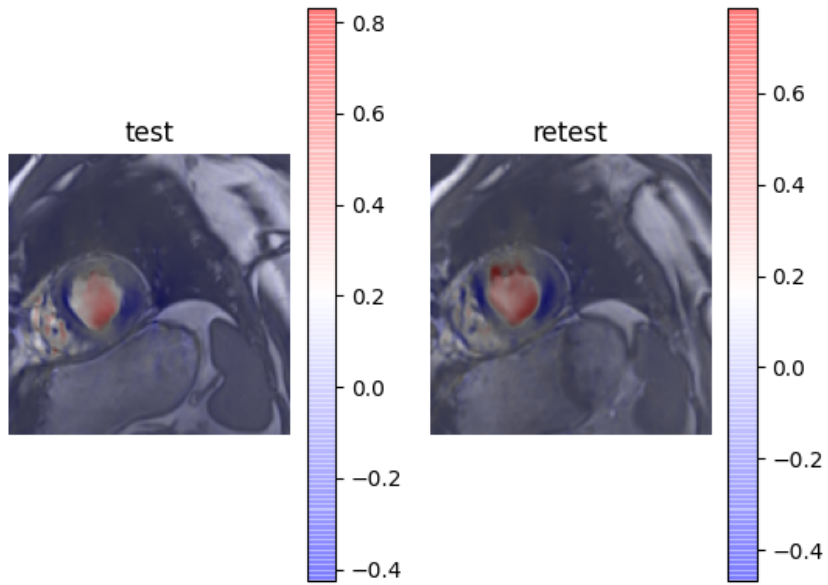

Figure B.26: Patient 6 test and retest acquisition longitudinal strain map

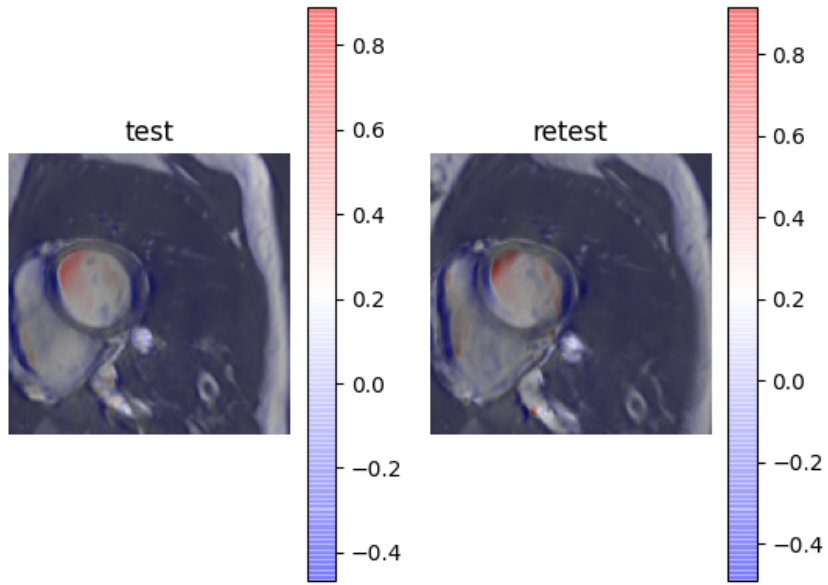

Figure B.27: Patient 7 test and retest acquisition longitudinal strain map

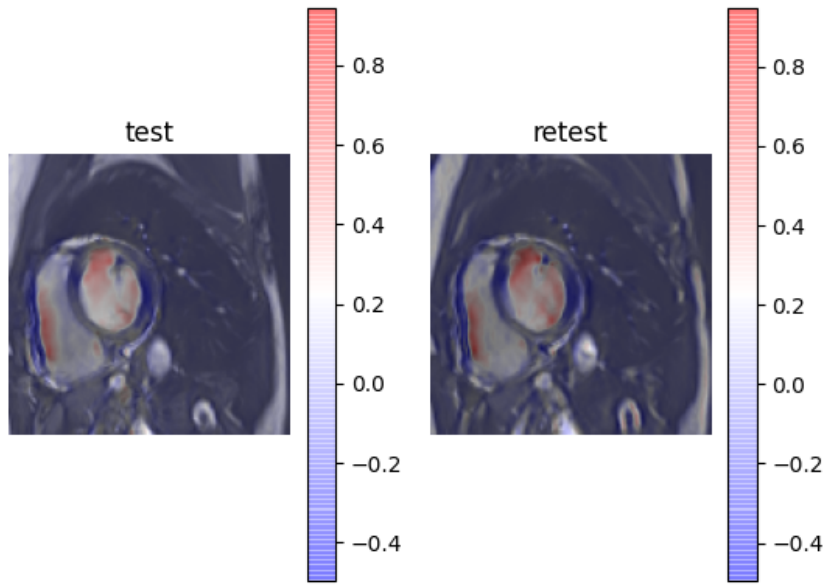

Figure B.28: Patient 8 test and retest acquisition longitudinal strain map

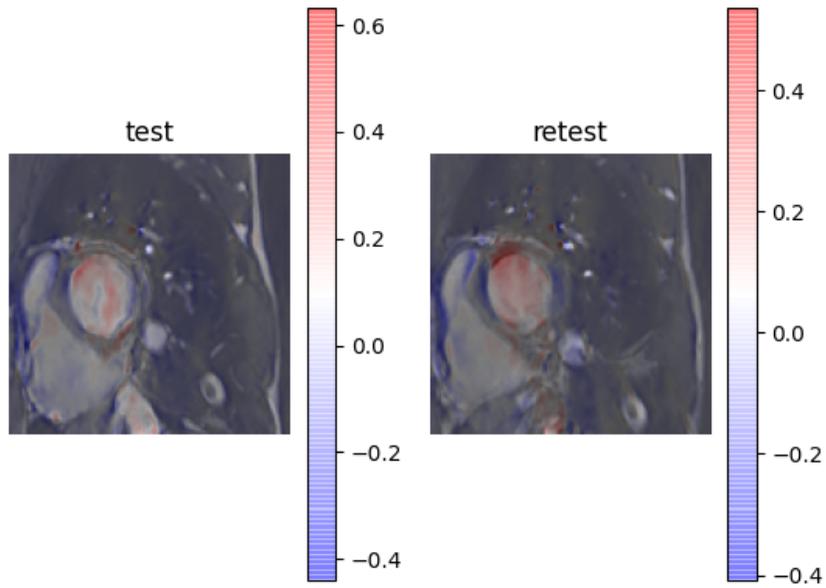

Figure B.29: Patient 9 test and retest acquisition longitudinal strain map

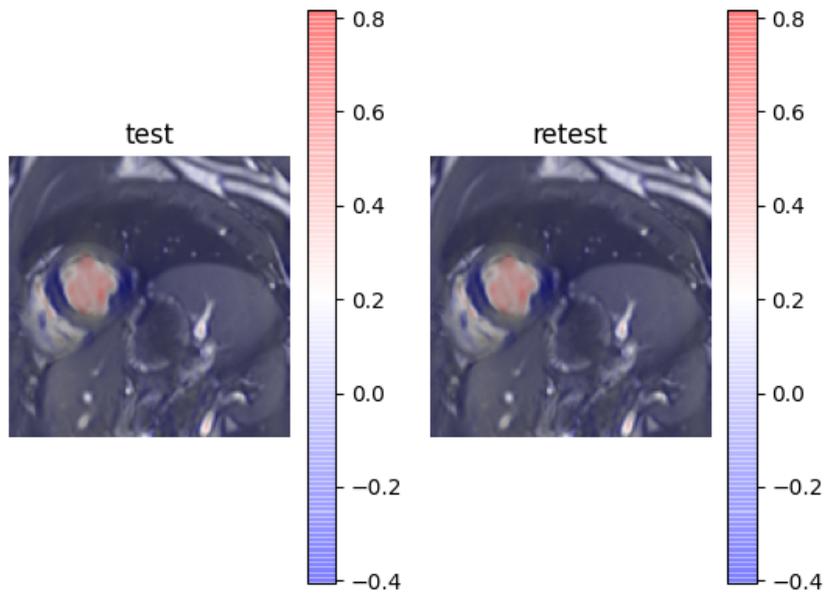

Figure B.30: Patient 10 test and retest acquisition longitudinal strain map